\documentclass[preprint2]{aastex631}
\usepackage{CJKutf8}
\usepackage{color,soul}

\accepted{for publication in The Planetary Science Journal on March 27, 2025}

\begin{document}
\begin{CJK*}{UTF8}{gbsn}

\title{Structure and Melting of Fe, MgO, SiO$_{2}$, and MgSiO$_{3}$ in Planets:\\Database, Inversion, and Phase Diagram}

\correspondingauthor{Junjie Dong}
\email{dong2j@caltech.edu}

\author[0000-0003-1114-9348]{Junjie Dong (董俊杰)}
\altaffiliation{Stanback Postdoctoral Fellow at the Caltech Center for Comparative Planetary Evolution}
\affiliation{Division of Geological and Planetary Sciences, California Institute of Technology, 1200 E California Blvd, Pasadena, CA 91125, USA}
\affiliation{Department of Earth, Planetary, and Space Sciences, University of California Los Angeles, 595 Charles E Young Dr E, Los Angeles, CA 90095, USA}
\affiliation{Department of Earth and Planetary Sciences, Harvard University, 20 Oxford St, Cambridge, MA 02138, USA}
\affiliation{Department of the History of Science, Harvard University, 1 Oxford St, Cambridge, MA 02138, USA}

\author[0000-0003-0932-9905]{Gabriel-Darius Mardaru}
\affiliation{Department of Earth and Planetary Sciences, Harvard University, 20 Oxford St, Cambridge, MA 02138, USA}

\author[0000-0001-6025-8925]{Paul D. Asimow}
\affiliation{Division of Geological and Planetary Sciences, California Institute of Technology, 1200 E California Blvd, Pasadena, CA 91125, USA}

\author[0000-0003-3778-2432]{Lars P. Stixrude}
\affiliation{Department of Earth, Planetary, and Space Sciences, University of California Los Angeles, 595 Charles E Young Dr E, Los Angeles, CA 90095, USA}

\author[0000-0001-7965-897X]{Rebecca A. Fischer}
\affiliation{Department of Earth and Planetary Sciences, Harvard University, 20 Oxford St, Cambridge, MA 02138, USA}



\begin{abstract}
We present globally inverted pressure--temperature ($P$--$T$) phase diagrams  up to 5,000 GPa for four fundamental planetary  materials, Fe, MgO, SiO$_{2}$, and MgSiO$_{3}$, derived from logistic regression and supervised learning, together with an experimental phase equilibria database. These new $P$--$T$ phase diagrams provide a solution to long-standing disputes about their melting curves. Their implications extend to the melting and freezing of rocky materials in the interior of giant planets and super-Earth exoplanets, contributing to the refinement of their internal structure models.
\end{abstract}



\section{Introduction} \label{sec:intro}

Self-gravitation and internal heating of a planet induce high pressures and temperatures that alter the structure and physical properties of its interior. Knowledge of the phase equilibria of a planet's constituent materials, including solid-solid phase transitions and melting, is essential, as these fundamental thermodynamic processes can regulate convective dynamics, thermal evolution, chemical differentiation, and magnetic field generation \citep[cf.][and references therein]{schubert_mantle_2001,lewis_physics_2004,seager_exoplanets_2010,perryman_exoplanet_2018}. In light of the rapid discovery of exoplanets by ground- and space-based telescopes \citep[cf.][and references therein]{seager_exoplanets_2010,perryman_exoplanet_2018}, as well as the emerging geophysical constraints on giant planets in the outer Solar System from existing and planned missions \citep[e.g.][]{baines_interior_2018,simon_uranus_2021,rymer_neptune_2021,helled_revelations_2022}, there is a growing need to better characterize the material properties of giant planet and exoplanet interiors in order to better understand our observations. While extensive research has focused on the behavior of hydrogen \citep[cf.][and references within]{helled_understanding_2020} and water \citep[e.g.,][]{millot_nanosecond_2019,huang_stability_2020,cheng_phase_2021,prakapenka_structure_2021,weck_evidence_2022} at high pressure and temperature due to their stellar abundance and suitability for testing fundamental material theories, considerably less attention has been paid to rocky materials (MgO, SiO$_{2}$, and MgSiO$_{3}$) and iron (Fe), except in the context of studying the Earth's interior \citep{liu_elements_1986,stixrude_thermodynamics_2011}. However, these materials also constitute the deep interiors of giant planets \citep{lewis_physics_2004,mazevet_fate_2019} and exoplanets \citep{duffy_mineralogy_2015}.

The pressure--temperature ($P$--$T$) phase diagrams of these materials, especially their melting curves, remain poorly understood under the relevant conditions of giant planets and exoplanets. The centers of giant planets in our Solar System can reach pressures of 500--5,000 GPa \citep{nettelmann_uranus_2016,mazevet_fate_2019,stixrude_thermal_2021-1,militzer_relation_2023,militzer_study_2024,james_thermal_2024}, while super-Earth exoplanets have central pressures of up to 5,000 GPa at 10 Earth masses \citep{stixrude_melting_2014,duffy_mineralogy_2015,boujibar_superearth_2020}. In contrast to the low melting temperature and relatively uniform fluid behavior of hydrogen, rocky materials and iron metal are relatively refractory between 1,000--5,000 GPa \citep{root_shock_2015,soubiran_anharmonicity_2020,hansen_melting_2021,gonzalez-cataldo_melting_2016,fei_melting_2021,geng_ab_2024,deng_melting_2023,morard_melting_2011,stixrude_structure_2012,kraus_measuring_2022,gonzalez-cataldo_ab_2023}, and exhibits a wide range of exotic chemical behaviors at ultrahigh pressures potentially as the core of gaint planets or the mantle of super-Earth and sub-Neptune exoplanets, including the metallization of oxide solids \citep[e.g.,][]{mcwilliams_phase_2012}, the diverse polymorphism of silicate solids \citep[e.g.,][]{umemoto_phase_2017}, and the enhanced electrical conductivity of the silicate liquid \citep{francis_dragulet_electrical_2023,lherm_thermal_2024}. These results highlight the profound implications of solid-solid phase transitions and melting in the interiors of giant planets and exoplanets.

Previous efforts to directly determine the $P$--$T$ phase boundaries of rocks and iron have typically focused on one phase boundary at a time, using a single set of phase equilibrium data consisting of 10--100 points near the phase boundary to bracket the transition \citep[e.g.][]{chanyshev_depressed_2022}. 
Alternatively, equations of state for the materials are determined experimentally and then combined with extrapolated thermodynamic properties, such as enthalpy ($\Delta H$) and entropy ($\Delta S$), to construct a thermodynamic model of their phase equilibria \citep[e.g.][]{stixrude_thermodynamics_2011,kojitani_precise_2016}. The phase boundaries determined by the former approach often rely on visual inspection and freehand drawing based on a small data set and do not directly show interlaboratory discrepancies; the topology of the phase equilibria computed by the latter approach is very sensitive to the choice of thermodynamic parameters, which are often measured experimentally only at low temperatures and pressures and then extrapolated to the pressure and temperature ranges of planetary interiors \citep{kojitani_precise_2016}.

Here, we present an alternative approach (Section \ref{method}) that combines all available static compression data from 1 bar to 100--300 GPa, supplemented by select dynamic compression and computational data obtained at much higher pressures up to 5,000 GPa (Section \ref{result}). We applied a global inversion algorithm based on logistic regression and supervised learning (Section \ref{inversion}) to determine the phase transition boundaries (including melting curves) of MgO, SiO$_{2}$, MgSiO$_{3}$, and Fe. This global approach, based on an extensive database (Section \ref{database}), allows us to conveniently infer the solid--solid phase transitions and melting curves of rocky materials and iron at pressure conditions relevant to giant planets and super-Earth exoplanets (1,000--5,000 GPa) without the need for complete information on their thermodynamic properties. We then performed thermodynamic modeling of the phase equilibria for MgO, SiO$_{2}$, MgSiO$_{3}$, and Fe between 1 bar to 200 GPa to validate our global inversion results. Finally, we discuss their implications for melting and freezing in planetary interiors (Section \ref{implications}).

\section{Methods} \label{method}

\subsection{Phase Equilibria Database} \label{database}

We have compiled a comprehensive phase equilibria database for Fe, MgO, SiO$_{2}$, and MgSiO$_{3}$, including a wide range of experimental and theoretical data produced over $\sim$80 years, from the early 1940s to the present. The database includes $\sim$3,300 entries for Fe, $\sim$1,100 for MgO, $\sim$1,600 for SiO$_2$, and $\sim$800 for MgSiO$_3$. The data are categorized into three main types: 
\begin{enumerate}
    \item Experimental data from static compression techniques, such as multi-anvil press (MA) and diamond anvil cell (DAC) \citep{ito_theory_2007,mao_solids_2018};
    \item Experimental data from dynamic compression techniques, such as gas gun compression and laser and magnetic pulse compression \citep{duffy_ultra-high_2019};
    \item Computationally-simulated data from first-principles quantum mechanical methods, such as density functional theory (DFT) calculations combined with molecular dynamics (MD) simulations as well as quantum Monte Carlo (QMC) simulations \citep{wentzcovitch_theoretical_2010}.
\end{enumerate}

Most of the data entries are compiled from experimental observations of phase equilibria in static compression experiments, and among them we further distinguish two main types: 
\begin{enumerate}
\item Direct observations: crossing of a phase transition or reaction boundary;
\item Indirect observations: identification of a stable phase.
\end{enumerate}

 Direct observations involve the identification of a phase transition or reaction in a single experiment. We note that such observations can be influenced by kinetic barriers, making it difficult to unambiguously identify the equilibrium phase boundaries. For example, the sluggish kinetics of MgSiO$_3$ at low temperature often leads to conflicting observations of its transition or reaction boundary \citep{kuroda_determination_2000}, and therefore we only include data collected $>$1,000 K to ensure reliable interpretation of these observations (Section \ref{mgsio3}).
 
 Indirect observations involve the identification of a phase at a given $P$--$T$ without crossing a phase transition or reaction boundary. For example, experimental data of the equation of state of a single phase may not directly inform us of its phase equilibrium \citep[e.g.][]{pigott_highpressure_2015}, but they contain valuable information about the $P$--$T$ conditions at which its polymorphs are likely to be \textit{not} stable. While these phases provide valuable information, they do not guarantee that the observed phase is the stable phase under the given conditions. A notorious example is SiO$_2$, whose metastable phases are very common and often persist at low temperatures well beyond the expected phase boundary \citep{prakapenka_high_2004,akaogi_calorimetric_2022}. Careful consideration must therefore be given to which data sets to include. Those with clear evidence of metastable phases are excluded from our inversion. In addition, we also include some miscellaneous measurements, such as experimental data on the physical properties of the Fe liquid \citep[e.g.][]{ohta_experimental_2016}, which, although not primarily focused on determining the melting curve, provide additional information on the conditions under which the solid phase is likely \textit{not} stable. 

These indirect observations, which have not historically been used to construct $P$--$T$ phase diagrams, are not, in our view, ancillary to direct observations of phase transitions. On the contrary, the statistical framework we build here uses information about both the conditions under which a phase is stable and those under which it is not (see details in Section \ref{inversion}). 

Furthermore, melting experiments sometimes only report the ``incipient melting'' temperature, which is the lowest temperature at which this detection method could identify liquid \citep[e.g.][]{zhang_temperature_2016}. We include these data only if they are reported with uncertainties, and we treat each ``incipient melting'' temperature as two data points, using the lower temperature limit of the uncertainty range as a solid phase observation and the upper limit as a liquid phase observation. 

A small fraction of the data entries are compiled from dynamic compression experiments and computational simulations. They occupy a very different region of $P$--$T$ space than the static compression data and provide critical information under otherwise inaccessible conditions. Therefore, we supplement the database with selected dynamic compression and computational data because they are key to anchoring the phase diagrams beyond the multimegabar conditions and guide the quality of the inversion at lower $P$--$T$. For example, in MgSiO$_{3}$, dynamic compression data are the only experimental constraints on melting at $>$300 GPa (Section \ref{mgsio3}).

The results of different dynamic compression techniques and computational simulations can have different degrees of fidelity. In dynamic compression, for example, the resolution of temperatures is poor because the temperature is not directly constrained by the Hugoniot equations \citep{duffy_ultra-high_2019}. We have tried to include all dynamic compression data where stable phases can be inferred. In computational simulations, for example, systematic shifts are expected from the DFT when different electron exchange-correlation functionals are assumed. We only include the computational data performed at finite temperature and where the stable phases are directly identified from two-phase simulations with DFT--MD or inferred from thermodynamic integration calculations if uncertainties are provided.

\begin{figure*}
\centering
\noindent\includegraphics[width=0.9\textwidth]{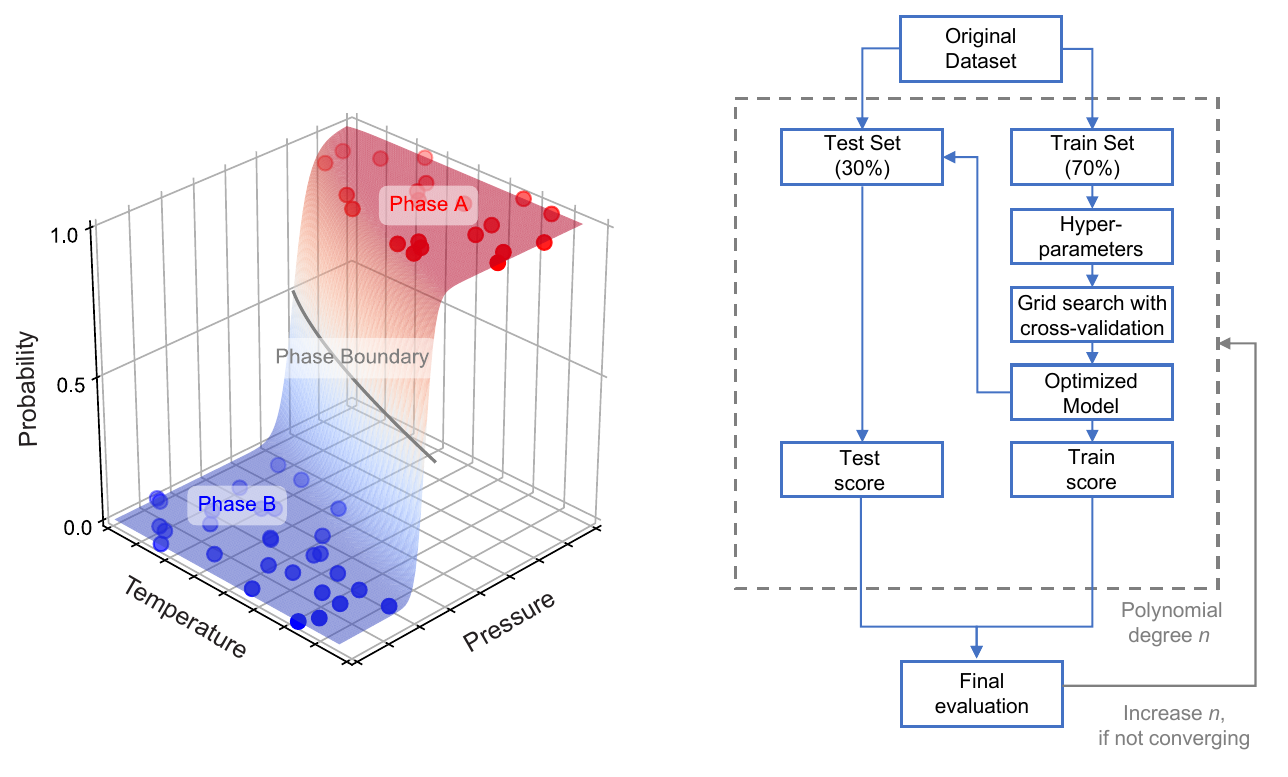}
\caption{\textbf{Global Inversion Algorithm.} a) This global inversion algorithm uses a multi-class logistic model with supervised learning for phase diagram determination \citep{dong_nonlinearity_2025}. Polynomial functions of $P$--$T$ are used as log-odds, $f(P,T) = \sum_{i,j=0}^{n} \beta_{i,j}^{k} P^{i} T^{j}$, to estimate the probability of phase stability. The model then selects the phase with the highest probability of being stable at a given $P$--$T$. b) To avoid overfitting, we tune the hyperparameters using supervised learning with L$_{1}$ \textit{lasso} regularization. The data set is divided into training and test sets, and the hyperparameters are optimized using a grid search and $k$-fold cross-validation. The model with the highest $F_{1}$ score is selected to fit the entire dataset and construct the final phase diagram. }
\label{fig:inversion}
\end{figure*}

\subsection{Global Inversion Algorithm} \label{inversion}
We now present a machine learning-based approach for inverting phase equilibria observations over a wide $P$--$T$ range spanning multiple stable phases. Our algorithm uses a combination of multi-class logistic regression and supervised learning, applied to the $P$--$T$ phase diagrams of Fe, MgO, SiO$_{2}$, and MgSiO$_{3}$. This method not only identifies phase and reaction boundaries, but also quantifies their associated uncertainties. A detailed discussion of the method can be found in \cite{dong_nonlinearity_2025}.

\subsubsection{Multi-Class Logistic Model} \label{logistic}

In constructing our statistical model, we treat the prediction of stable phases as a classification problem. The response variable (stable phase) is categorical, related to specific $P$--$T$ conditions. This is an extension of the binary logistic regression method used by \cite{aoki_statistical_2000}, which was limited to two phases. Our model allows for multiple stable phases and does not require pre-specified shapes for phase boundaries.

The multi-class logistic model is formulated as follows \citep[cf.][and references therein]{dong_nonlinearity_2025}:

\begin{enumerate}
\item The probability of observing a particular phase $Y = k$ at a given $P$--$T$ is expressed as a logistic function with two variables, $P$ and $T$:
\begin{equation}\label{eq-ch4-2}
    \hat{p}(Y=k|P,T) = \frac{e^{f(P,T)}}{1 + {e^{f(P,T)}}} \\ 
    = \frac{e^{\sum_{i,j=0}^{n} \beta_{i,j}^{k} P^{i} T^{j}}}{1 + e^{\sum_{i,j=0}^{n} \beta_{i,j}^{k} P^{i} T^{j}}}
\end{equation}

\item A multi-class generalization of the logistic function, the \textit{softmax} function, is used to transform separate logistic model probabilities of different stable phases into a unified set:
\begin{equation}\label{eq-ch4-4}
    \hat{p}(Y=k|P,T) = \frac{e^{\sum_{i,j=0}^{n} \beta_{i,j}^{k} P^{i} T^{j}}}{\sum_{h=1}^{K} e^{\sum_{i,j=0}^{n} \beta_{i,j}^{h} P^{i} T^{j}}}
\end{equation}

\item The rescaled probability estimates sum to one. The phase with the highest probability is considered stable, and the
triple point is taken to be the $P$--$T$ condition at which $\hat{p}(Y = \textsc{phase a}) = \hat{p}(Y = \textsc{phase b}) = \hat{p}(Y = \textsc{phase c}) = \frac{1}{3}$. 

\item The coefficients, $\beta_{i,j}^{k}$, are estimated by minimizing a combined negative log-likelihood function, or total cross entropy $-L$. In this expression, $t_{m,l}(y_{i}=k)$ equals one if and only if the observation $m$ belongs to the phase $k$. $p_{m}(y_{m}= k)$ is the estimated probability that  observation $m$ belongs to phase $k$, and $M$ is the total number of observations.
\begin{equation}\label{eq-ch4-5}
    -L = - \frac{1}{M} \sum_{m=1}^{M} \sum_{k=1}^{K}\left\{ t_{m,k}(y_{m}=k) \cdot \ln \left [ p_{m}(y_{m}=k) \right ] + t_{m,l}(y\neq k) \cdot \ln \left [ 1 - p_{m}(y_{m}\neq k) \right ] \right\}
\end{equation}
\end{enumerate}

Phase boundaries and triple points can be determined numerically using the ``scikit-learn'' Python package \citep{pedregosa_scikit-learn_2011}. The multi-class logistic model allows us to incorporate nonlinear phase boundaries as logistic functions of $P$ and $T$ and identify the triple points among them, while avoiding the complexities of using inexplicit thermodynamic function derivations \citep{ghiorso_equation_2004}.

\subsubsection{Supervised Learning} \label{learning}
To avoid model overfitting, we use supervised learning with L$_{1}$ \textit{lasso} regularization \citep{james_introduction_2021}. The dataset is divided, with 70\% of the data in a training set and 30\% in a test set. We tune the hyperparameters of the model, including the inverse of the regularization strength $C$ and other parameters, using a grid search combined with $k$-fold cross-validation. We consider logistic models of $P$ and $T$ with polynomials of up to degree ten, $f(P,T) = \sum_{i,j=0}^{n} \beta_{i,j}^{k} P^{i} T^{j}$ (where $n = 1-10$). Model performance is then evaluated using the $F_{1}$ score, $ 2\times\frac{\textrm{precision} \times \textrm{recall}}{\textrm{precision} + \textrm{recall}}$, which is the harmonic mean of precision, $\frac{\textrm{true positive}}{\textrm{true positive + false positive}}$, and recall, $\frac{\textrm{true positive}}{\textrm{true positive + false negative}}$. The final model is fit to the training and test sets combined after selecting the polynomial degree with the highest $F_{1}$ score for the test set, to generate a globally-optimized phase diagram.

\subsection{Thermodynamic Calculation} 

We have calculated the solid phase equilibria of Fe, MgO, SiO$_{2}$, and MgSiO$_{3}$ over pressures from 1 bar to 200 GPa and temperatures from 300 K to 5,000 K, with a resolution of 0.1 GPa and 10 K, using the thermodynamic code HeFESTo with the latest self-consistent parameter set\footnote{In HeFESTo, the species flags are shown in parentheses and may differ from the mineral abbreviations used in Section \ref{result}.} \citep{stixrude_thermodynamics_2005,stixrude_thermodynamics_2011,stixrude_thermal_2021,stixrude_thermodynamics_2024}. Three species are considered for Fe: $\alpha$-Fe (fea), $\gamma$-Fe (feg), and $\epsilon$-Fe (fee); one species is considered for MgO: periclase (pe); four species are considered for SiO$_{2}$: quartz (qtz), coesite (coes), stishovite (st), and seifertite (apbo); and fourteen species are considered for MgSiO${3}$: enstatite (en), HP clinoenstatite (mgc2), clinoenstatite (cen), Mg-wadsleyite (mgwa), Mg-ringwoodite (mgri), Mg-akimotoite (mgil), Mg-bridgmanite (mgpv), Mg-post-perovskite (mppv), pe, qtz, coes, st, apbo, and Mg-majorite (mgmj). These calculations aim to compare and contrast the phase diagrams generated by the global inversion results presented below. The parameter set used is optimized specifically for the Earth’s mantle conditions, with no liquid phases included in the model. Further details on the code and the parameter set can be found in \citep{stixrude_thermodynamics_2005,stixrude_thermodynamics_2011,stixrude_thermal_2021,stixrude_thermodynamics_2024}.

\begin{figure}
\centering

\noindent\includegraphics[width=0.5\textwidth]{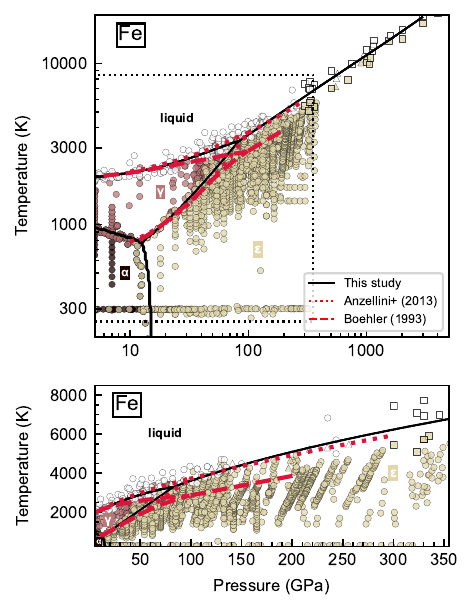}
\caption{\textbf{\textit{P--T} Phase Diagram of Fe up to 5,000 GPa.} Fe is a primary component of Earth and super-Earth cores. It is stable as the body-centered cubic phase ($\alpha$-Fe) below 820--1,000 K at 3--10 GPa, transitioning to the face-centered cubic phase ($\gamma$-Fe) above these temperatures. The transformation from $\gamma$-Fe to the hexagonal close-packed phase ($\epsilon$-Fe) occurs at 12--82 GPa; the $\gamma$--$\epsilon$--liquid triple point is found at 82 GPa and 3,300 K. For comparison, phase boundaries from \cite{anzellini_melting_2013} (dotted line) and \cite{boehler_temperatures_1993} (dashed line) are also shown. Data points contributed by static compression experiments, dynamic compression experiments, and computer simulations are marked with circles ($\bigcirc$), triangles ($\bigtriangleup$), and squares ($\Box$), respectively, and coded with different colors based on the stable phase observed at that $P$--$T$ condition.}
\label{fig:fe}
\end{figure}

\section{Results and Discussion} \label{result}

\subsection{Fe} \label{fe}

Iron (Fe) is the primary component of the cores of the Earth, of other terrestrial planets in the Solar System, and presumably of super-Earth exoplanets. The three major solid phases of Fe are $\alpha$-Fe ferrite (body-centered cubic, bcc), $\gamma$-Fe austenite (face-centered cubic, fcc), and $\epsilon$-Fe hexaferrum (hexagonal close-packed, hcp). In our global inversion (Figure \ref{fig:fe}), we find that the $\alpha$ $\leftrightarrow$ $\gamma$ transition occurs at 1,000 K and 3 GPa and at 830 K and 10 GPa with a slope of approximately –39 MPa/K, while the $\alpha$ $\leftrightarrow$ $\epsilon$ transition occurs at 700 K at 13 GPa and at 230 K at 15 GPa with a negative slope of –4 MPa/K. The $\alpha$--$\gamma$--$\epsilon$ and $\gamma$--$\epsilon$--liquid triple points are located at 12 GPa and 770 K, and at 82 GPa and 3,300 K, respectively. The slope of the $\gamma$ $\leftrightarrow$ $\epsilon$ phase boundary is +28 MPa/K.

\begin{figure*}
\centering
\noindent\includegraphics[width=0.75\textwidth]{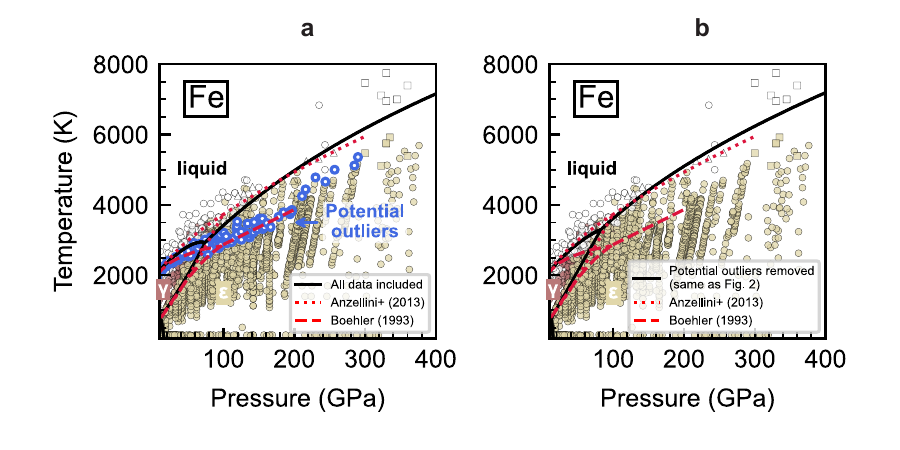}
\caption{\textbf{Global Inversion of the Fe Phase Diagram (a) with and (b) without Potential Outliers.} Discrepancies in the high-pressure melting curve of Fe have led to ongoing debate over its $P$–$T$ phase diagram. Our inversion identifies five potential outlier datasets (blue circles in \textbf{a}), which report systematically lower melting temperatures. Excluding these datasets (\textbf{b}) results in a melting curve that aligns more closely with thermodynamic models \citep{komabayashi_thermodynamics_2014,dorogokupets_thermodynamics_2017} and eliminates an unphysical cusp near the $\gamma$--$\epsilon$--liquid triple point.}
\label{fig:outliers}
\end{figure*}

The current debate over the $P$--$T$ phase diagram of Fe concerns discrepancies in its high-pressure melting curve \citep[e.g.][red curves in Figure \ref{fig:outliers}]{boehler_temperatures_1993,anzellini_melting_2013}. We addressed this discrepancy of nearly 1000 K at 100--200 GPa by identifying and removing a small subset of data that was internally inconsistent with the rest. We found that 5 of the 33 liquid or melting studies included in the inversion, \cite{boehler_melting_1990}, \cite{boehler_temperatures_1993}, \cite{saxena_temperatures_1994}, \cite{aquilanti_melting_2015}, and \cite{sinmyo_melting_2019} (highlighted with blue circles in Figure \ref{fig:outliers}a), are potential outliers. These datasets are systematically lower in temperature than the rest of the datasets that report observations of liquid or melting \citep{strong_experimental_1959,strong_iron_1973,liu_melting_1975,williams_melting_1987,yoo_phase_1995,shen_melting_1998,terasaki_viscosity_2002,rutter_viscosity_2002,ma_situ_2004,shen_structure_2004,sola_melting_2009,morard_melting_2011,deng_high_2013,jackson_melting_2013,kono_high-pressure_2015,ohta_experimental_2016,zhang_temperature_2016,secco_thermal_2017,morard_solving_2018,pommier_influence_2018,silber_electrical_2018,yong_iron_2019,li_shock_2020,kuwayama_equation_2020,hou_melting_2021,kraus_measuring_2022,gonzalez-cataldo_ab_2023,sun_ab_2023}.

In Figures \ref{fig:fe} and \ref{fig:outliers}, we find that $\gamma$-Fe melts at 2,100 K and 10 GPa, at 2,550 K and 30 GPa, and at 3,050 K and 60 GPa, with a slope of 16--23 K/GPa\footnote{Conventionally, the slope of solid-solid transitions is expressed in MPa/K because they are often pressure-induced, steep, and plotted with P on the $x$-axis in $P$--$T$ phase diagrams. In contrast, the melting curve, which is temperature driven and generally flatter, is expressed in K/GPa.}; $\epsilon$-Fe melts at 3,950 K and 120 GPa, at 7,200 K and 400 GPa, and at 10,800 K at 1,000 GPa, with a slope of up to 5--12 K/GPa.

We present the inversion with the exclusion of the potential outliers (Figures \ref{fig:fe} and \ref{fig:outliers}b) as our optimal $P$--$T$ phase diagram of Fe for three reasons:

\begin{enumerate}
    \item The five excluded datasets have been contested by later researchers, and their relatively low melting temperatures can be explained alternatively by carbon contamination \citep{morard_solving_2018}, fast recrystallization \citep{anzellini_melting_2013} and/or deformation of iron \citep{hou_melting_2021}, and instrumental discrepancy \citep{sinmyo_melting_2019}. 
    
    \item When the outlier datasets are excluded (Figure \ref{fig:outliers}b), our inversion provides a $\epsilon$-Fe melting curve of about 12 K/GPa between 100 GPa and 400 GPa and a $\gamma$--$\epsilon$--liquid triple point of 83 GPa and 3,300 K in closer agreement with the two state-of-the-art thermodynamic models for Fe: \cite{komabayashi_thermodynamics_2014} (97 GPa, 3,330 K with a slope of 14 K/GPa) and \cite{dorogokupets_thermodynamics_2017} (107 GPa and 3,790 K with a slope of 14 K/GPa). These models were constructed independently by extrapolating selected equations of state for the solid and liquid phases of Fe using fundamental thermodynamic relations.
    
    \item When the outlier datasets are excluded (Figure \ref{fig:outliers}b), the melting curves of $\gamma$-Fe and $\epsilon$-Fe rise continuously across the $\gamma$--$\epsilon$--liquid triple point. When all data were included in the inversion (Figure \ref{fig:outliers}a), a small cusp appears on the melting curve where the $\gamma$-Fe melting curve suddenly flattens prior to the subsolidus transition of $\gamma$ $\leftrightarrow$ $\epsilon$. A melting curve slope, $\frac{dT}{dP} = \frac{\Delta V}{\Delta S}$, can change over a small pressure range, if $\frac{\Delta V}{\Delta S}$ varies rapidly. $\Delta S$ typically remains similar, such a cusp, when found, is often observed in materials where the liquid is highly compressible or where a solid--solid phase transition involves volume collapse \citep[][Section \ref{sio2}]{Kechin_1995,kechin_melting_2001,dong_melting_2019}. For Fe, however, with its densely packed phases (including the liquid), a rapid change in $\Delta V$ is unlikely. Therefore, the triple point shift and cusp observed in Figure \ref{fig:outliers}a are assumed to result from the inclusion of the potential outliers; when the outliers were included, the inversion attempted to reconcile the two divergent trends in the $\epsilon$-Fe melting curve.

\end{enumerate}

We made an additional intervention to the inversion by disregarding the entropy-driven re-entrant bcc-Fe fields just below the melting curve. Its stability field is either too small to be resolved by our method in the case of the low-pressure $\delta$ phase \citep{strong_iron_1973}, or not well established by experimental evidence in the case of its potential second re-entrant field near the Earth's inner core conditions predicted by the DFT results \citep{belonoshko_stabilization_2017}. By lumping the additional bcc phases into their adjacent solid phases, we sacrifice resolution in the inversion in favor of its overall performance, particularly in constructing the melting curve. 

\begin{figure*}
\centering
\noindent\includegraphics[width=1\textwidth]{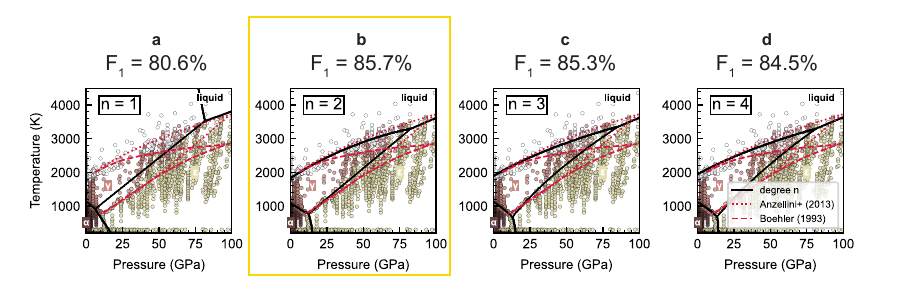}
\caption{\textbf{Model Selection and Performance of Varying Complexity for Fe.} The phase equilibria data for Fe are plotted against our phase boundary predictions for degrees 1 to 4 (solid black lines). $F_{1}$ scores indicate the the performance of the models on the test set, with the degree-2 polynomial model achieving an optimal balance between complexity and performance (highest $F_{1}$ score). Higher degree models show diminishing returns due to overfitting, as indicated by lower $F_{1}$ scores and minimal improvements in phase boundary predictions.}
\label{fig:modelselection}
\end{figure*}

Figure \ref{fig:modelselection} demonstrates the model selection process in supervised learning. It shows the performance of four logistic models for Fe of varying complexity, expressed as polynomials of $P$ and $T$ from degree $n$=1--4: $f(P,T) = \sum_{i,j=0}^{n} \beta_{i,j}^{k} P^{i} T^{j}$. Each model undergoes hyperparameter optimization based on the 70\% training set through supervised learning. For $n=1$, the phase boundaries of the model fit the data poorly. The model with $n=2$ registers the highest $F_{1}$ score on the 30\% unseen test set, indicating accurate classification with minimal overfitting. As $n$ increases beyond 2, the $F_{1}$ score on the test set decreases, signaling a decline in performance on unseen data that is likely due to overfitting. Furthermore, the topology of the phase diagram reaches stability as $n \geq 2$, confirming that an optimal balance between model complexity and performance is reached at $n=2$.

\begin{figure}
\centering
\noindent\includegraphics[width=0.5\textwidth]{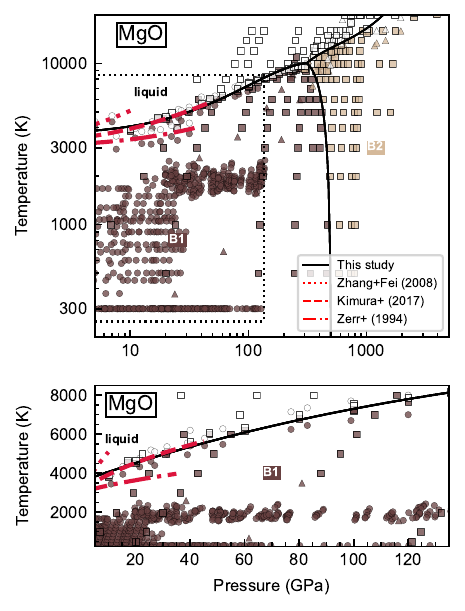}
\caption{\textbf{\textit{P--T} Phase Diagram of MgO up to 5,000 GPa.} MgO is the magnesium end member of ferropericlase, a key mineral phase that is abundant in the Earth's mantle and likely present in the mantle of super-Earths. Its B1--B2 phase transition occurs at 400--500 GPa in the interiors of super-Earths, and a B1--B2--liquid triple point is located around 310 GPa and 10,000 K.
For comparison, phase boundaries from \cite{zerr_constraints_1994} (dash-dotted line), \cite{zhang_melting_2008} (dotted line), and \cite{kimura_melting_2017} (dashed line) are also shown. Data points contributed by static compression experiments, dynamic compression experiments, and computer simulations are marked with circles ($\bigcirc$), triangles ($\bigtriangleup$), and squares ($\Box$), respectively, and coded with different colors based on the stable phase observed at that $P$--$T$ condition.}
\label{fig:mgo}
\end{figure}

\subsection{MgO}\label{mgo}

Magnesium oxide (MgO) is the magnesium endmember of the (Mg,Fe)O solid solution, a major mineral of the Earth's mantle known for its B1 structure as ferropericlase. A super-Earth, especially one with a higher Mg/Si ratio, may have free ferropericlase throughout its mantle \citep{tsuchiya_prediction_2011,niu_prediction_2015,umemoto_phase_2017}. At the pressure conditions of super-Earth mantles or giant planet cores ($\sim$500--4,000 GPa), MgO would instead have a B2 structure. The B1--B2 phase boundary of MgO remain underconstrained. The reported location of the B1--B2--liquid triple point varies from 260 GPa and 9,450 K \citep{root_shock_2015} (DFT), to 360 GPa and 10,400 K \citep{mcwilliams_phase_2012} (shock compression). 

In our global inversion (Figure \ref{fig:mgo}), we find that the B1--B2 phase boundary of MgO occurs at 6,000 K and 415 GPa, and 300 K at 495 GPa, with a slope of about $-14$ MPa/K. The B1--B2--liquid triple point is found to be at 310 GPa and 10,000 K, in closer agreement with recent shock experiments \citep{wicks_b1-b2_2024} and theoretical predictions \citep{soubiran_anharmonicity_2020}. No static compression is yet available to constrain the triple point, so its location may still have considerable uncertainty.

Furthermore, the melting temperatures of MgO at low pressures vary substantially by about 100 K at 1 bar and as much as 1,000 K \citep{zhang_melting_2008,kimura_melting_2017,zerr_constraints_1994}: Three completely different zero pressure slopes have been reported in the literature from a 1 bar melting temperature of 3,100--3,200 K \citep[e.g. ][]{dubrovinsky_thermal_1997,chernyshev_thermal_1993}, to 5 GPa melting at 3,220 K \citep{zerr_constraints_1994}, and 3,560 K \citep{kimura_melting_2017}, and 4,200 K 

\begin{figure*}
\centering
\noindent\includegraphics[width=0.75\textwidth]{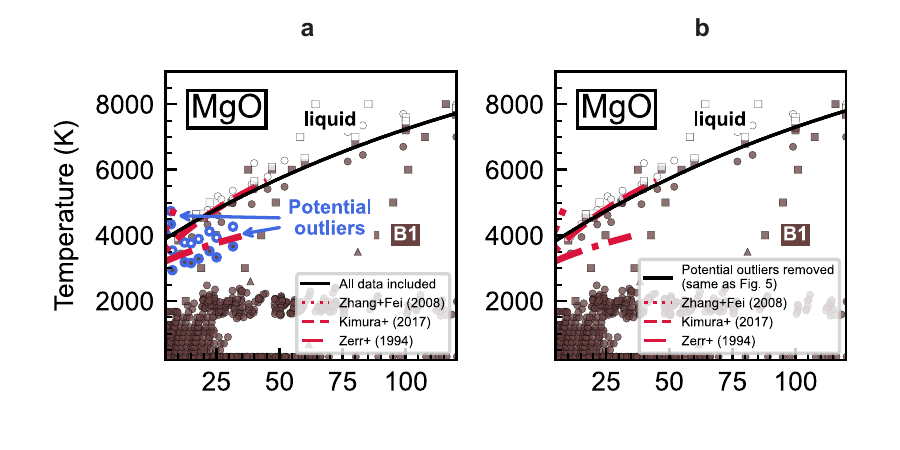}
\caption{\textbf{Global Inversion of the MgO Phase Diagram (a) with and (b) without Potential Outliers.} Two potential outlier datasets for MgO melting were identified and removed from our inversion, \cite{zerr_constraints_1994} and \cite{zhang_melting_2008}, which were internally inconsistent with the rest (\textbf{a}). The low melting temperatures in \cite{zerr_constraints_1994} were likely misinterpreted pre-melting signals \citep{kimura_melting_2017,hou_melting_2021}, while the extrapolated curve in \cite{zhang_melting_2008} was biased by FeO-rich liquidus data \citep{fatyanov_melting_2024}. Their exclusion has minimal impact on the inversion (\textbf{b}), likely due to limited low-pressure data. Our preferred melting curve (Figure \ref{fig:mgo} and \textbf{a}) places B1-MgO melting at 3,900 K (5 GPa), 7,300 K (100 GPa), and 9,200 K (200 GPa) with a slope of $+20–40$ K/GPa, while B2-MgO reaches 12,500 K (500 GPa) and 16,500 K (1,000 GPa) with a slope of $+9$ K/GPa. Further experiments are needed to validate the low-pressure melting curve (1 bar to 30 GPa).}
\label{fig:mgo-outliers}
\end{figure*}

Similar to the approach we took for the Fe melting curve, we identified and removed 2 of the 20 melting or liquid studies on MgO, \cite{zerr_constraints_1994}, \cite{zhang_melting_2008}, that were internally inconsistent with the rest \citep{mcnally_laboratory_1961,riley_determination_1966, chernyshev_thermal_1993,dubrovinsky_thermal_1997,ronchi_melting_2001,assael_reference_2006,mcwilliams_phase_2012,du_highpressure_2014,cebulla_ab_2014, root_shock_2015,bolis_decaying_2016,kimura_melting_2017,FU20181,musella_physical_2019,wisesa_machine-learning_2023,wicks_b1-b2_2024} for the following reasons:

\begin{enumerate}
    \item The significantly lower melting temperatures reported by \cite{zerr_constraints_1994} were contested by later researchers as a misinterpreted pre-melting signal due to deformation of the solid MgO \citep{kimura_melting_2017}, similar arguments were made for the anomalously low Fe melting curve \citep{boehler_temperatures_1993,hou_melting_2021}.
    \item The melting curve in \cite{zhang_melting_2008} was extrapolated from the melting temperatures of several MgO--FeO solid solutions with different stiochiometries, and its significantly high values were likely due to its poor fit to the liquidus data in the FeO-rich region \citep{fatyanov_melting_2024}. 
\end{enumerate}  

The inclusion of the outlier datasets has almost no effect on the inversion result for MgO (Figure \ref{fig:mgo-outliers}). It is likely that the divergent trends from \cite{zerr_constraints_1994} and \cite{zhang_melting_2008} have been averaged due to the limited data in the low pressure region. We caution that only two experimental studies, \cite{du_highpressure_2014} and \cite{kimura_melting_2017}, have been used here to constrain the low-pressure melting curve, although we consider the \cite{kimura_melting_2017} results to be the best constraints to date, as their melting experiments were performed directly on MgO and melting was also confirmed by microtexture analysis \citep{kimura_melting_2017}. It is imperative that further experiments be performed to validate our inversion. This is not a trivial task because the low-pressure melting temperature of MgO is outside the typical range of two conventional static compression techniques: the temperature is too high for a MA experiment and no robust thermocouple measurement is available at the temperature of the MgO liquid, and the pressure is too low for a laser-heated DAC experiment to achieve reliable heating \citep[see Methods in][and references therein]{dong_nonlinearity_2025}.

We present the inversion excluding \cite{zerr_constraints_1994} and \cite{zhang_melting_2008} as the optimal $P$--$T$ diagram of MgO (Figure \ref{fig:mgo} and \ref{fig:mgo-outliers}b). Its B1 melting curve reaches at 3,900 K at 5 GPa, 7,300 K at 100 GPa and 9,200 K at 200 GPa, with a slope of 20--40 K/GPa. The melting curve of B2 MgO is based mainly on theoretical data and reaches at 12,500 K at 500 GPa and 16,500 at 1,000 GPa, with a slope of about $+9$ K/GPa.

\begin{figure}
\centering
\noindent\includegraphics[width=0.5\textwidth]{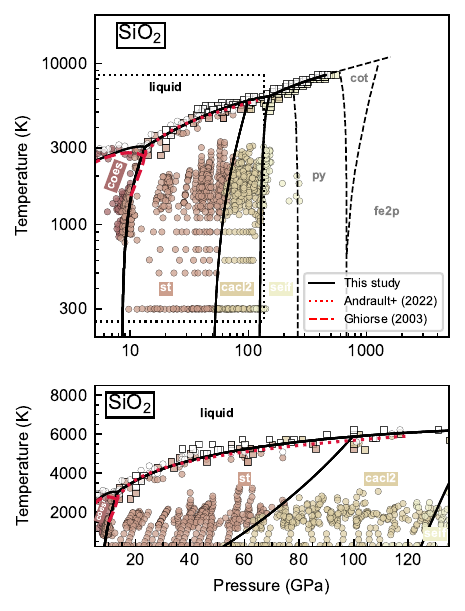}
\caption{\textbf{\textit{P--T} Phase Diagram of SiO$_{2}$ up to 5,000 GPa.} SiO$_{2}$ is a major oxide component of the mantle of Earth and super-Earths. Free SiO$_{2}$ minerals exist in subducted slabs on Earth and in the mantle of some super-Earth exoplanets. SiO$_{2}$ has several polymorphs, and its well-studied major phase transitions include coesite to stishovite ($coes$$\leftrightarrow$$st$) at 9--13 GPa, $st$ to post-stishovite or CaCl$_{2}$-type SiO$_{2}$ ($st$$\leftrightarrow$$cacl2$, 50--100 GPa), and $cacl2$ to seifertite ($cacl2$$\leftrightarrow$$seif$) at 125--145 GPa. For comparison, phase boundaries from \cite{ghiorso_equation_2004} (dashed line) and \cite{andrault_comment_2022} (dotted line) are also shown. Data points contributed by static compression experiments, dynamic compression experiments, and computer simulations are marked with circles ($\bigcirc$), triangles ($\bigtriangleup$), and squares ($\Box$), respectively, and coded with different colors based on the stable phase observed at that $P$--$T$ condition.}
\label{fig:sio2}
\end{figure}

\subsection{SiO$_{2}$}\label{sio2}

Silica (SiO$_{2}$) is another major oxide component of the mantle. On Earth, free SiO$_{2}$ minerals are present in subducted mid-ocean ridge basalt (MORB) materials within slabs \citep{stixrude_geophysics_2012}. A super-Earth mantles with low Mg/Si may also have free SiO$_{2}$ as a distinct phase \citep{tsuchiya_prediction_2011,umemoto_phase_2017,niu_prediction_2015}. SiO$_{2}$ has a large number of high-pressure polymorphs, including six stable polymorphs that have been well studied experimentally: quartz (\textit{qt}), coesite (\textit{coes}), stishovite (\textit{st}), post-stishovite (\textit{cacl2}, CaCl$_{2}$-type SiO$_{2}$), and seifertite (\textit{seif}). 

In our global inversion (Figure \ref{fig:sio2}), we find that the \textit{coes}--\textit{st} transition occurs at 800 K and 9 GPa, and at 2,000 K and 11 GPa, with a positive slope of about $+2$ MPa/K. The \textit{st}--\textit{cacl2} transition occurs at 1,500 K and 65 GPa, and at 3,100 K and 80 GPa, with a positive slope of about $+9$ MPa/K. The \textit{cacl2}--\textit{seif} transition occurs at 1,350 K and 128 GPa, and at 3,500 K and 135 GPa, with a positive slope of about $+3$ MPa/K.

Due to the simultaneous slow kinetics and structural versatility, metastable polymorphs of SiO$_{2}$ are often observed experimentally outside their $P$--$T$ stability fields under varying compression and heating paths \citep{prakapenka_high_2004}. We present the inversion with a database of SiO$_{2}$ phase stability from 34 studies \citep{boyd_quartz-coesite_1960,lewis_h_cohen_high-low_1967,akimoto_coesite-stishovite_1969,jackson_melting_1976,yagi_direct_1976,suito_phase_1977,mirwald_lowhigh_1980,bohlen_quartz_1982,kanzaki_melting_1990,zhang_melting_1993,bose_quartz-coesite_1995,serghiou_coesitestishovite_1995,shen_measurement_1995,zhang_situ_1996,andrault_pressure-induced_1998,dubrovinsky_pressure-induced_2001,ono_post-stishovite_2002,hudon_melting_2002,murakami_stability_2003,shieh_x-ray_2005,kuwayama_pyrite-type_2005,nomura_precise_2010,usui_ab_2010,wang_pvt_2012,grocholski_stability_2013,yamazaki_over_2014,pigott_highpressure_2015,buchen_equation_2018,fischer_equations_2018,sun_high_2019,andrault_comment_2022,geng_ab_2024}, with the exclusion of any metastable phases identified in previous studies:

\begin{enumerate}
    \item For example, \cite{nishihara_p-v-t_2005} and \cite{andrault_equation_2003} observed the \textit{st} phase extending well into the lower pressure stability field of \textit{coes}. Such data were excluded, and only direct phase transition observations were included in our inversion to recover the \textit{coes}--\textit{st} boundary.

    \item  Similarly, the \textit{cacl2}--\textit{seif} transition appears susceptible to metastability, as mixed phases of \textit{cacl2} + \textit{seif} were observed over a wide pressure range of 40 GPa, violating the Gibbs phase rule and indicating metastability. These mixed phase observations in \cite{sun_high_2019} were also excluded from our inversion.
    
\end{enumerate}

The slopes of the \textit{coes}--\textit{st} and \textit{cacl2}--\textit{seif} phase boundaries determined by us agree well with those reported by \cite{fischer_equations_2018} and \cite{ono_precise_2017}. Despite significant scatter in observations, our \textit{cacl2}--\textit{seif} boundary is generally consistent with the majority of existing static experimental data \citep{grocholski_stability_2013, shen_structure_2004, murakami_stability_2003, sun_high_2019}. Additional high-pressure polymorphs of SiO$_{2}$ at multimegabar conditions, including pyrite-type, cotunnite-type, and Fe$_{2}$P-type structures, have been predicted computationally \citep[e.g.][]{tsuchiya_prediction_2011,gonzalez-cataldo_melting_2016} but not yet directly confirmed or sufficiently constrained by experiments. Therefore, the relevant phase boundaries are indicated with dashed lines in Figure \ref{fig:sio2}.

\begin{figure*}
\centering
\noindent\includegraphics[width=0.75\textwidth]{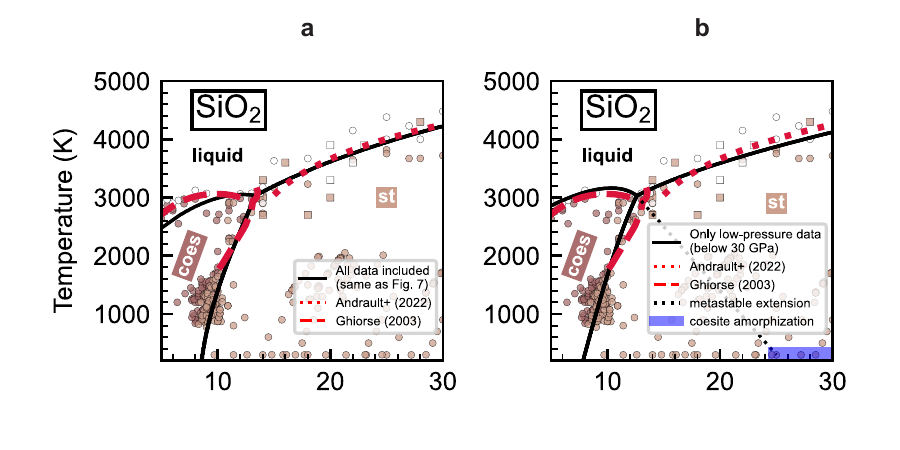}
\caption{\textbf{Global (a) and Local (b) Inversion of the SiO$_2$ Phase Diagram.} The SiO$_2$ melting curve has been debated, especially at low pressures ($\leq$10 GPa). (\textbf{a}) Our global inversion identifies a melting curve maximum at 12 GPa and 3,070 K, followed by a cusp at the \textit{coes}-\textit{st}--liquid triple point at 13.4 GPa and 3,050 K (Figure \ref{fig:sio2}a), consistent with the thermodynamic model of \cite{ghiorso_equation_2004}. This negative melting slope is attributed to a significant volume collapse ($\sim$30\%) during the $coes$-$st$ transition. (\textbf{b}) A focused local inversion limited to data below 30 GPa a melting curve maximum at 10 GPa and 3,150 K, followed by a cusp at the \textit{coes}-\textit{st}--liquid triple point at 12.5 GPa and 3,050 K. This local inversion supports a more pronounced negative $coes$ melting curve and better agrees with the data in this $P$--$T$ range. Amorphization of metastable \textit{coes} at 25--30 GPa indirectly supports the negative slope of the melting curve, as its extrapolation to 300 K (black dotted line) aligns with the pressures of \textit{coes} amorphization (blue area) \citep{hemley_pressure-induced_1988,luo_polymorphism_2003}.}
\label{fig:sio2_compare}
\end{figure*}

The melting curve of SiO$_{2}$ has been debated, particularly at low pressures (5--10 GPa), for reasons similar to MgO: the melting temperature in this range is largely inaccessible in both MA and DAC experiments (Section \ref{mgo}). In our global inversion, we identify a maximum at 12 GPa and 3,070 K, followed by a cusp at the \textit{coes}-\textit{st}--liquid triple point around 13.4 GPa and 3,050 K (Figure \ref{fig:sio2}). This melting curve shape was hypothesized by \cite{ghiorso_equation_2004} using thermodynamic modeling with a general equation-of-state model for silicate melts. 

Unlike Fe (Section \ref{fe}), the melting curve maximum and cusp are expected and attributed to a significant volume collapse (over 30\%) during the $coes$--$st$ transition \citep[e.g.][]{liu_elements_1986}: Between $coes$ and $st$, the volume change is abrupt, but in the liquid, the volume change occurs gradually. As a result, just below the triple point pressure, the liquid phase occupies a smaller volume than the solid ($V_{\text{liquid}} < V_{coes}$). Assuming that entropy increases during melting ($\Delta S_{\text{liquid}-coes} > 0$), the volume change during melting is negative between the melting curve maximum and the triple point ($\Delta V_{\text{liquid}-coes} < 0$), but becomes positive at higher pressures ($\Delta V_{\text{liquid}-st} > 0$). Amorphization of metastable $coes$ at 25--30 GPa indirectly supports the negative slope of the melting curve, as its extension to 300 K coincides with the pressures of $coes$ amorphization \citep[Figure \ref{fig:sio2_compare},][]{hemley_pressure-induced_1988,luo_polymorphism_2003}.

Negative melting curves, although observed in other systems \citep[e.g.][]{young_phase_1991,petrenko_physics_2002,li_determination_2017,dong_melting_2019}, are generally difficult to resolve experimentally due to their often subtle nature. To look further into the \textit{coes}--\textit{st}--liquid triple point, we performed another inversion limited to pressures below 30 GPa. This analysis yielded a more pronounced negative $coes$ melting curve (a melting curve maximum at 10 GPa and 3,150 K, followed by a cusp at the \textit{coes}-\textit{st}--liquid triple point at 12.5 GPa and 3,050 K), which is more consistent with the data in this range (Figure \ref{fig:sio2_compare}b) and the \cite{ghiorso_equation_2004} model, and is likely a better representation of the $coes$ melting curve.

We emphasize that our global inversion method provides a new way to determine melting curves when they intersect first-order transitions. Existing equations for extrapolating melting curves, such as the Simon-Glatzel equation \citep{simon_bemerkungen_1929}, Lindemann's law \citep{lindemann_uber_1910}, the Kraut-Kennedy equation \citep{kraut_new_1966}, and the Kechin equation \citep{kechin_melting_2001}, cannot account for the effects of triple points. However, accurately capturing the slope of the melting curve, not just its sign, depends on the density and coverage of the data, as explored with synthetic data in Appendix \ref{app_test}. Current data on $coes$ are likely insufficient to determine the precise slope of its melting (Figure \ref{fig:test}).

Another complication arises from the $st$--$cacl2$ phase transition, which is second-order and lacks the discontinuous volume or entropy changes across the boundary that are the first derivative of the Gibbs function \citep{kingma_transformation_1995}; only its second derivatives, such as the bulk modulus, exhibit finite, discontinuous changes \citep{callen_thermodynamics_1985}. Thus, at the $st$--$cacl2$--liquid triple point, the slope of the melting curve remains constant $\left( \frac{dT}{dP}\right)_{st} = \left(\frac{dT}{dP}\right)_{cacl2}$, without a cusp. This occurs because the volume of the solid phases changes continuously across the transition, resulting in a constant volume change during melting, $\Delta V_{\text{liquid}-st} = \Delta V_{\text{liquid}-cacl2}$. To ensure that the $st$--$cacl2$--liquid triple point satisfies thermodynamic constraints, we separately invert first- and higher-order phase transitions, then combine a nest of multiple inversions into a final phase diagram (Figure \ref{fig:sio2}).

\begin{figure}
\centering
\noindent\includegraphics[width=0.5\textwidth]{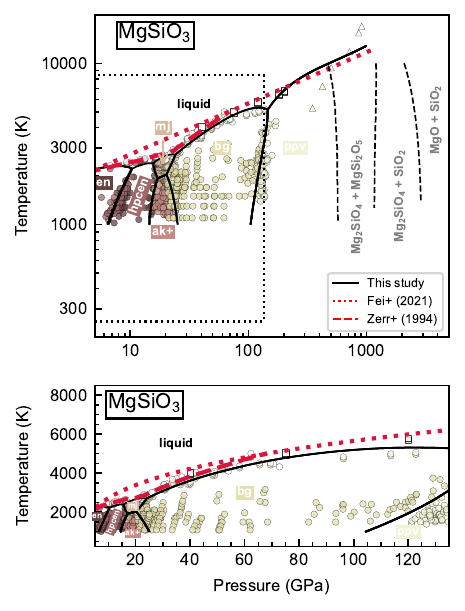}
\caption{\textbf{\textit{P--T} Phase Diagram of MgSiO$_{3}$ up to 5,000 GPa.} MgSiO$_{3}$ is a first-order approximation of the mantle of Earth and super-Earth exoplanets, since its Mg/Si ratio is close to the solar and bulk Earth values. MgSiO$_{3}$ has six known major high-pressure polymorphs: orthopyroxene (enstatite, $en$), clinoenstatite ($hpcen$), akimotoite ($ak$), majorite garnet ($mj$), bridgmanite ($bg$), and post-perovskite ($ppv$). Our inversion places the \textit{bg}–\textit{ppv} transition at 120 GPa, 2,000 K, and 135 GPa, 3,100 K, with a slope of +14 MPa/K. The \textit{mj}–\textit{bg} transition occurs at 21 GPa, 2,150 K, and 23 GPa, 2,550 K, with a slope of +5 MPa/K. Other transitions include \textit{en}–\textit{hpcen} (+3 MPa/K), \textit{hpcen}–\textit{mj} (–3 MPa/K), and \textit{rw}+\textit{st} $\leftrightarrow$ \textit{bg} (–4 MPa/K). Data points contributed by static compression experiments, dynamic compression experiments, and computer simulations are marked with circles ($\bigcirc$), triangles ($\bigtriangleup$), and squares ($\Box$), respectively, and coded with different colors based on the stable phase observed at that $P$--$T$ condition.}
\label{fig:mgsio3}
\end{figure}

\subsection{MgSiO$_{3}$}\label{mgsio3}

MgSiO$_{3}$ is a prototypical silicate mineral for the mantle of Earth and super-Earth exoplanets, and six of its polymorphs have been extensively studied experimentally: orthopyroxene (enstatite, $en$), clinoenstatite ($hpcen$), akimotoite ($ak$), majorite garnet ($mj$), bridgmanite ($bg$), and post-perovskite ($ppv$). In addition, clinoenstatite dissociates into wadsleyite plus stishovite ($wa$+$st$) and ringwoodite plus stishovite ($ri$+$st$), recombining into $ak$ at 15--30 GPa. These transition and reaction boundaries have been studied extensively because of their influence on mantle dynamics and their seismic visibility \citep[e.g.][]{schubert_mantle_2001,stixrude_thermodynamics_2005,tackley_mantle_2013,miyagoshi_thermal_2015,faccenda_role_2017,shahnas_penetrative_2018,shahnas_focused_2021}. However, the exact slopes and locations of these boundaries in \textit{P--T} space remain controversial due to experimental difficulties, such as slow kinetics in the recombination of dissociated phases \citep{chanyshev_depressed_2022,dong_nonlinearity_2025}, or difficulty in reliably accessing the ultra-high pressures required for $ppv$\citep[e.g.][]{hirose_postperovskite_2006}. 

 In our inversion (Figure \ref{fig:mgsio3}), the \textit{bg}--\textit{ppv} transition occurs at 120 GPa and 2,000 K, and at 135 GPa and 3,100 K, with a positive slope of approximately $+14$ MPa/K; the \textit{mj}--\textit{bg} transition occurs at 21 GPa and 2,150 K, and at 23 GPa and 2,550 K, with a positive slope of approximately $+5$ MPa/K. Our $mj$--$bg$ slope falls between those reported in the literature \cite[$+0.8$ to $+6.4$ MPa/K;][]{faccenda_role_2017}, and our $bg$--$ppv$ slope is also consistent with the latest state-of-the-art DAC experiments \cite[$+13\pm1$ MPa/K;][]{tateno_determination_2009}, indicating that the $ppv$ phase is likely to form at the Earth's core--mantle boundary when the temperature cools to 2,500--3,000 K. We also find that the \textit{en}--\textit{hpcen} transition occurs at 7 GPa and 1,200 K, and at 10 GPa and 2,100 K, with a positive slope of approximately $+3$ MPa/K; the \textit{rw}+\textit{st}$\leftrightarrow$\textit{bg} transition occurs at 23 GPa and 1,450 K, and at 21 GPa and 1,950 K, with a negative slope of approximately $-4$ MPa/K; and the \textit{hpcen}--\textit{mj} transition occurs at 15.5 GPa and 2,300 K, and at 16.5 GPa and 2,000 K, with a negative slope of approximately $-3$ MPa/K. 

We present the inversion with a database of MgSiO$_{3}$ phase stability from 34 studies \citep{boyd_rhombic_1959,boyd_effects_1964,akimoto_pyroxenegarnet_1977,ito_mgsio3_1985,kato_garnet_1985,angel_structure_1989,ito_postspinel_1989,manghnani_phase_1987,pacalo_reversals_1990,presnall_melting_1990,kanzaki_orthoclinoenstatite_1991,syono_melting_1992,zerr_melting_1993,shen_measurement_1995,shinmei_situ_1999,kuroda_determination_2000,ono_situ_2001,oganov_theoretical_2004,ono_situ_2005,guignot_thermoelastic_2007,katsura_pvt_2009,akashi_orthoenstatiteclinoenstatite_2009,tateno_determination_2009,yoshiasa_single-crystal_2013,tateno_melting_2014,ono_reaction_2017,ono_decomposition_2018,kulka_bridgmaniteakimotoitemajorite_2020,fei_melting_2021,zhou_sound_2021,PIERRU2022117770,chanyshev_depressed_2022,deng_melting_2023,okuda_electrical_2024}, with two additional interventions:

\begin{enumerate}
    \item We have combined the two-phase fields ($wd$ + $st$ and $ri$ + $st$) and $ak$ into one field \citep{akimoto_pyroxenegarnet_1977,manghnani_phase_1987,ito_mgsio3_1985,ono_reaction_2017,ono_decomposition_2018}, due to the difficulty of obtaining a satisfactory inversion when the two-phase fields are treated separately.
    \item In some cases, this may be due to the kinetics in dissociation reactions being more sluggish than in polymorphic phase transitions, so we perform the global inversion with the exclusion of the data collected below 1,000 K \citep{boyd_effects_1964,manghnani_phase_1987,shinmei_situ_1999,oganov_theoretical_2004,ono_equation_2006,guignot_thermoelastic_2007,katsura_pvt_2009,tateno_melting_2014}, where the kinetic barrier is likely much more severe \citep{kuroda_determination_2000}.
\end{enumerate}
Similar to the relabeling of re-entrant bcc-Fe
fields (section \ref{fe}), these two interventions in MgSiO$_3$ again highlight the need for data relabeling that sacrifices resolution for the overall performance of the inversion.

 At mantle conditions of super-Earth exoplanets ($>$1,000--2,000 GPa), $ppv$ is predicted to completely dissociate into oxides \citep{tsuchiya_prediction_2011,niu_prediction_2015,umemoto_phase_2017}. Additional dissociation reactions of $ppv$ have been hypothesized to occur in super-Earth mantles (above 800 GPa) before it completely dissociates into B2-MgO + Fe$_{2}$P-type SiO$_{2}$, including $I\overline{4}2d$-type Mg$_{2}$SiO$_{4}$ + $P2_{1}/c$-type MgSi$_{2}$O$_{5}$ or $I\overline{4}2d$-type Mg$_{2}$SiO$_{4}$ + Fe$_{2}$P-type SiO$_{2}$ \citep{niu_prediction_2015,umemoto_phase_2017}, but these reactions have not been confirmed by experiments. Therefore, the relevant phase boundaries are indicated with dashed lines in Figure \ref{fig:mgsio3}.

The melting curve of MgSiO$_{3}$ is less constrained than those of Fe, MgO, and SiO$_{2}$ at similar pressures (Figure \ref{fig:mgsio3}). The data coverage for its melting curve is very sparse for $hpcen$ and $mj$ between 5 and 20 GPa and for $bg$ and $ppv$ above 60 GPa. In our inversion, we find the melting temperature of $bg$ to be 3,200 K at 30 GPa and 4,400 K at 60 GPa with a slope of $+40$ K/GPa, while that of $ppv$ is 8,500 K at 300 GPa and 10,000 K at 600 GPa with a slope of approximately $+8$ K/GPa. From 1 bar to 40 GPa, our melting curves of low-pressure MgSiO$_{3}$ polymorphs generally agree with the exisiting experimental and computational data \citep{boyd_effects_1964,kato_garnet_1985,presnall_melting_1990,syono_melting_1992,zerr_melting_1993,shen_measurement_1995,PIERRU2022117770,deng_melting_2023}. The slope of our $bg$  melting curve falls between 30 K/GPa \citep{syono_melting_1992} and 60--80 K/GPa \citep{zerr_melting_1993,deng_melting_2023}, and our $ppv$ melting curve closely reproduces the shock compression experiments up to 1,000 GPa \citep{fei_melting_2021}.

Although our inversion captures both the low and high pressure portions of the MgSiO$_3$ melting curve reasonably well to first order and is likely the most faithful representation of the existing data, we are unable to evaluate the robustness of the triple points. The data used to constrain these regions are extremely sparse and scattered. For example, we observe a negative melting curve near the $bg$--$ppv$ liquid triple point \ref{fig:mgsio3}. While such a negative slope for $bg$ could be justified by the potential diminishing volume of MgSiO$_3$ liquid \citep[e.g.][]{stixrude_structure_2005} and the similar behavior observed in some petrological systems \citep[e.g. in the MgSiO$_3$--MgCO$_3$ system,][]{thomson_experimental_2014}, the possibility that this negative curve is a numerical artifact cannot be ruled out, as the coverage and quality of the data near these triple points remain too sparse to allow a robust inversion (See Appendix \ref{app_test}). Additional high-precision experimental data on the melting of $bg$ and $ppv$ are essential to better constrain the $bg$--$ppv$--liquid triple point.

\begin{figure*}
\centering
\noindent\includegraphics[width=0.75\textwidth]{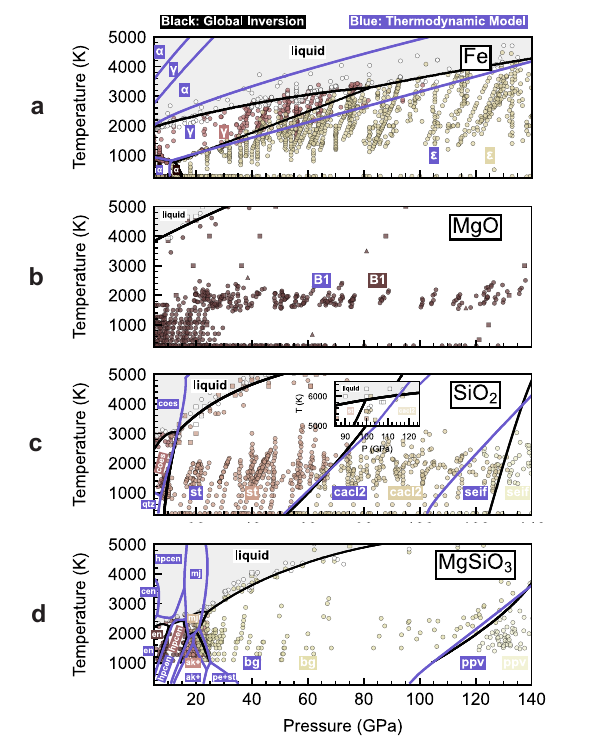}
\caption{\textbf{Comparison Between the Global Inversion and the Thermodynamic Model}. The solid--solid phase boundaries predicted by the thermodynamic code, HeFESTo (blue lines) and our results (black lines) are in strong agreement, validating the global inversion method. The main differences include the slope of the $\alpha$--$\epsilon$ transition in Fe and the \textit{st}--\textit{cacl2} and \textit{cacl2}--\textit{seif} boundaries in SiO$_{2}$, which are due to the inclusion of additional high-temperature data to our inversion.}
\label{fig:hefesto}
\end{figure*}

\subsection{Validation with Thermodynamic Calculations} \label{compare_hefesto}

In Figure \ref{fig:hefesto}, the solid phase equilibria of Fe, MgO, SiO$_{2}$, and MgSiO$_{3}$ are predicted by the thermodynamic model HeFESTo (shown in blue) to validate our global inversion results (shown in black). In the liquid stability fields of our results, since no liquid phase is included in the current parameter set for HeFESTo, additional supersolidus (and fictitious) phases are predicted: the $\alpha$ and $\gamma$ phases in Fe and the \textit{cen} and \textit{hpcen} phases in MgSiO$_{3}$.

Overall, there is strong agreement between the phase relations determined by our global inversion using extensive experimental and computational data (Section \ref{method}) and those predicted by HeFESTo, which has its current parameter set anchored to a minimal set of carefully selected data \citep[see Appendix \ref{app_data} in][]{stixrude_thermodynamics_2011}. Nevertheless, some differences are found in Fe and SiO$_{2}$ (Figure \ref{fig:hefesto}a--c). 

For Fe, HeFESTo predicts a slightly gentler slope for the $\gamma$--$\epsilon$ transition compared to the global inversion. This small difference could cause a significant shift in the liquid--$\gamma$--$\epsilon$ triple point, from $\sim$70 GPa to $\sim$100 GPa, given that the melting curve of iron has a similar slope. However, the majority of the $\gamma$-Fe enrties in our database are stable up to 80 GPa at most, so we consider the global inversion to be more faithful to the existing experimental constraints. The $\alpha$--$\epsilon$ transition also differs between the global inversion and the HeFESTo prediction. All direct experimental constraints on this boundary show a slightly negative slope \citep{johnson_temperature_1962,bundy_pressuretemperature_1965,giles_high-pressure_1971}; the global inversion suggests a more gentle negative slope, while HeFESTo predicts a steep positive slope. Both methods were unable to reproduce the experimental data, which we attribute to the confusion caused by the discrepancy of the metastable $\alpha$-Fe at 300 K. 

For SiO$_{2}$, the \textit{cacl2}--\textit{seif} boundary differs between the two methods. This is because the current parameter set for HeFESTo is mainly based on the the \textit{cacl2}--\textit{seif} transition data at 300 K, while the global inversion incorporates part of the experimental data at 1000--3000 K reported by \cite{sun_high_2019}. A smaller difference in the slope is also seen in the \textit{st}--\textit{cacl2} boundary at high temperatures. We consider the global inversion a better presentation of the phase relation of SiO$_{2}$ because additional liquid data at higher temperatures are incorporated and the global inversion better reproduces the topology of the \textit{st}--\textit{cacl2}--liquid triple point (Figure \ref{fig:hefesto}, inset). The \textit{qtz} stability field from the global inversion differs from HeFESTo, which is inaccurate but expected since the global inversion focuses on the higher pressure region and is not specifically optimized for the \textit{qtz} field.

\section{Implications} \label{implications}

\begin{figure*}
\centering
\noindent\includegraphics[width=0.8\textwidth]{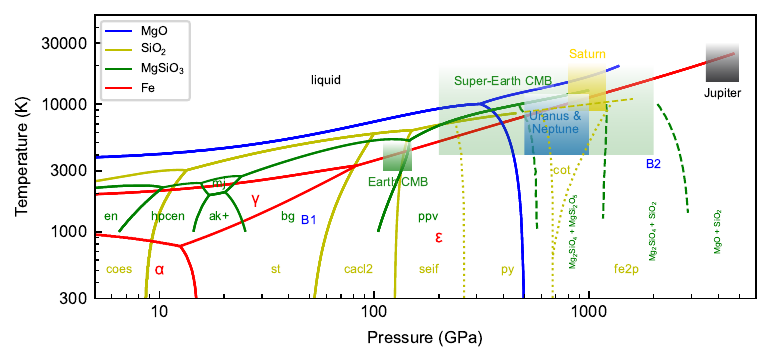}
\caption{\textbf{\textit{P--T} Phase Diagrams of Fe, MgO, SiO$_{2}$, and MgSiO$_{3}$ up to 5,000 GPa.} This figure highlights the relative refractoriness of each material with respect to the $P$--$T$ conditions at the core-mantle boundaries of Earth and super-Earth exoplanets, as well as the core conditions of giant planets (Earth CMB: forest green, super-Earth CMB: light forest green, Uranus and Neptune: sky blue, Saturn: gold, and Jupiter: midnight black). Up to 500 GPa, Fe has the lowest melting temperatures compared to MgO, SiO$_{2}$, and MgSiO$_{3}$. Among rocky components, MgSiO$_{3}$ has the lowest melting temperature below 200 GPa, but above 200 GPa, SiO$_{2}$ becomes the least refractory. Up to 1,000 GPa, MgO remains significantly more refractory than Fe, MgO, MgSiO$_{3}$. Solid lines indicate globally inverted phase boundaries; dashed and dotted lines represent less constrained boundaries, presented to guide the eye based on the literature.}
\label{fig:cores}
\end{figure*}
\begin{figure*}
\centering
\noindent\includegraphics[width=\textwidth]{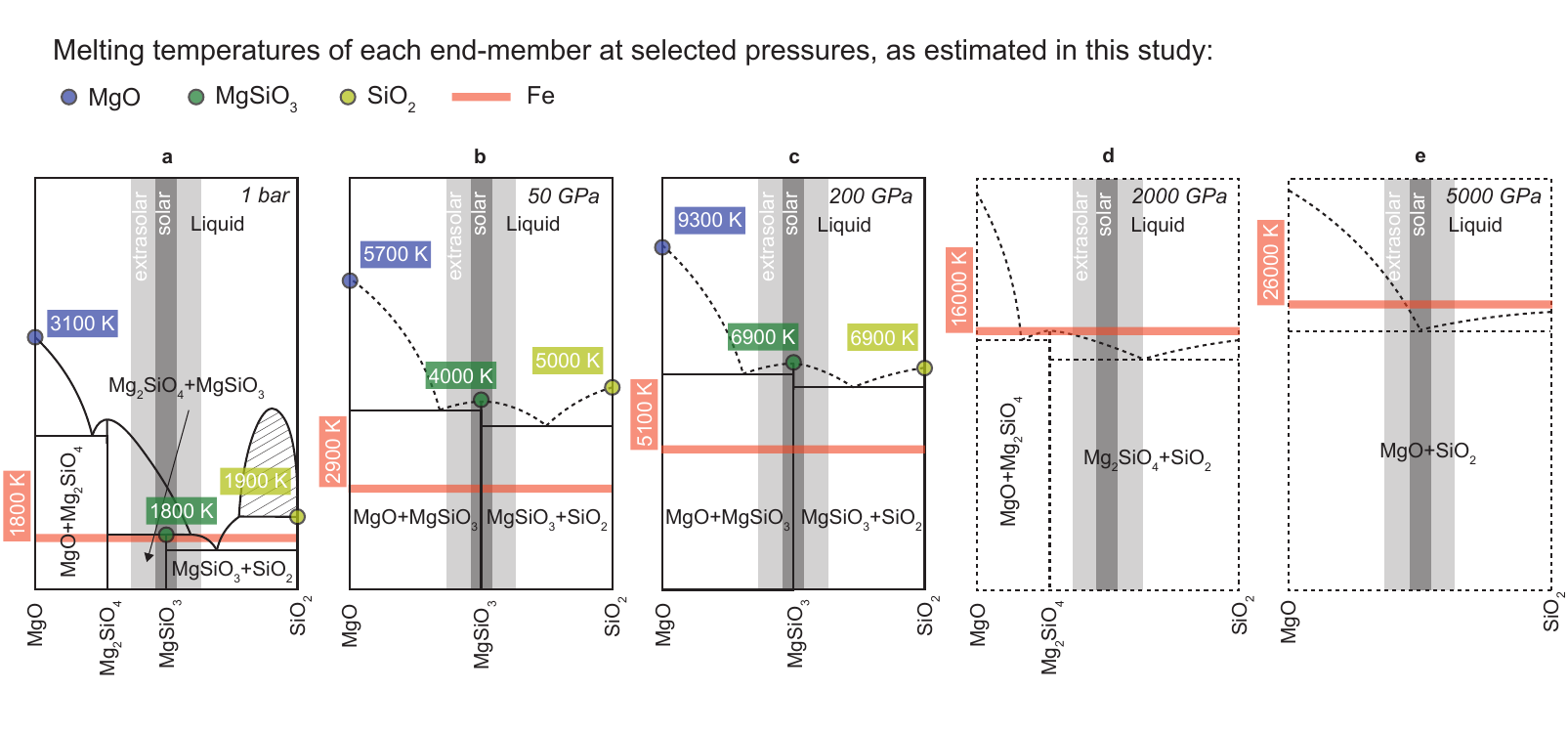}
\caption{\textbf{Proposed Melting Relations for the MgO--SiO$_{2}$ Binary System Up to 5,000 GPa.} The pressure--temperature--composition ($P$--$T$--$X$)  phase diagrams of the MgO--SiO$_{2}$ binary system show its melting behavior, in comparison to that of Fe at (\textbf{a}) 1 bar, (\textbf{b}) 50 GPa, (\textbf{c}) 200 GPa, (\textbf{d}) 2,000 GPa, and (\textbf{e}) 5,000 GPa. Eutectic points in the $T$--$X$ diagram indicate the compositions required for the lowest possible melting temperature for the mixture of the two components of a binary joint. (\textbf{a}--\textbf{c}) Between 1 bar and 200 GPa, their eutectic temperatures and compositions are constrained experimentally and through thermodynamic modeling \citep[e.g.][]{bowen_evolution_1928,liebske_melting_2012,de_koker_thermodynamics_2013,ohnishi_melting_2017,baron_experimental_2017,ozawa_experimental_2018,yao_lower_2021}. (\textbf{d}--\textbf{e}) At higher pressures (200--5000 GPa), the phase relations are largely unconstrained, potentially with new silicate phases forming \citep[e.g., MgSi$_2$O$_5$ and Mg$_2$SiO$_4$,][]{tsuchiya_prediction_2011,niu_prediction_2015,umemoto_phase_2017,dutta_ultrahigh-pressure_2022,dutta_high-pressure_2023}. The inferred upper bound of the eutectic temperatures for these assemblages, based on the end members (Figure \ref{fig:cores}), indicates that the MgO--SiO$_2$ eutectics and the Fe melting curve are likely to intersect between 1,000 and 2,000 GPa. (Section \ref{melting}).}
\label{fig:binary}
\end{figure*}

\subsection{Melting Inside Super-Earths} \label{melting}

The melting behavior of silicate rocks at high pressures, even at the Earth's core-mantle boundary (CMB) pressures of about 140 GPa, is not well understood, much less at super-Earth CMB pressures of 1,000--2,000 GPa. This challenge arises not only from the lack of constraints on the melting temperature of simple rock components, such as oxides and MgSiO$_{3}$ (Figures \ref{fig:mgo}, \ref{fig:sio2}, and \ref{fig:mgsio3}), but also from the complexities of mixing between these components, which causes melting temperature depression in multicomponent systems. For example, even fundamental details such as the eutectic temperatures and compositions of the MgO--SiO$_{2}$ system remain unexplored beyond 200 GPa \citep[e.g.,][Figure \ref{fig:binary}]{bowen_evolution_1928,liebske_melting_2012,de_koker_thermodynamics_2013,ohnishi_melting_2017,baron_experimental_2017,ozawa_experimental_2018,yao_lower_2021}. 

\begin{figure}
\centering
\noindent\includegraphics[width=0.45\textwidth]{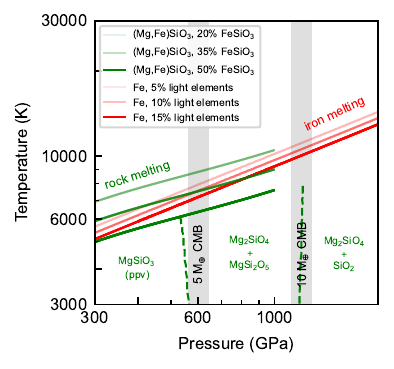}
\caption{\textbf{Prevalence of Basal Magma Oceans in the Interior of Super-Earth Exoplanets.} Small amounts of iron lower the MgSiO$_3$ melting curve \citep[e.g.][]{stixrude_melting_2014}, and iron prefers to partition into the silicate melt with increasing pressure, making a basal magma ocean (BMO) more likely on super-Earths \citep{dragulet_partitioning_2024,lherm_thermal_2024}. For example, in a super-Earth of 10 M$_{\oplus}$, the MgSiO$_3$ mantle with $>30$\% iron could become less refractory than the Fe core, favoring BMO formation. However, the emergence of new silicate phases at $>500$ GPa (Figure \ref{fig:mgsio3}, dashed lines) could suppress BMO formation by steepening the MgSiO$_3$ melting curve \citep[e.g.][]{tsuchiya_prediction_2011,umemoto_phase_2017,dutta_ultrahigh-pressure_2022,dutta_high-pressure_2023}.}
\label{fig:bmo}
\end{figure}

Although we cannot yet study the melting and crystallization processes of a super-Earth exoplanet with petrological resolution, some understanding can be gained by considering the melting behavior of MgSiO$_{3}$ as a prototypical mantle silicate and comparing it with that of Fe. In Figures \ref{fig:cores}--\ref{fig:bmo}, we show that the melting temperature of Fe is consistently lower than that of MgSiO$_{3}$ from 1 bar to 1,000--2,000 GPa \citep[the likely CMB pressures of a 5--10 M$_{\oplus}$ super-Earth, e.g.][]{stixrude_melting_2014,duffy_mineralogy_2015,boujibar_superearth_2020}. This relative refractoriness of MgSiO$_{3}$ to Fe becomes fundamental to the ability of a super-Earth to self-regulate its secular cooling, and a simple physical picture of which can be understood as follows \citep[cf.][and references therein]{stixrude_melting_2014}: A super-Earth exoplanet is likely to be completely molten after accretion, with the kinetic energy of accretion converted to heat sufficient to melt a super-Earth several times over, especially from giant impacts \citep{tonks_magma_1993,nakajima_scaling_2021}. The initially molten planet cools rapidly until it reaches the rock--iron solvus closure temperature \citep[e.g.][]{wahl_high-temperature_2015,insixiengmay_mgo_2025}, at which point phase separation begins, with iron-rich liquid exolving from the silicate-rich phase. Since we have shown that MgSiO$_{3}$ remains more refractory than Fe at all likely CMB pressures of a 5--10 M$_{\oplus}$ super-Earth (Figures \ref{fig:cores}--\ref{fig:bmo}), assuming effective separation of iron-rich liquid and silicate-rich liquid \citep[e.g.][]{solomatov_magma_2007} as the temperature continues to cool toward the melting temperature of MgSiO$_{3}$, rapid crystallization of silicate solids is expected while the Fe remains completely molten. The sudden increase in silicate viscosity from liquid to solid makes it easier for the mantle and core to convect separately \citep{newsom_physics_1990,newsom_fluid_1990,solomatov_magma_2007}, establishing a thermal boundary layer between them that slows heat loss from the core \citep{newsom_heat_1990,solomatov_magma_2007,stixrude_melting_2014}. While the convecting mantle continues to cool, likely along its isentropes, the core remains hot and largely molten \citep{newsom_heat_1990,stixrude_melting_2014,wahl_high-temperature_2015}. 

Comparing the melting curves of MgSiO$_{3}$ and Fe provides a first-order approximation for the internal structure of a super-Earth exoplanet. In such a simple idealization, super-Earth exoplanets, regardless of their size, are likely to have an Earth-like structure, with a solid silicate mantle and a partially or fully molten iron core. However, the mantle of a super-Earth exoplanet would not consist entirely of MgSiO$_{3}$. This simple idealization is less relevant in reality. The composition of extrasolar systems exhibits a wide range of Mg/Si ratios, approximately from 0.5 to 1.5, as observed in nearby stars \citep{hinkel_stellar_2014,buder_galah_2021}. The super-Earth mantle could therefore have either excess MgO or excess SiO$_{2}$, which would allow for complex mineralogical structures \citep{tsuchiya_prediction_2011,niu_prediction_2015,umemoto_phase_2017}. 

Here we qualitatively construct the $P$--$T$--$X$ phase diagrams of the MgO--SiO$_2$ system and compare its melting behavior with that of Fe (Figure \ref{fig:binary}). At pressures between 50 and 200 GPa (Figure \ref{fig:binary}b--c), corresponding to the Earth's lower mantle and the upper part of a super-Earth mantle, two polymorphs of MgSiO$_3$, $bg$ and $ppv$, are the stable intermediates that split the MgO--SiO$_2$ system into two smaller binary joints: MgO--MgSiO$_3$ and MgSiO$_3$--SiO$_2$. The eutectic temperatures and compositions of both binary joints have been estimated from experimental constraints and thermodynamic modeling \citep{liebske_melting_2012,de_koker_thermodynamics_2013,ohnishi_melting_2017,baron_experimental_2017,ozawa_experimental_2018,yao_lower_2021}. In this pressure range, the eutectic temperatures of the MgO--MgSiO$_3$ and MgSiO$_3$--SiO$_2$ joints are very similar and slightly lower than the melting temperature of MgSiO$_3$ \citep[e.g.,][Figure \ref{fig:binary}d--e]{yao_lower_2021}. As a result, the melting temperature of MgSiO$_3$ serves as an upper limit for mantle melting (for Mg/Si ratios between 0.5 and 1.5) throughout the stability field of $ppv$, up to about 500 GPa.

From 500 to 5,000 GPa, the phase relations of the MgO--SiO$_2$ system are virtually unconstrained due to the lack of direct experimental observations of several proposed magnesium silicates of different stoichiometries and their assemblages: MgSi$_2$O$_5$ \citep[][the decomposition products of MgSiO$_3$ predicted between 500 and 1,000 GPa]{niu_prediction_2015,umemoto_phase_2017}, and Mg$_2$SiO$_4$ \citep[][suggested by both theoretical prediction and shock experiments, the congruently melting intermediate between 1,000 and 2,000 GPa]{dutta_ultrahigh-pressure_2022,dutta_high-pressure_2023} may be present instead of $ppv$ or a higher pressure polymorph of MgSiO$_3$. Given the expected Mg/Si range in the super-Earth mantle, the dominant mineralogies at these pressures may include MgSi$_2$O$_5$ + Mg$_2$SiO$_4$ (1,000 GPa), Mg$_2$SiO$_4$ + SiO$_2$ (2,000 GPa), and MgO + SiO$_2$ (5,000 GPa) (Figure \ref{fig:binary}d--e). Since the melting temperatures of these intermediates are also unknown, we can again use the extrapolated melting temperatures of MgSiO$_3$ as a generous upper bound for mantle melting (Figure \ref{fig:cores}), ranging from about 12,000--14,000 K at 1,000 GPa to 15,000--17,000 K at 2,000 GPa (Figure \ref{fig:cores}), although this is likely an overestimate. Since the difference in melting temperature between MgSiO$_3$ and Fe starts diminishing with increasing pressure beyond 200 GPa, the eutectic temperatures of the actual mineralogical assemblages (such as Mg$_2$SiO$_4$ + MgSi$_2$O$_5$ or Mg$_2$SiO$_4$ + SiO$_2$) are likely to become lower than the melting temperature of the Fe core near 1,000--2,000 GPa, close to the core-mantle boundary pressure of a 5--10 M$_{\oplus}$ super-Earth. The pressure at which the eutectics of MgO--SiO$_2 $ and the Fe melting curve intersect remains uncertain, since the subsolidus mineral assemblages and their corresponding melting behavior above 200 GPa are not yet known (Figure \ref{fig:binary}d--e).

Last but not least, small amounts of iron in mantle silicates can further lower their melting temperatures \citep{PIERRU2022117770}. We can extend the prototypical MgSiO$_{3}$ mantle to a pseudo-binary system of MgSiO$_{3}$--FeSiO$_{3}$ to explore the potential prevalence of basal magma oceans (BMO) in super-Earth exoplanets. Crystallization of the silicate mantle tends to enrich the BMO with up to 20\% iron at the base of the Earth's mantle \citep{tateno_melting_2014,dragulet_partitioning_2024}, and the extent of iron enrichment in the silicate liquid becomes more pronounced as iron becomes less compatible with silicate solids with increasing pressure and tends to partition into the coexisting liquid \citep{francis_dragulet_electrical_2023,lherm_thermal_2024}. The increasing iron enrichment would further decrease the melting temperature of the rock in a super-Earth mantle, and this effect can be approximated as the effect of iron on the melting depression of MgSiO$_{3}$ in the MgSiO$_{3}$--FeSiO$_{3}$ pseudo-binary using the cryoscopic equation \citep[e.g.][]{stixrude_melting_2014}: $T_{(\textrm{Mg,Fe})\textrm{SiO}_{3}} = T_{\textrm{Mg}\textrm{SiO}_{3}} \left(1 - \ln x_{\textrm{Mg}\textrm{SiO}_{3}} \right)^{-1}$, where $T_{(\textrm{Mg,Fe})\textrm{SiO}_{3}}$ is the actual (Mg,Fe)SiO$_{3}$ melting temperature, $T_{\textrm{Mg}\textrm{SiO}_{3}}$ is that of the pure MgSiO$_{3}$, and $x_{\textrm{Mg}\textrm{SiO}_{3}}$ is the mole fraction of the pure MgSiO$_{3}$. With $x_{\textrm{Mg}\textrm{SiO}_{3}}$ = 80\% at the bottom of Earth's CMB and 50\% at the bottom of a 10 M$_{\oplus}$ super-Earth, the silicate mantle could potentially become less refractory than the Fe core (Figure \ref{fig:bmo}). In addition, more massive planets tend to retain heat longer \citep{boujibar_superearth_2020}, preserving such a BMO. Since the melting temperatures of iron-enriched (Mg,Fe)SiO$_{3}$ and Fe could intersect at 500--1,000 GPa, a BMO is more likely to form within a large super-Earth exoplanet. Unlike the fully solidified silicate mantle of Earth, the BMO inside super-Earths, especially those with 5--10 M$_{\oplus}$, would play a critical role in the geometry and evolution of their magnetic fields. This iron-enriched BMO could generate its own dynamo action due to the significant increase in electrical conductivity with pressure and iron enrichment \citep{francis_dragulet_electrical_2023,dragulet_partitioning_2024,lherm_thermal_2024}.

An alternative scenario that could suppress BMO formation in a large super-Earth involves the stability of solid silicate phases. Post-perovskite, stable at 100--200 GPa, could transform or dissociate into new, more refractory phases at higher pressures, steepening the MgSiO$_{3}$ melting curve sufficiently to make the silicate mantle more refractory than the Fe core. While these new solid phases are not well characterized thermodynamically, they are expected to occur in the range of 500--1,000 GPa based on new evidence from recent experimental and computational studies \citep{tsuchiya_prediction_2011,niu_prediction_2015,umemoto_phase_2017,dutta_ultrahigh-pressure_2022,zurkowski_synthesis_2022,dutta_high-pressure_2023}.

Although our estimate of the FeO anti-freeze effect using the cryoscopic equation is in agreement with experimental observations and thermodynamic modeling of the MgO--FeO system \citep{boukare_thermodynamics_2015,miyazaki_timescale_2019}, a self-consistent thermodynamic model requires at least considering the solidi and liquidi of the MgO--SiO$_{2}$--FeO system to better approximate the melting and crystallization of a realistic super-Earth mantle. We have not yet been able to perform a global inversion for the $P$--$T$ phase diagrams of FeO and FeSiO$_3$ due to the scarcity and inconsistency of experimental data, as pure FeO is difficult to synthesize and often exists as non-stoichiometric compounds, Fe$_{1-x}$O \citep[e.g.][]{fischer_equation_2011}, while FeSiO$_3$ does not occur naturally as a pure mineral phase and cannot be synthesized at ambient pressure \citep[e.g.][]{lindsley_ferrosilite_1964}, limiting high pressure studies. Despite the lack of experimental data on the iron end members, our conceptual analysis of the FeO effect on silicates is heuristic and shows how iron enrichment could alter silicate phase stability and melting behavior at pressures relevant to 5--10 M$_{\oplus}$ super-Earths.

\subsection{Frozen Cores of Giant Planets} \label{core}

Another class of planets for which the melting curves determined in this study are relevant are giant planets. Unlike terrestrial planets, giant planets are characterized by an abundance of hydrogen plus helium and/or icy envelopes \citep[cf.][and references therein]{lewis_physics_2004}. Their deep interiors are enriched in ``metals''---elements heavier than hydrogen and helium. These heavier elements are most abundant in Uranus and Neptune, less abundant in Saturn, and least abundant in Jupiter \citep{swain_planet_2024}.

Two leading models describe the interiors of giant planets: the fuzzy core model and the compact core model. Their primary difference is expressed in the structure, or more specifically the density profile of the interior, which largely correlate with the temperature of the planet. The fuzzy core model proposes a compositional gradient, with extensive mixing of light and heavy materials in the fluid state, which would least require a fully molten interior \citep[e.g.,][]{stevenson_formation_1982,muller_challenge_2020,vazan_new_2022,helled_fuzzy_2024}. 

In contrast, the compact core model suggests a layered structure similar to terrestrial planets, with chemically distinct regions. One way to form such chemically distinct regions is by crystallization of a solid core \citep[e.g.][]{nettelmann_uranus_2016,stixrude_thermal_2021-1,militzer_relation_2023,james_thermal_2024,militzer_study_2024}.

Observations of a compact core, or its absence, via moments of inertia or non-zonal gravity fields, serve as key diagnostics for the potential existence of such a solid core, and help to distinguish these competing models. A solid core implies an adiabatic interior supporting effective heat transport \citep[e.g.][]{stixrude_thermal_2021-1,militzer_relation_2023}. Conversely, evidence for a compositional gradient or the absence of a compact core suggests a super-adiabatic interior, possibly with inhibited convective mixing \citep[e,g,][]{stevenson_formation_1982,helled_fuzzy_2024}.

Using Saturn as an example, if we extrapolate the surface temperature of the planet along an adiabatic temperature profile to the base of the hydrogen-helium layer at 880 GPa, the temperature is 9,900 K \citep{militzer_relation_2023}. The detection of a compact core could imply a solid core, with its top limited to temperatures at least below the melting curve of the most refractory components in the silicate system, specifically the melting temperature of MgO under these conditions (Figure \ref{fig:mgo} and \ref{fig:cores}). This core surface temperature would most likely be even lower due to the melting temperature depression caused by complex compositions in the rock, as discussed in Section \ref{melting}. The evidence for a compact or sluggish core in Saturn based on tides and ring seismology \citep{lainey_new_2017,mankovich_diffuse_2021} may indicate that the planet has cooled sufficiently to freeze out a solid core and that the interior temperature distribution is close to adiabatic. Similar tests can be applied to future measurements for the Uranus Orbiter and Probe (UOP) mission \citep{simon_uranus_2021,hofstadter_uranus_2024}.

While the geophysical constraints on giant planets remain limited to date, and their availability is highly constrained by the long intervals between giant planet missions, we emphasize the crucial role of understanding the material properties of rocky materials and iron up to 5000 GPa in the interim. For example, as we have shown here, the melting temperature of rocks is fundamental to constraining their interior temperatures and can be studied through experimental laboratory work or computer simulations. In particular, further research on the melting temperature of realistic compositions of rocky materials, as well as the rock-ice mixture between 500 and 5,000 GPa, has profound implications for advancing our understanding of giant planets \citep[see][and references therein]{hofstadter_uranus_2024}.

\section{Conclusions} \label{conclusions}

In conclusion, we have presented a comprehensive phase equilibria database for Fe, MgO, SiO$_{2}$, and MgSiO$_{3}$ and applied a global inversion algorithm to generate $P$--$T$ phase diagrams for these materials up to 5,000 GPa. Our approach allows the mapping of their melting curves, solid-solid phase transition and reaction boundaries based on extensive experimental and computational phase stability data. Complementary to the conventional thermodynamic framework, this work demonstrates a convenient and effective method for determining high-pressure phase diagram by integrating extensive datasets with advanced statistical methods. Using our melting curves as a first-order approximation to the high-pressure melting temperature of rocky materials, we find that the potential crossover of the rock and iron melting curves may favor the formation of basal magma oceans inside massive super-Earth exoplanets. In addition, a sufficiently cooled giant planet could grow a frozen core at its center, and the melting temperatures of rocky materials can be used with geophysical observations to infer its internal temperature.




\newpage
\appendix
\section{Database Summary}\label{app_data}

We have compiled the experimental data available in the literature on Fe, MgO, SiO$_{2}$, and MgSiO$_{3}$ that satisfy the criteria defined in Section \ref{database}, supplemented by selected computational data, and present a summary of the database for each material in Tables \ref{tab:t_fe}–\ref{tab:t_mgsio3} in Appendix \ref{app_data}. Each table includes the following columns: 1) reference, 2) year, 3) phase(s) observed, 4) number of observations, 5) pressure range (P${_\textrm{min}}$, P${_\textrm{max}}$), 6) temperature range (T${_\textrm{min}}$, T${_\textrm{max}}$), and 7) method. The complete database is available in the High Pressure Phase Equilibria Database for Planetary Materials (HP-PEDPM) project on Zenodo (\url{https://doi.org/10.5281/zenodo.14853852}) and GitHub (\url{https://github.com/dong2j/HP-PEDPM.git}). While we have attempted to include as much literature data as possible, the database may not be exhaustive due to the large number of publications on these materials. This provisional version will be updated regularly as new data or previously overlooked literature becomes available. Updated datasets will be accessible via the aforementioned repositories on Zenodo and GitHub or by direct request to the authors.

\section{Synthetic Data Benchmark}\label{app_test}

\begin{figure}[!b]
\centering
\noindent\includegraphics[width=1\textwidth]{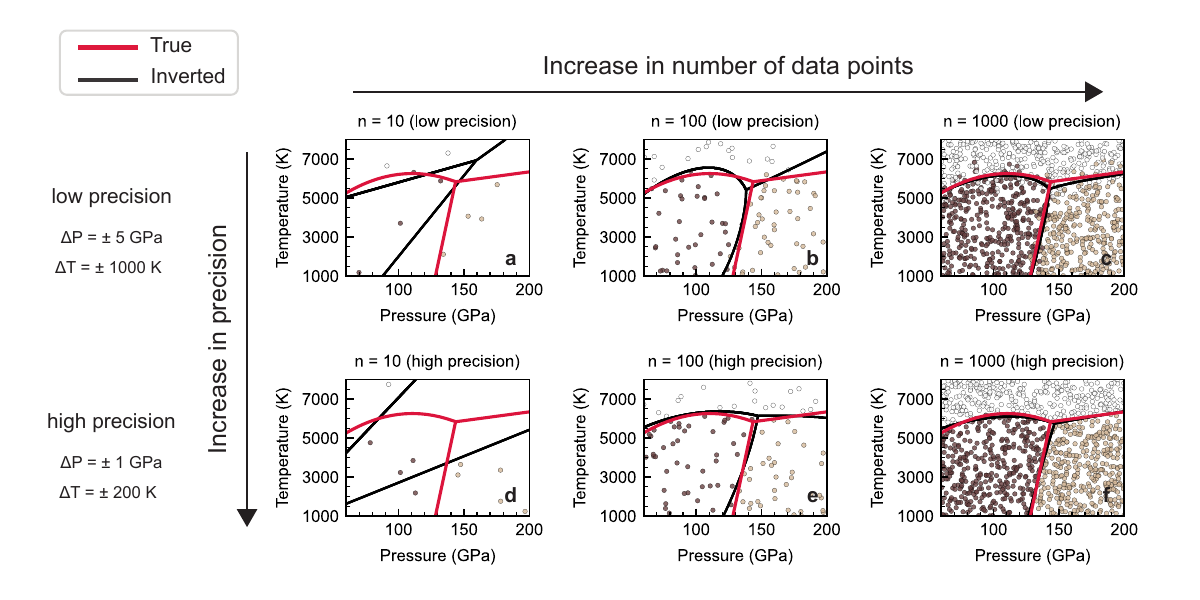}
\caption{\textbf{Comparison of True and Inverted Phase Boundaries for Synthetic Data Sets of Varying Size and Fidelity}. The rows illustrate the effect of simulated experimental precision, with the top row representing low-precision data ($\Delta$P = ±5 GPa, $\Delta$T = ±1000 K) and the bottom row representing high-precision data ($\Delta$P = ±1 GPa, $\Delta$T = ±200 K). The columns show increasing data size from left to right ($n$ = 10, 100, 1000). Red curves represent the true phase boundaries, while black curves represent the inverted boundaries. As precision and data size increase, the inverted boundaries more accurately recover the true locations of the phase boundaries and their slopes.}
\label{fig:test}
\end{figure}

To demonstrate the methodology, we performed a benchmark test using synthetic datasets of varying size and fidelity. We focued a pressure range of 60 to 200 GPa and a temperature range of 1,000 to 8,000 K, typical of planetary interiors where high-pressure polymorphism and melting are expected for many planetary materials. The true pressures and temperatures of the solid A and solid B melting curves and an A--B transition were predefined based on three independent and arbitrary equations. The liquid--solid A melting curve was set to have a slight negative slope, which may or may not be present in reality, depending on the material of interest. Random sampling was used to generate data sets of 10, 100, and 1,000 data points, each within a specified range of pressure (60--200 GPa) and temperature (1,000--8,000 K). Each data point was assigned a phase label based on its position relative to predefined phase boundaries. Data points close to the phase boundaries were probabilistically assigned to one of the two phases, introducing the maximum variability expected in high-pressure experiments. For high precision (typical or smaller for the MA data experiments), ΔT = ±200 K and ΔP = ±1 GPa, and for low precision (typical or smaller for the DAC data), ΔT = ±1,000 K and ΔP = ±5 GPa. This approach models the probabilistic nature of phase transition observations. We then applied global inversion algorithms, as described in the Methods section, to evaluate the effects of data point density and associated uncertainty on the quality of our inversion.

We examined datasets with different precisions (ΔP = ±5 GPa, ΔT = ±1,000 K for low precision, and ΔP = ±1 GPa, ΔT = ±200 K for high precision) and data densities (n = 10, 100, 1,000) to assess how these factors affect the fidelity of inverted phase boundaries. Our results indicate that for a range of 100 GPa and 7,000 K, 100 randomly distributed data points are sufficient to accurately recover the topology of the phase relations, assuming no internal inconsistency. Increased data density and precision both contribute to more accurate recovery of the true phase boundary slopes. At low precision, the inverted boundaries deviate significantly from the true boundaries, especially at lower data densities. However, as the precision improves, the inverted boundaries become closer to the true boundaries, even with fewer data points. In addition, higher data density improves the recovery of the true phase boundary slopes.

For the four materials in our dataset, we found that Fe, MgO, and SiO$_{2}$ have sufficient data to recover the topology and slope for most of their phase boundaries. For MgSiO$_{3}$, however, the current data may only be sufficient to constrain the general topology of its phase relations but not specific features such as triple points. For example, the $bg$--$ppv$--liquid triple point, where a negative slope found in the $bg$ melting curve may be an artifact of our inversion, while the melting curve away from the triple points remains robust. Future phase stability experiments targeting the $P$--$T$ space with sparse data coverage would be crucial to improve upon the existing phase boundary constraints presented in this work.

Assuming that the random errors of each individual experimental data set are generally limited to a few GPa and a few hundred K, and that no significant systematic error exists between data sets, we believe that the use of these uncorrected $P$--$T$ values is sufficient to recover the general topology of these phase diagrams (Appendix \ref{app_test}) and to infer the first-order structures of planetary interiors. Therefore, we have used the original $P$ and $T$ values reported in the literature without correcting them to a self-consistent pressure and/or temperature calibration. However, the use of different pressure scales and temperature measurement techniques can introduce systematic errors and biases the inversion when the number of data is small and the fidelity is poor. To recover the true slope of the phase boundary with better precision, all experimental data would have to be recalibrated to self-consistent pressure and temperature scales, which may or may not be feasible for all materials, and a detailed discussion can be found in \cite{dong_nonlinearity_2025}.

In general, the inversion method presented here operates under the simple assumption that the majority of our existing empirical evidence for phase stability reflects the actual stable phase at equilibrium. All phase boundaries found by our global inversion method are constructed to be as consistent as possible with all currently available empirical data on relevant phase relations. Although the scenario that the majority of our current experiments are flawed or contain unrecognized systematic errors is most likely counterfactual, we cannot rule out this hypothetical possibility.  The inverted phase boundaries may shift if new empirical evidence, contrary to the existing data, overwhelms the latter, and such a shift should not be unexpected.  This global inversion method does not claim absolute correctness of the phase diagram, but rather reflects an alignment with our current knowledge of the empirical evidence. Furthermore, successful inversion would require a sufficient amount of correctly labeled data with accurately reported values in pressure and temperature (Appendix \ref{app_test}), which is not guaranteed for the high pressure data given its limited quantity and scattering nature.
                  
\begin{longrotatetable} 
\begin{deluxetable*}{lcccccccc}
\tabletypesize{\scriptsize}
\tablewidth{0pt} 
\tablecaption{Summary of the Equilibria Database for Fe \label{tab:Fe}}
\tablehead{
\textbf{Reference} &
\textbf{Year} &
\textbf{Phase(s) observed} &
  \textbf{Number of observations} &
  \textbf{P$_{\textrm{min}}$ (GPa)} &
  \textbf{P$_{\textrm{max}}$ (GPa)} &
  \textbf{T$_{\textrm{min}}$ (K)} &
  \textbf{T$_{\textrm{max}}$ (K)} & \textbf{Method}
}
\startdata 
\cite{strong_experimental_1959} & 1959 & $\alpha$, $\gamma$, liquid & 61 & 0.3 & 9.8 & 457 & 1993 & static\\
\cite{claussen_detection_1960} & 1960 & $\alpha$, $\gamma$ & 68 & 0.9 & 10.1 & 804 & 1092 & static\\
\cite{johnson_temperature_1962} & 1962 & $\alpha$, $\gamma$, $\epsilon$ & 42 & 3.7 & 15 & 66 & 1024 & static\\
\cite{kaufman_lattice_1963} & 1963 & $\alpha$, $\gamma$ & 20 & 2.6 & 9.5 & 753 & 1010 & static\\
\cite{kennedy_solid-liquid_1963} & 1963 & $\alpha$, $\gamma$ & 20 & 0 & 4.6 & 916 & 1179 & static\\
\cite{clendenen_effect_1964} & 1964 & $\alpha$, $\epsilon$ & 52 & 3.9 & 39.9 & 298 & 298 & static\\
\cite{bundy_pressuretemperature_1965} & 1965 & $\alpha$, $\gamma$, $\epsilon$ & 25 & 6.2 & 18 & 418 & 948 & static\\
\cite{mao_effect_1967} & 1967 & $\alpha$, $\epsilon$ & 27 & 5 & 30.9 & 296 & 296 & static\\
\cite{giles_high-pressure_1971} & 1971 & $\alpha$, $\epsilon$ & 37 & 0 & 16.3 & 298 & 298 & static\\
\cite{strong_iron_1973} & 1973 & $\alpha$, $\gamma$, $\delta$, liquid & 99 & 1.4 & 5.7 & 278 & 2006 & static\\
\cite{liu_melting_1975} & 1975 & liquid & 8 & 2.9 & 19.7 & 1912 & 2347 & static\\
\cite{williams_melting_1987} & 1987 & $\gamma$, $\epsilon$, liquid & 26 & 1.7 & 102.2 & 1768 & 4362 & static \& dynamic\\
\cite{manghnani_internally-heated_1987} & 1987 & $\alpha$, $\gamma$, $\epsilon$ & 10 & 1.7 & 29.2 & 1029 & 1316 & static\\
\cite{manghnani_experimental_1987} & 1987 & $\gamma$, $\epsilon$ & 8 & 19.2 & 35.3 & 1071 & 1652 & static\\
\cite{secco_electrical_1989} & 1989 & $\alpha$, $\gamma$ & 156 & 2.5 & 3.3 & 346 & 1865 & static\\
\cite{boehler_melting_1990} & 1990 & liquid & 25 & 7.4 & 113.6 & 2037 & 3006 & static\\
\cite{boehler_temperatures_1993} & 1993 & $\gamma$, $\epsilon$, liquid & 57 & 7.3 & 196.6 & 1605 & 3855 & static\\
\cite{saxena_temperatures_1994} & 1994 & $\gamma$, $\epsilon$, liquid & 26 & 0.6 & 143.4 & 1802 & 3510 & static\\
\cite{yoo_phase_1995} & 1995 & $\gamma$, $\epsilon$, liquid & 64 & 7.6 & 132.6 & 517 & 3046 & static\\
\cite{shen_measurement_1995} & 1998 & $\gamma$, $\epsilon$, liquid & 52 & 12 & 84 & 1395 & 2991 & static\\
\cite{dubrovinsky_high-pressure_1998}& 1998 & $\gamma$, $\epsilon$ & 83 & 18 & 62 & 277 & 1678 & static\\
\cite{rutter_viscosity_2002}& 2002 & liquid & 3 & 1.6 & 5.5 & 2050 & 2050 & static\\
\cite{terasaki_viscosity_2002}& 2002 & liquid & 6 & 2.8 & 7 & 1965 & 2173 & static\\
\cite{shen_structure_2004}& 2004 & $\gamma$, $\epsilon$, liquid & 10 & 26.9 & 57.9 & 2047 & 2859 & static\\
\cite{ma_situ_2004}& 2004 & $\gamma$, $\epsilon$, liquid, mix & 85 & 28.9 & 243.4 & 263 & 6835 & static\\
\cite{assael_reference_2006}& 2006 & liquid & 14 & 0 & 0 & 1850 & 2500 & static\\
\cite{kuwayama_phase_2008}& 2008 & $\epsilon$ & 53 & 60.3 & 319 & 300 & 2300 & static\\
\cite{komabayashi_-situ_2009}& 2009 & $\gamma$, $\epsilon$, mix & 101 & 20.8 & 68.7 & 300 & 2409 & static\\
\cite{sola_melting_2009} & 2009 & $\epsilon$, liquid & 2 & 330 & 330 & 5750 & 6950 & theory\\
\cite{tateno_structure_2010}& 2010 & $\epsilon$ & 60 & 134.6 & 376.3 & 1959 & 5687 & static\\
\cite{konopkova_thermal_2011}& 2011 & $\epsilon$ & 30 & 38 & 70 & 990 & 1996 & static\\
\cite{morard_melting_2011}& 2011 & $\epsilon$, liquid & 11 & 335 & 1585 & 5920 & 14960 & theory\\
\cite{deng_high_2013}& 2013 & $\alpha$, $\gamma$, liquid & 61 & 5 & 7 & 294 & 2100 & static\\
\cite{jackson_melting_2013}& 2013 & liquid & 8 & 20 & 82.1 & 2812 & 4039 & static\\
\cite{anzellini_melting_2013}& 2013 & $\gamma$, $\epsilon$, mix & 385 & 49.9 & 206 & 1771 & 5335 & static\\
\cite{aquilanti_melting_2015}& 2015 & $\gamma$, $\epsilon$, liquid & 51 & 69.9 & 119.5 & 1670 & 3524 & static\\
\cite{kono_high-pressure_2015}& 2015 & liquid & 6 & 1.6 & 6.4 & 1873 & 2043 & static\\
\cite{zhang_temperature_2016}& 2016 & $\gamma$, liquid, mix & 48 & 19 & 110 & 1521 & 3412 & static\\
\cite{ohta_experimental_2016}& 2016 & $\gamma$, $\epsilon$, liquid & 80 & 26 & 184.8 & 300 & 4497 & static\\
\cite{secco_thermal_2017}& 2017 & $\alpha$, $\gamma$, liquid & 213 & 0 & 2.5 & 300 & 1881 & static\\
\cite{silber_electrical_2018}& 2018 & liquid & 9 & 3 & 12 & 1857 & 2038 & static\\
\cite{pommier_influence_2018}& 2018 & $\alpha$, $\gamma$, liquid & 41 & 4.5 & 4.5 & 701 & 1957 & static\\
\cite{morard_solving_2018}& 2018 & $\gamma$, $\epsilon$, liquid & 40 & 41.9 & 133 & 1856 & 4710 & static\\
\cite{sinmyo_melting_2019}& 2019 & $\gamma$, $\epsilon$, liquid & 425 & 6 & 290 & 1420 & 5360 & static\\
\cite{yong_iron_2019} & 2019 & $\epsilon$, liquid & 12 & 14 & 24 & 825 & 2177 & static\\
\cite{kuwayama_equation_2020}& 2020 & liquid & 16 & 16 & 116.1 & 2200 & 4340 & static\\
\cite{zhang_reconciliation_2020}& 2020 & $\epsilon$ & 93 & 82 & 177 & 292 & 3000 & static\\
\cite{li_shock_2020}& 2020 & $\epsilon$, liquid & 8 & 120.4 & 256 & 4258 & 5484 & dynamic\\
\cite{hou_melting_2021}& 2021 & $\gamma$, $\epsilon$, liquid & 375 & 35.3 & 135 & 100 & 4306 & static\\
\cite{kraus_measuring_2022}& 2022 & $\epsilon$, liquid, mix & 7 & 547.5 & 991.3 & 7191 & 11511 & dynamic\\
\cite{gonzalez-cataldo_ab_2023}& 2023 & $\epsilon$, liquid & 16 & 300 & 5000 & 5469 & 25996 & theory\\
\cite{sun_ab_2023}& 2023 & $\epsilon$, liquid & 4 & 323 & 360 & 5112 & 7393& theory
\enddata
\tablecomments{Abbreviations for mineral names: $\alpha$=body-centered cubic (bcc) $iron$, $\gamma$=face-centered cubic (fcc) $iron$, $\epsilon$=hexagonal close-packed (hcp) $iron$, and $\delta$=re-entrant body-centered cubic $iron$.}
\end{deluxetable*}
\label{tab:t_fe}
\end{longrotatetable}

\begin{longrotatetable}
\begin{deluxetable*}{lcccccccc}
\tabletypesize{\scriptsize}
\tablewidth{0pt} 
\tablecaption{Summary of the Equilibria Database for MgO \label{tab:MgO}}
\tablehead{
\textbf{Reference} &
\textbf{Year} &
  \textbf{Phase(s) observed} &
  \textbf{Number of observations} &
  \textbf{P$_{\textrm{min}}$ (GPa)} &
  \textbf{P$_{\textrm{max}}$ (GPa)} &
  \textbf{T$_{\textrm{min}}$ (K)} &
  \textbf{T$_{\textrm{max}}$ (K)} & \textbf{Method}
}
\startdata 
\cite{mcnally_laboratory_1961} & 1961 & B1, liquid & 2 & 0 & 0 & 3078 & 3118 & static\\
\cite{riley_determination_1966} & 1966 & B1, liquid & 2 & 0 & 0 & 3030 & 3070 & static\\
\cite{chernyshev_thermal_1993} & 1993 & B1, liquid & 2 & 0 & 0 & 3185 & 3245 & static\\
\cite{zerr_constraints_1994} & 1994 & B1, liquid & 16 & 0 & 31.6 & 2740 & 4264 & static\\
\cite{dubrovinsky_thermal_1997}& 1997 & liquid & 1 & 0 & 0 & 3098 & 3098 & static\\
\cite{manghnani_volume_1998} & 1998 & B1 & 68 & 0 & 9.5 & 300 & 1673 & static\\
\cite{fei_effects_1999}& 1999 & B1 & 62 & 0 & 66.3 & 300 & 1100 & static\\
\cite{fiquet_high-temperature_1999} & 1999 & B1 & 37 & 0 & 0 & 298 & 2973 & static\\
\cite{dewaele_pvt_2000} & 2000 & B1 & 61 & 0 & 53 & 300 & 2474 & static\\
\cite{speziale_quasihydrostatic_2001} & 2001 & B1 & 32 & 0.8 & 52.2 & 300 & 300 & static\\
\cite{ronchi_melting_2001} & 2001 & B1, liquid & 2 & 0 & 0 & 3230 & 3270 & static\\
\cite{alfe_melting_2005} & 2001 & B1, liquid & 8 & -0.4 & 135.6 & 3020 & 8184 & theory\\
\cite{zhang_melting_2008} & 2008 & B1, liquid & 6 & 3 & 7 & 3600 & 4740 & static\\
\cite{mcwilliams_phase_2012} & 2012 & B1, B2, liquid & 4 & 400 & 650 & 9000 & 15100 & dynamic\\
\cite{cebulla_ab_2014} & 2014 & B1, B2, liquid & 167 & 0.5 & 876.3 & 500 & 20000 & theory\\
\cite{du_highpressure_2014} & 2014 & B1, liquid & 8 & 3 & 40 & 3200 & 6200 & static\\
\cite{root_shock_2015} & 2015 & B1, B2, liquid & 35 & 0 & 1334 & 370 & 42000 & dynamic\\
\cite{bolis_decaying_2016} & 2016 & B2, liquid & 2 & 470 & 470 & 9050 & 10670 & dynamic\\
\cite{kimura_melting_2017} & 2017 & B1, liquid & 16 & 0 & 45.3 & 2870 & 5790 & static\\
\cite{ye_intercomparison_2017} & 2017 & B1 & 257 & 17.3 & 137.8 & 300 & 2503 & static\\
\cite{FU20181} & 2018 & B1, liquid & 10 & 52 & 120 & 5480 & 8000 & static\\
\cite{taniuchi_melting_2018} & 2018 & B1, B2, liquid & 16& 0 & 3900 & 2800 & 21100  & theory\\
\cite{musella_physical_2019} & 2019 & liquid & 12& 355 & 11416 & 15000 & 45000  & theory\\
\cite{hansen_melting_2021} & 2021 &  B2, liquid & 8 & 1170 & 2109 & 12900 & 26200 & dynamic\\
\cite{wisesa_machine-learning_2023} & 2023 & B1, liquid & 34& 0& 300 & 2612 & 9730  & theory\\
\cite{wicks_b1-b2_2024} & 2023 & B1, B2, liquid & 10& 390& 649 & 9614 & 14152  & dynamic
\enddata
\tablecomments{Abbreviations for mineral names: B1=NaCl-type (face-centered cubic) MgO and B2=CsCl-type (body-centered cubic) MgO.}
\end{deluxetable*}
\end{longrotatetable}

\begin{longrotatetable}
\begin{deluxetable*}{lcccccccc}
\tabletypesize{\scriptsize}
\tablewidth{0pt} 
\tablecaption{Summary of the Equilibria Database for SiO$_{2}$ \label{tab:sio2}}
\tablehead{
\textbf{Reference} &
\textbf{Year} &
  \textbf{Phase(s) observed} &
  \textbf{Number of observations} &
  \textbf{P$_{\textrm{min}}$ (GPa)} &
  \textbf{P$_{\textrm{max}}$ (GPa)} &
  \textbf{T$_{\textrm{min}}$ (K)} &
  \textbf{T$_{\textrm{max}}$ (K)} & \textbf{Method}
}
\startdata 
\cite{akimoto_coesite-stishovite_1969} & 1969 & coes, st, mix & 26 & 8.2 & 9.8 & 823 & 1473 & static\\
\cite{jackson_melting_1976} & 1976 & qtz, crist, liquid, mix & 47 & 0.3 & 2.5 & 1873 & 2523 &static\\
\cite{yagi_direct_1976} & 1976 & coes, st, mix & 21 & 7.9 & 9.8 & 773 & 1373 &static\\
\cite{suito_phase_1977} & 1977 & coes, st, mix & 16 & 7.7 & 12.4 & 839 & 1272 &static\\
\cite{bohlen_quartz_1982} & 1982 & qtz, coes & 23 & 2.3 & 2.9 & 573 & 1173 &static\\
\cite{kanzaki_melting_1990} & 1990 & qtz, coes, st, liquid, mix & 20 & 3 & 11 & 1800 & 2850 &static\\
\cite{zhang_melting_1993} & 1993 & coes, st, liquid, mix & 18 & 9.2 & 14 & 1273 & 3128 &static\\
\cite{shen_measurement_1995} & 1995 & liquid & 9 & 5.6 & 36.4 & 2750 & 4250 &static\\
\cite{serghiou_coesitestishovite_1995} & 1995 & coes, st & 5 & 9.9 & 11.4 & 2606 & 2908 &static\\
\cite{zhang_situ_1996} & 1996 & coes, st & 24 & 7.7 & 11.4 & 800 & 1803 &static\\
\cite{andrault_pressure-induced_1998} & 1998 & st, cacl2 & 16 & 50.5 & 128 & 298 & 298 &static\\
\cite{dubrovinsky_pressure-induced_2001} & 2001 & mix & 13 & 60.9 & 76 & 780 & 2557 &static\\
\cite{ono_post-stishovite_2002} & 2002 & st, cacl2 & 41 & 46.7 & 91.8 & 300 & 2142 &static\\
\cite{andrault_equation_2003} & 2003 & st & 23 & 1.2 & 52.7 & 298 & 298 &static\\
\cite{murakami_stability_2003} & 2003 & cacl2, seif & 8 & 101 & 151 & 2101 & 2501 &static\\
\cite{nishihara_p-v-t_2005} & 2005 & st & 44 & 4.8 & 22.5 & 300 & 1073 &static\\
\cite{shieh_x-ray_2005} & 2005 & cacl2 & 66 & 72.9 & 131.3 & 1299 & 3052 &static\\
\cite{kuwayama_pyrite-type_2005} & 2005 & seif, pyrite & 6 & 194 & 271 & 1400 & 2000 &static\\
\cite{usui_ab_2010} & 2010 & st, cacl2, seif, liquid & 39 & 14 & 155 & 2700 & 6000 &theory\\
\cite{nomura_precise_2010} & 2010 & st, cacl2, mix & 18 & 44.6 & 74.5 & 300 & 2490 &static\\
\cite{wang_pvt_2012} & 2012 & st & 56 & 16.8 & 54.5 & 300 & 1700 &static\\
\cite{grocholski_stability_2013} & 2013 & cacl2, seif, mix & 21 & 102 & 155.3 & 300 & 3101 &static\\
\cite{yamazaki_over_2014} & 2014 & st, cacl2, mix & 41 & 14.5 & 109.6 & 600 & 1500 &static\\
\cite{pigott_highpressure_2015} & 2015 & st & 146 & 12.6 & 46.8 & 300 & 2445 &static\\
\cite{ono_precise_2017} & 2017 & coes, st & 31 & 6.4 & 11.4 & 1200 & 1700 &static\\
\cite{buchen_equation_2018} & 2018 & st, cacl2, mix & 30 & 9.7 & 72.9 & 298 & 298 &static\\
\cite{fischer_equations_2018} & 2018 & st, cacl2 & 131 & 20.9 & 88.9 & 1065 & 3272 &static\\
\cite{sun_high_2019} & 2019 & st, cacl2, seif, mix & 239 & 56 & 144 & 300 & 3606 &static\\
\cite{andrault_comment_2022} & 2020 & st, cacl2, liquid & 36 & 14.4 & 118.6 & 2758 & 6261 &static\\
\cite{geng_ab_2024} & 2024 & st, cacl2, seif, pyrite, liquid & 62 & 37 & 449 & 5000 & 8500 & theory
\enddata
\tablecomments{Abbreviations for mineral names: qtz=$quartz$, crist=$cristobalite$, coes=$coesite$, st=$stishovite$, cacl2=CaCl$_{2}$-type $silica$, seif=$seifertite$, and pyrite=$pyrite$-type $silica$.}
\end{deluxetable*}
\end{longrotatetable}

\begin{longrotatetable}
\begin{deluxetable*}{lcccccccc}
\tabletypesize{\scriptsize}
\tablewidth{0pt} 
\tablecaption{Summary of the Equilibria Database for MgSiO$_{3}$ \label{tab:mgsio3}}
\tablehead{
\textbf{Reference} &
\textbf{Year} &
  \textbf{Phase(s) observed} &
  \textbf{Number of observations} &
  \textbf{P$_{\textrm{min}}$ (GPa)} &
  \textbf{P$_{\textrm{max}}$ (GPa)} &
  \textbf{T$_{\textrm{min}}$ (K)} &
  \textbf{T$_{\textrm{max}}$ (K)} & \textbf{Method}
}

\startdata 
\cite{boyd_rhombic_1959} & 1959 & en, lpcen & 19 & 0.5 & 4 & 898 & 1051 & static \\
\cite{boyd_effects_1964} & 1964 & pten, en, liquid & 29 & 0 & 4.7 & 1823 & 2223 & static\\
\cite{akimoto_pyroxenegarnet_1977} & 1977 & hpcen, ri+st, wa+st & 15 & 13.2 & 19.5 & 1273 & 1573 & static\\
\cite{ito_mgsio3_1985} & 1985 & hpcen, ri+st, wa+st, ak & 18 & 14.5 & 20.1 & 1276 & 1881 & static\\
\cite{kato_garnet_1985} & 1985 & mj, ak, liquid & 3 & 20 & 20 & 2033 & 2523 & static\\
\cite{manghnani_phase_1987} & 1987 & hpcen, ri+st, wa+st, mj, ak, bm & 52 & 0 & 24 & 273 & 2373 & static\\
\cite{eiji_ito_mgsio3_1985}& 1989 & ak, bm & 16 & 22.2 & 24.8 & 1273 & 1873 & static\\
\cite{pacalo_reversals_1990} & 1990 & en, hpcen & 21 & 6.4 & 14.2 & 1173 & 1973 & static\\
\cite{presnall_melting_1990} & 1990 & en, hpcen, mj, liquid & 14 & 10 & 17 & 2413 & 2643 & static\\
\cite{kanzaki_orthoclinoenstatite_1991} & 1991 & en, hpcen & 18 & 7 & 9.4 & 1273 & 1674 & static\\
\cite{syono_melting_1992}& 1992 & mj, ak, bm, liquid & 17 & 21 & 25 & 2123 & 2923 & static\\
\cite{zerr_melting_1993}& 1993 & mj, bm, liquid & 26 & 22.5 & 62.3 & 2460 & 5016 & static\\
\cite{shen_measurement_1995} & 1995 & en, bm, liquid & 14 & 6.8 & 42 & 2220 & 4050 & static\\
\cite{shinmei_situ_1999} & 1999 & en, lpcen, hpcen & 39 & 0 & 11.9 & 300 & 1473 & static\\
\cite{kuroda_determination_2000} & 2000 & mj, ak, bm & 37 & 20.6 & 22.3 & 1173 & 2273 & static\\
\cite{ono_situ_2001} & 2001 & ak, bm & 28 & 23.6 & 24.5 & 1099 & 1651 & static\\
\cite{oganov_theoretical_2004} & 2004 & bm, ppv & 22 & 80.7 & 128.4 & 300 & 2500 & theory\\
\cite{ono_situ_2005} & 2005 & bm, ppv & 42 & 112.5 & 133.3 & 1510 & 3350 & static\\
\cite{ono_equation_2006} & 2006 & ppv & 6 & 121.4 & 151.1 & 300 & 300 & static\\
\cite{guignot_thermoelastic_2007} & 2007 & ppv & 48 & 111.1 & 144.5 & 300 & 2535 & static\\
\cite{akashi_orthoenstatiteclinoenstatite_2009}& 2009 & en, hpcen & 17 & 6.2 & 9.5 & 1023 & 1873 & static\\
\cite{katsura_pvt_2009} & 2009 & bm & 89 & 18.9 & 52.6 & 300 & 2500 & static\\
\cite{tateno_determination_2009}& 2009 & bm, ppv & 17 & 119 & 171 & 1640 & 4380 & static\\
\cite{tange_pvt_2012} & 2012 & bm & 42 & 27.8 & 108.3 & 300 & 2430 & static\\
\cite{ono_reaction_2017}& 2017 & ak, ri+st & 15 & 19 & 21.2 & 1100 & 1450 & static\\
\cite{ono_decomposition_2018}& 2018 & hpcen, ri+st, wa+st & 10 & 15 & 18.7 & 1150 & 1400 & static\\
\cite{kulka_bridgmaniteakimotoitemajorite_2020}& 2020 & mj,ak,bm & 45 & 16.8 & 30.3 & 1506 & 2349 & static\\
\cite{fei_melting_2021} & 2021 & ppv, liquid & 6 & 349.3 & 910.6 & 5289 & 17020 & dynamic\\
\cite{chanyshev_depressed_2022} & 2023 & ak, bm & 40 & 20.3 & 24.4 & 1239 & 2084 & static\\
\cite{PIERRU2022117770} & 2022 & bm, liquid & 20 & 40 & 137.2 & 3454 & 5217 & static\\
\cite{deng_melting_2023} & 2023 & bm, ppv, liquid & 10 & 40 & 200 & 3950 & 6800 & theory \\
\cite{okuda_electrical_2024} & 2023 & bm, liquid & 20 & 36.2 & 72.8 & 1830 & 4490 & static
\enddata
\tablecomments{Abbreviations for mineral names: en=$orthoenstatite$, pten=$protoenstatite$, lpcen=low pressure $clinoenstatite$, hpcen=high pressure $clinoenstatite$, mj=$majorite$, ak=$akimotoite$, wa+st=$wadsleyite$+$stishovite$, ri+st=$ringwoodite$+$stishovite$, bm=$bridgmanite$, and ppv=$post$-$perovskite$.}
\end{deluxetable*}
\label{tab:t_mgsio3}
\end{longrotatetable}



\begin{thebibliography}{}
\expandafter\ifx\csname natexlab\endcsname\relax\def\natexlab#1{#1}\fi
\providecommand{\url}[1]{\href{#1}{#1}}
\providecommand{\dodoi}[1]{doi:~\href{http://doi.org/#1}{\nolinkurl{#1}}}
\providecommand{\doeprint}[1]{\href{http://ascl.net/#1}{\nolinkurl{http://ascl.net/#1}}}
\providecommand{\doarXiv}[1]{\href{https://arxiv.org/abs/#1}{\nolinkurl{https://arxiv.org/abs/#1}}}

\bibitem[{Akaogi(2022)}]{akaogi_calorimetric_2022}
Akaogi, M. 2022, in High-{Pressure} {Silicates} and {Oxides} (Singapore: Springer Nature Singapore), 47--70, \dodoi{10.1007/978-981-19-6363-6_4}

\bibitem[{Akashi {et~al.}(2009)Akashi, Nishihara, Takahashi, Nakajima, Tange, \& Funakoshi}]{akashi_orthoenstatiteclinoenstatite_2009}
Akashi, A., Nishihara, Y., Takahashi, E., {et~al.} 2009, Journal of Geophysical Research: Solid Earth, 114, 2008JB005894, \dodoi{10.1029/2008JB005894}

\bibitem[{Akimoto(1977)}]{akimoto_pyroxenegarnet_1977}
Akimoto, S. 1977, Physics of the Earth and Planetary Interiors, 15, 90

\bibitem[{Akimoto \& Syono(1969)}]{akimoto_coesite-stishovite_1969}
Akimoto, S.-i., \& Syono, Y. 1969, Journal of Geophysical Research, 74, 1653, \dodoi{10.1029/JB074i006p01653}

\bibitem[{Alfè(2005)}]{alfe_melting_2005}
Alfè, D. 2005, Physical Review Letters, 94, 235701, \dodoi{10.1103/PhysRevLett.94.235701}

\bibitem[{Andrault {et~al.}(2003)Andrault, Angel, Mosenfelder, \& Le~Bihan}]{andrault_equation_2003}
Andrault, D., Angel, R.~J., Mosenfelder, J.~L., \& Le~Bihan, T. 2003, American Mineralogist, 88, 301, \dodoi{10.2138/am-2003-2-307}

\bibitem[{Andrault {et~al.}(1998)Andrault, Fiquet, Guyot, \& Hanfland}]{andrault_pressure-induced_1998}
Andrault, D., Fiquet, G., Guyot, F., \& Hanfland, M. 1998, Science, 282, 720, \dodoi{10.1126/science.282.5389.720}

\bibitem[{Andrault {et~al.}(2022)Andrault, Pison, Morard, Garbarino, Mezouar, Bouhifd, \& Kawamoto}]{andrault_comment_2022}
Andrault, D., Pison, L., Morard, G., {et~al.} 2022, Physics and Chemistry of Minerals, 49, 3, \dodoi{10.1007/s00269-021-01174-2}

\bibitem[{Angel {et~al.}(1989)Angel, Finger, Hazen, Kanzaki, Weidner, Liebermann, \& Veblen}]{angel_structure_1989}
Angel, R.~J., Finger, L.~W., Hazen, R.~M., {et~al.} 1989, American Mineralogist, 74, 509

\bibitem[{Anzellini {et~al.}(2013)Anzellini, Dewaele, Mezouar, Loubeyre, \& Morard}]{anzellini_melting_2013}
Anzellini, S., Dewaele, A., Mezouar, M., Loubeyre, P., \& Morard, G. 2013, Science, 340, 464, \dodoi{10.1126/science.1233514}

\bibitem[{Aquilanti {et~al.}(2015)Aquilanti, Trapananti, Karandikar, Kantor, Marini, Mathon, Pascarelli, \& Boehler}]{aquilanti_melting_2015}
Aquilanti, G., Trapananti, A., Karandikar, A., {et~al.} 2015, Proceedings of the National Academy of Sciences, 112, 12042, \dodoi{10.1073/pnas.1502363112}

\bibitem[{Assael \& Kakosimos(2006)}]{assael_reference_2006}
Assael, M.~J., \& Kakosimos, K. 2006, J. Phys. Chem. Ref. Data, 35

\bibitem[{Baron {et~al.}(2017)Baron, Lord, Myhill, Thomson, Wang, Trønnes, \& Walter}]{baron_experimental_2017}
Baron, M.~A., Lord, O.~T., Myhill, R., {et~al.} 2017, Earth and Planetary Science Letters, 472, 186, \dodoi{10.1016/j.epsl.2017.05.020}

\bibitem[{Belonoshko {et~al.}(2017)Belonoshko, Lukinov, Fu, Zhao, Davis, \& Simak}]{belonoshko_stabilization_2017}
Belonoshko, A.~B., Lukinov, T., Fu, J., {et~al.} 2017, Nature Geoscience, 10, 312, \dodoi{10.1038/ngeo2892}

\bibitem[{Boehler(1993)}]{boehler_temperatures_1993}
Boehler, R. 1993, Nature, 363, 534, \dodoi{10.1038/363534a0}

\bibitem[{Boehler {et~al.}(1987)Boehler, Nicol, \& Johnson}]{manghnani_internally-heated_1987}
Boehler, R., Nicol, M., \& Johnson, M.~L. 1987, in Geophysical {Monograph} {Series}, ed. M.~H. Manghnani \& Y.~Syono, Vol.~39 (Washington, D. C.: American Geophysical Union), 173--176, \dodoi{10.1029/GM039p0173}

\bibitem[{Boehler {et~al.}(1990)Boehler, Von~Bargen, \& Chopelas}]{boehler_melting_1990}
Boehler, R., Von~Bargen, N., \& Chopelas, A. 1990, Journal of Geophysical Research: Solid Earth, 95, 21731, \dodoi{10.1029/JB095iB13p21731}

\bibitem[{Bohlen \& Boettcher(1982)}]{bohlen_quartz_1982}
Bohlen, S.~R., \& Boettcher, A.~L. 1982, Journal of Geophysical Research: Solid Earth, 87, 7073, \dodoi{10.1029/JB087iB08p07073}

\bibitem[{Bolis {et~al.}(2016)Bolis, Morard, Vinci, Ravasio, Bambrink, Guarguaglini, Koenig, Musella, Remus, Bouchet, Ozaki, Miyanishi, Sekine, Sakawa, Sano, Kodama, Guyot, \& Benuzzi‐Mounaix}]{bolis_decaying_2016}
Bolis, R.~M., Morard, G., Vinci, T., {et~al.} 2016, Geophysical Research Letters, 43, 9475, \dodoi{10.1002/2016GL070466}

\bibitem[{Bose \& Ganguly(1995)}]{bose_quartz-coesite_1995}
Bose, K., \& Ganguly, J. 1995, American Mineralogist, 80, 231, \dodoi{10.2138/am-1995-3-404}

\bibitem[{Boujibar {et~al.}(2020)Boujibar, Driscoll, \& Fei}]{boujibar_superearth_2020}
Boujibar, A., Driscoll, P., \& Fei, Y. 2020, Journal of Geophysical Research: Planets, 125, e2019JE006124, \dodoi{10.1029/2019JE006124}

\bibitem[{Boukaré {et~al.}(2015)Boukaré, Ricard, \& Fiquet}]{boukare_thermodynamics_2015}
Boukaré, C., Ricard, Y., \& Fiquet, G. 2015, Journal of Geophysical Research: Solid Earth, 120, 6085, \dodoi{10.1002/2015JB011929}

\bibitem[{Bowen(1928)}]{bowen_evolution_1928}
Bowen, N.~L. 1928, The {Evolution} of the {Igneous} {Rocks} (Princeton, N.J.: Princeton University Press)

\bibitem[{Boyd \& England(1959)}]{boyd_rhombic_1959}
Boyd, F.~R., \& England, J.~L. 1959, in The {Carnegie} {Institution} of {Washington} {Year} {Book}, Vol.~64 (The {Carnegie} {Institution} of {Washington}), 117--120

\bibitem[{Boyd \& England(1960)}]{boyd_quartz-coesite_1960}
---. 1960, Journal of Geophysical Research, 65, 749, \dodoi{10.1029/JZ065i002p00749}

\bibitem[{Boyd {et~al.}(1964)Boyd, England, \& Davis}]{boyd_effects_1964}
Boyd, F.~R., England, J.~L., \& Davis, B. T.~C. 1964, Journal of Geophysical Research, 69, 2101, \dodoi{10.1029/JZ069i010p02101}

\bibitem[{Buchen {et~al.}(2018)Buchen, Marquardt, Schulze, Speziale, Boffa Ballaran, Nishiyama, \& Hanfland}]{buchen_equation_2018}
Buchen, J., Marquardt, H., Schulze, K., {et~al.} 2018, Journal of Geophysical Research: Solid Earth, 123, 7347, \dodoi{10.1029/2018JB015835}

\bibitem[{Buder {et~al.}(2021)Buder, Sharma, Kos, Amarsi, Nordlander, Lind, Martell, Asplund, Bland-Hawthorn, Casey, De Silva, D’Orazi, Freeman, Hayden, Lewis, Lin, Schlesinger, Simpson, Stello, Zucker, Zwitter, Beeson, Buck, Casagrande, Clark, Čotar, Da Costa, de Grijs, Feuillet, Horner, Kafle, Khanna, Kobayashi, Liu, Montet, Nandakumar, Nataf, Ness, Spina, Tepper-García, Ting(丁源森), Traven, Vogrinčič, Wittenmyer, Wyse, Žerjal, \& {GALAH Collaboration}}]{buder_galah_2021}
Buder, S., Sharma, S., Kos, J., {et~al.} 2021, Monthly Notices of the Royal Astronomical Society, 506, 150, \dodoi{10.1093/mnras/stab1242}

\bibitem[{Bundy(1965)}]{bundy_pressuretemperature_1965}
Bundy, F.~P. 1965, Journal of Applied Physics, 36, 616, \dodoi{10.1063/1.1714038}

\bibitem[{Callen(1985)}]{callen_thermodynamics_1985}
Callen, H.~B. 1985, Thermodynamics and an {Introduction} to {Thermostatistics}, 2nd edn. (New York: Wiley)

\bibitem[{Cebulla \& Redmer(2014)}]{cebulla_ab_2014}
Cebulla, D., \& Redmer, R. 2014, Physical Review B, 89, 134107, \dodoi{10.1103/PhysRevB.89.134107}

\bibitem[{Chanyshev {et~al.}(2022)Chanyshev, Ishii, Bondar, Bhat, Kim, Farla, Nishida, Liu, Wang, Nakajima, Yan, Tang, Chen, Higo, Tange, \& Katsura}]{chanyshev_depressed_2022}
Chanyshev, A., Ishii, T., Bondar, D., {et~al.} 2022, Nature, 601, 69, \dodoi{10.1038/s41586-021-04157-z}

\bibitem[{Cheng {et~al.}(2021)Cheng, Bethkenhagen, Pickard, \& Hamel}]{cheng_phase_2021}
Cheng, B., Bethkenhagen, M., Pickard, C.~J., \& Hamel, S. 2021, Nature Physics, 17, 1228, \dodoi{10.1038/s41567-021-01334-9}

\bibitem[{Chernyshev {et~al.}(1993)Chernyshev, Petrov, Titov, \& Vorobyev}]{chernyshev_thermal_1993}
Chernyshev, A., Petrov, V., Titov, V., \& Vorobyev, A. 1993, Thermochimica Acta, 218, 195, \dodoi{10.1016/0040-6031(93)80422-7}

\bibitem[{Claussen(1960)}]{claussen_detection_1960}
Claussen, W.~F. 1960, Review of Scientific Instruments, 31, 878, \dodoi{10.1063/1.1717076}

\bibitem[{Clendenen \& Drickamer(1964)}]{clendenen_effect_1964}
Clendenen, R., \& Drickamer, H. 1964, Journal of Physics and Chemistry of Solids, 25, 865, \dodoi{10.1016/0022-3697(64)90098-8}

\bibitem[{Davies(1990)}]{newsom_heat_1990}
Davies, G.~F. 1990, in Origin of the {Earth}, ed. H.~E. Newsom \& J.~H. Jones (Oxford University PressNew York, NY), 175--194, \dodoi{10.1093/oso/9780195066197.003.0011}

\bibitem[{De~Koker {et~al.}(2013)De~Koker, Karki, \& Stixrude}]{de_koker_thermodynamics_2013}
De~Koker, N., Karki, B.~B., \& Stixrude, L. 2013, Earth and Planetary Science Letters, 361, 58, \dodoi{10.1016/j.epsl.2012.11.026}

\bibitem[{Deng {et~al.}(2023)Deng, Niu, Hu, Chen, \& Stixrude}]{deng_melting_2023}
Deng, J., Niu, H., Hu, J., Chen, M., \& Stixrude, L. 2023, Physical Review B, 107, 064103, \dodoi{10.1103/PhysRevB.107.064103}

\bibitem[{Deng {et~al.}(2013)Deng, Seagle, Fei, \& Shahar}]{deng_high_2013}
Deng, L., Seagle, C., Fei, Y., \& Shahar, A. 2013, Geophysical Research Letters, 40, 33, \dodoi{10.1029/2012GL054347}

\bibitem[{Dewaele {et~al.}(2000)Dewaele, Fiquet, Andrault, \& Hausermann}]{dewaele_pvt_2000}
Dewaele, A., Fiquet, G., Andrault, D., \& Hausermann, D. 2000, Journal of Geophysical Research: Solid Earth, 105, 2869, \dodoi{10.1029/1999JB900364}

\bibitem[{Dong {et~al.}(2019)Dong, Li, Zhu, Li, \& Farawi}]{dong_melting_2019}
Dong, J., Li, J., Zhu, F., Li, Z., \& Farawi, R. 2019, American Mineralogist, 104, 671, \dodoi{10.2138/am-2019-6891}

\bibitem[{Dong {et~al.}(2025)Dong, Fischer, Stixrude, Brennan, Daviau, Suer, Turner, Meng, \& Prakapenka}]{dong_nonlinearity_2025}
Dong, J., Fischer, R.~A., Stixrude, L.~P., {et~al.} 2025, Nature Communications, 16, 1039, \dodoi{10.1038/s41467-025-56231-z}

\bibitem[{Dorogokupets {et~al.}(2017)Dorogokupets, Dymshits, Litasov, \& Sokolova}]{dorogokupets_thermodynamics_2017}
Dorogokupets, P.~I., Dymshits, A.~M., Litasov, K.~D., \& Sokolova, T.~S. 2017, Scientific Reports, 7, 41863, \dodoi{10.1038/srep41863}

\bibitem[{Dragulet \& Stixrude(2024)}]{dragulet_partitioning_2024}
Dragulet, F., \& Stixrude, L. 2024, Geophysical Research Letters, 51, e2023GL107979, \dodoi{10.1029/2023GL107979}

\bibitem[{Dragulet \& Stxirude(2023)}]{francis_dragulet_electrical_2023}
Dragulet, F., \& Stxirude, L.~P. 2023, in American {Geophysical} {Union} 2023 {Annual} {Meeting} {Abstracts}, San Francisco

\bibitem[{Du \& Lee(2014)}]{du_highpressure_2014}
Du, Z., \& Lee, K. K.~M. 2014, Geophysical Research Letters, 41, 8061, \dodoi{10.1002/2014GL061954}

\bibitem[{Dubrovinsky {et~al.}(2001)Dubrovinsky, Dubrovinskaia, Saxena, Tutti, Rekhi, Le~Bihan, Shen, \& Hu}]{dubrovinsky_pressure-induced_2001}
Dubrovinsky, L., Dubrovinskaia, N., Saxena, S., {et~al.} 2001, Chemical Physics Letters, 333, 264, \dodoi{10.1016/S0009-2614(00)01147-7}

\bibitem[{Dubrovinsky \& Saxena(1997)}]{dubrovinsky_thermal_1997}
Dubrovinsky, L.~S., \& Saxena, S.~K. 1997, Physics and Chemistry of Minerals, 24, 547, \dodoi{10.1007/s002690050070}

\bibitem[{Dubrovinsky {et~al.}(1998)Dubrovinsky, Saxena, \& Lazor}]{dubrovinsky_high-pressure_1998}
Dubrovinsky, L.~S., Saxena, S.~K., \& Lazor, P. 1998, Physics and Chemistry of Minerals, 25, 434, \dodoi{10.1007/s002690050133}

\bibitem[{Duffy {et~al.}(2015)Duffy, Madhusudhan, \& Lee}]{duffy_mineralogy_2015}
Duffy, T.~S., Madhusudhan, N., \& Lee, K. 2015, in Treatise on {Geophysics} (Elsevier), 149--178, \dodoi{10.1016/B978-0-444-53802-4.00053-1}

\bibitem[{Duffy \& Smith(2019)}]{duffy_ultra-high_2019}
Duffy, T.~S., \& Smith, R.~F. 2019, Frontiers in Earth Science, 7, 23, \dodoi{10.3389/feart.2019.00023}

\bibitem[{Dutta {et~al.}(2023)Dutta, Tracy, \& Cohen}]{dutta_high-pressure_2023}
Dutta, R., Tracy, S.~J., \& Cohen, R.~E. 2023, Physical Review B, 107, 184112, \dodoi{10.1103/PhysRevB.107.184112}

\bibitem[{Dutta {et~al.}(2022)Dutta, Tracy, Cohen, Miozzi, Luo, Yang, Burnley, Smith, Meng, Chariton, Prakapenka, \& Duffy}]{dutta_ultrahigh-pressure_2022}
Dutta, R., Tracy, S.~J., Cohen, R.~E., {et~al.} 2022, Proceedings of the National Academy of Sciences, 119, e2114424119, \dodoi{10.1073/pnas.2114424119}

\bibitem[{{Eiji Ito} \& {Alexandar Navrotsky}(1985)}]{eiji_ito_mgsio3_1985}
{Eiji Ito}, \& {Alexandar Navrotsky}. 1985, American Mineralogist, 70, 1020

\bibitem[{Faccenda \& Dal~Zilio(2017)}]{faccenda_role_2017}
Faccenda, M., \& Dal~Zilio, L. 2017, Lithos, 268-271, 198, \dodoi{10.1016/j.lithos.2016.11.007}

\bibitem[{Fat’yanov \& Asimow(2024)}]{fatyanov_melting_2024}
Fat’yanov, O.~V., \& Asimow, P.~D. 2024, Journal of Geophysical Research: Solid Earth

\bibitem[{Fei(1999)}]{fei_effects_1999}
Fei, Y. 1999, American Mineralogist, 84, 272, \dodoi{10.2138/am-1999-0308}

\bibitem[{Fei {et~al.}(2021)Fei, Seagle, Townsend, McCoy, Boujibar, Driscoll, Shulenburger, \& Furnish}]{fei_melting_2021}
Fei, Y., Seagle, C.~T., Townsend, J.~P., {et~al.} 2021, Nature Communications, 12, 876, \dodoi{10.1038/s41467-021-21170-y}

\bibitem[{Fiquet {et~al.}(1999)Fiquet, Richet, \& Montagnac}]{fiquet_high-temperature_1999}
Fiquet, G., Richet, P., \& Montagnac, G. 1999, Physics and Chemistry of Minerals, 27, 103, \dodoi{10.1007/s002690050246}

\bibitem[{Fischer {et~al.}(2018)Fischer, Campbell, Chidester, Reaman, Thompson, Pigott, Prakapenka, \& Smith}]{fischer_equations_2018}
Fischer, R.~A., Campbell, A.~J., Chidester, B.~A., {et~al.} 2018, American Mineralogist, 103, 792, \dodoi{10.2138/am-2018-6267}

\bibitem[{Fischer {et~al.}(2011)Fischer, Campbell, Shofner, Lord, Dera, \& Prakapenka}]{fischer_equation_2011}
Fischer, R.~A., Campbell, A.~J., Shofner, G.~A., {et~al.} 2011, Earth and Planetary Science Letters, 304, 496, \dodoi{10.1016/j.epsl.2011.02.025}

\bibitem[{Fortney {et~al.}(2018)Fortney, Helled, Nettelmann, Stevenson, Marley, Hubbard, \& Iess}]{baines_interior_2018}
Fortney, J.~J., Helled, R., Nettelmann, N., {et~al.} 2018, in Saturn in the 21st {Century}, 1st edn., ed. K.~H. Baines, F.~M. Flasar, N.~Krupp, \& T.~Stallard (Cambridge University Press), 44--68, \dodoi{10.1017/9781316227220.003}

\bibitem[{Fu {et~al.}(2018)Fu, Yang, Zhang, Liu, Greenberg, Prakapenka, Okuchi, \& Lin}]{FU20181}
Fu, S., Yang, J., Zhang, Y., {et~al.} 2018, Earth and Planetary Science Letters, 503, 1, \dodoi{https://doi.org/10.1016/j.epsl.2018.09.014}

\bibitem[{Geng \& Mohn(2024)}]{geng_ab_2024}
Geng, M., \& Mohn, C.~E. 2024, Physical Review B, 109, 024106, \dodoi{10.1103/PhysRevB.109.024106}

\bibitem[{Ghiorso(2004)}]{ghiorso_equation_2004}
Ghiorso, M.~S. 2004, American Journal of Science, 304, 637, \dodoi{10.2475/ajs.304.8-9.637}

\bibitem[{Giles {et~al.}(1971)Giles, Longenbach, \& Marder}]{giles_high-pressure_1971}
Giles, P.~M., Longenbach, M.~H., \& Marder, A.~R. 1971, Journal of Applied Physics, 42, 4290, \dodoi{10.1063/1.1659768}

\bibitem[{González-Cataldo {et~al.}(2016)González-Cataldo, Davis, \& Gutiérrez}]{gonzalez-cataldo_melting_2016}
González-Cataldo, F., Davis, S., \& Gutiérrez, G. 2016, Scientific Reports, 6, 26537, \dodoi{10.1038/srep26537}

\bibitem[{González-Cataldo \& Militzer(2023)}]{gonzalez-cataldo_ab_2023}
González-Cataldo, F., \& Militzer, B. 2023, Physical Review Research, 5, 033194, \dodoi{10.1103/PhysRevResearch.5.033194}

\bibitem[{Grocholski {et~al.}(2013)Grocholski, Shim, \& Prakapenka}]{grocholski_stability_2013}
Grocholski, B., Shim, S., \& Prakapenka, V.~B. 2013, Journal of Geophysical Research: Solid Earth, 118, 4745, \dodoi{10.1002/jgrb.50360}

\bibitem[{Guignot {et~al.}(2007)Guignot, Andrault, Morard, Bolfancasanova, \& Mezouar}]{guignot_thermoelastic_2007}
Guignot, N., Andrault, D., Morard, G., Bolfancasanova, N., \& Mezouar, M. 2007, Earth and Planetary Science Letters, 256, 162, \dodoi{10.1016/j.epsl.2007.01.025}

\bibitem[{Hansen {et~al.}(2021)Hansen, Fratanduono, Zhang, Hicks, Suer, Sprowal, Huff, Gong, Henderson, Polsin, Zaghoo, Hu, Collins, \& Rygg}]{hansen_melting_2021}
Hansen, L.~E., Fratanduono, D.~E., Zhang, S., {et~al.} 2021, Physical Review B, 104, 014106, \dodoi{10.1103/PhysRevB.104.014106}

\bibitem[{Helled {et~al.}(2020)Helled, Mazzola, \& Redmer}]{helled_understanding_2020}
Helled, R., Mazzola, G., \& Redmer, R. 2020, Nature Reviews Physics, 2, 562, \dodoi{10.1038/s42254-020-0223-3}

\bibitem[{Helled \& Stevenson(2024)}]{helled_fuzzy_2024}
Helled, R., \& Stevenson, D.~J. 2024, AGU Advances, 5, e2024AV001171, \dodoi{10.1029/2024AV001171}

\bibitem[{Helled {et~al.}(2022)Helled, Stevenson, Lunine, Bolton, Nettelmann, Atreya, Guillot, Militzer, Miguel, \& Hubbard}]{helled_revelations_2022}
Helled, R., Stevenson, D.~J., Lunine, J.~I., {et~al.} 2022, Icarus, 378, 114937, \dodoi{10.1016/j.icarus.2022.114937}

\bibitem[{Hemley {et~al.}(1988)Hemley, Jephcoat, Mao, Ming, \& Manghnani}]{hemley_pressure-induced_1988}
Hemley, R.~J., Jephcoat, A.~P., Mao, H.~K., Ming, L.~C., \& Manghnani, M.~H. 1988, Nature, 334, 52, \dodoi{10.1038/334052a0}

\bibitem[{Hinkel {et~al.}(2014)Hinkel, Timmes, Young, Pagano, \& Turnbull}]{hinkel_stellar_2014}
Hinkel, N.~R., Timmes, F., Young, P.~A., Pagano, M.~D., \& Turnbull, M.~C. 2014, The Astronomical Journal, 148, 54, \dodoi{10.1088/0004-6256/148/3/54}

\bibitem[{Hirose(2006)}]{hirose_postperovskite_2006}
Hirose, K. 2006, Reviews of Geophysics, 44, 2005RG000186, \dodoi{10.1029/2005RG000186}

\bibitem[{Hofstadter {et~al.}(2024)Hofstadter, Helled, \& Stevenson}]{hofstadter_uranus_2024}
Hofstadter, M., Helled, R., \& Stevenson, D. 2024, Uranus {Study} {Report}, Tech. rep., W. M. Keck Institute for Space Studies (KISS), Pasadena, CA

\bibitem[{Hou {et~al.}(2021)Hou, Liu, Zhang, Du, Dong, Yan, Wang, Wang, \& Chen}]{hou_melting_2021}
Hou, M., Liu, J., Zhang, Y., {et~al.} 2021, Geophysical Research Letters, 48, e2021GL095739, \dodoi{10.1029/2021GL095739}

\bibitem[{Huang {et~al.}(2020)Huang, Liu, Lv, Li, Long, Wang, Chen, Hemley, \& Ma}]{huang_stability_2020}
Huang, P., Liu, H., Lv, J., {et~al.} 2020, Proceedings of the National Academy of Sciences, 117, 5638, \dodoi{10.1073/pnas.1921811117}

\bibitem[{Hudon {et~al.}(2002)Hudon, Jung, \& Baker}]{hudon_melting_2002}
Hudon, P., Jung, I.-H., \& Baker, D.~R. 2002, Physics of the Earth and Planetary Interiors, 130, 159, \dodoi{10.1016/S0031-9201(02)00005-5}

\bibitem[{Insixiengmay \& Stixrude(2025)}]{insixiengmay_mgo_2025}
Insixiengmay, L., \& Stixrude, L. 2025, Earth and Planetary Science Letters, 654, 119242, \dodoi{10.1016/j.epsl.2025.119242}

\bibitem[{Ito(2007)}]{ito_theory_2007}
Ito, E. 2007, in Treatise on {Geophysics} (Elsevier), 197--230, \dodoi{10.1016/B978-044452748-6.00036-5}

\bibitem[{Ito \& Katsura(1992)}]{syono_melting_1992}
Ito, E., \& Katsura, T. 1992, in High-{Pressure} {Research}: {Application} to {Earth} and {Planetary} {Sciences}, ed. Y.~Syono \& M.~H. Manghnani, Geophysical {Monograph} {Series} (Washington, D. C.: American Geophysical Union), 315--322, \dodoi{10.1029/GM067p0315}

\bibitem[{Ito \& Navrotsky(1985)}]{ito_mgsio3_1985}
Ito, E., \& Navrotsky, A. 1985, American Mineralogist, 70, 1020

\bibitem[{Ito \& Takahashi(1989)}]{ito_postspinel_1989}
Ito, E., \& Takahashi, E. 1989, Journal of Geophysical Research: Solid Earth, 94, 10637, \dodoi{10.1029/JB094iB08p10637}

\bibitem[{Jackson(1976)}]{jackson_melting_1976}
Jackson, I. 1976, Physics of the Earth and Planetary Interiors, 13, 218

\bibitem[{Jackson {et~al.}(2013)Jackson, Sturhahn, Lerche, Zhao, Toellner, Alp, Sinogeikin, Bass, Murphy, \& Wicks}]{jackson_melting_2013}
Jackson, J.~M., Sturhahn, W., Lerche, M., {et~al.} 2013, Earth and Planetary Science Letters, 362, 143, \dodoi{10.1016/j.epsl.2012.11.048}

\bibitem[{James \& Stixrude(2024)}]{james_thermal_2024}
James, D.~A., \& Stixrude, L. 2024, Space Science Reviews, 220, 21, \dodoi{10.1007/s11214-024-01053-6}

\bibitem[{James {et~al.}(2021)James, Witten, Hastie, \& Tibshirani}]{james_introduction_2021}
James, G., Witten, D., Hastie, T., \& Tibshirani, R. 2021, An introduction to statistical learning: with applications in {R}, second edition edn., Springer texts in statistics (New York, NY: Springer), \dodoi{10.1007/978-1-0716-1418-1}

\bibitem[{Johnson {et~al.}(1962)Johnson, Stein, \& Davis}]{johnson_temperature_1962}
Johnson, P.~C., Stein, B.~A., \& Davis, R.~S. 1962, Journal of Applied Physics, 33, 557, \dodoi{10.1063/1.1702465}

\bibitem[{Kanzaki(1990)}]{kanzaki_melting_1990}
Kanzaki, M. 1990, Journal of the American Ceramic Society, 73, 3706, \dodoi{10.1111/j.1151-2916.1990.tb04282.x}

\bibitem[{Kanzaki(1991)}]{kanzaki_orthoclinoenstatite_1991}
---. 1991, Physics and Chemistry of Minerals, 17, \dodoi{10.1007/BF00202244}

\bibitem[{Kato \& Kumazawat(1985)}]{kato_garnet_1985}
Kato, T., \& Kumazawat, M. 1985, Nature

\bibitem[{Katsura {et~al.}(2009)Katsura, Yokoshi, Kawabe, Shatskiy, Manthilake, Zhai, Fukui, Hegoda, Yoshino, Yamazaki, Matsuzaki, Yoneda, Ito, Sugita, Tomioka, Hagiya, Nozawa, \& Funakoshi}]{katsura_pvt_2009}
Katsura, T., Yokoshi, S., Kawabe, K., {et~al.} 2009, Geophysical Research Letters, 36, 2008GL035658, \dodoi{10.1029/2008GL035658}

\bibitem[{Kaufman {et~al.}(1963)Kaufman, Clougherty, \& Weiss}]{kaufman_lattice_1963}
Kaufman, L., Clougherty, E., \& Weiss, R. 1963, Acta Metallurgica, 11, 323, \dodoi{10.1016/0001-6160(63)90157-3}

\bibitem[{Kavner {et~al.}(2000)Kavner, Speed, \& Jeanloz}]{aoki_statistical_2000}
Kavner, A., Speed, T., \& Jeanloz, R. 2000, in Physics {Meets} {Mineralogy}, 1st edn., ed. H.~Aoki, Y.~Syono, \& R.~J. Hemley (Cambridge University Press), 71--80, \dodoi{10.1017/CBO9780511896590.006}

\bibitem[{Kechin(1995)}]{Kechin_1995}
Kechin, V.~V. 1995, Journal of Physics: Condensed Matter, 7, 531, \dodoi{10.1088/0953-8984/7/3/008}

\bibitem[{Kechin(2001)}]{kechin_melting_2001}
---. 2001, Physical Review B, 65, 052102, \dodoi{10.1103/PhysRevB.65.052102}

\bibitem[{Kennedy \& Newton(1963)}]{kennedy_solid-liquid_1963}
Kennedy, G.~C., \& Newton, R.~C. 1963, in Solids {Under} {Pressure} (New York: McGraw-Hill Book Co.), 11065--11070

\bibitem[{Kimura {et~al.}(2017)Kimura, Ohfuji, Nishi, \& Irifune}]{kimura_melting_2017}
Kimura, T., Ohfuji, H., Nishi, M., \& Irifune, T. 2017, Nature Communications, 8, 15735, \dodoi{10.1038/ncomms15735}

\bibitem[{Kingma {et~al.}(1995)Kingma, Cohen, Hemley, \& Mao}]{kingma_transformation_1995}
Kingma, K.~J., Cohen, R.~E., Hemley, R.~J., \& Mao, H.-k. 1995, Nature, 374, 243, \dodoi{10.1038/374243a0}

\bibitem[{Kojitani {et~al.}(2016)Kojitani, Inoue, \& Akaogi}]{kojitani_precise_2016}
Kojitani, H., Inoue, T., \& Akaogi, M. 2016, Journal of Geophysical Research: Solid Earth, 121, 729, \dodoi{https://doi.org/10.1002/2015JB012211}

\bibitem[{Komabayashi(2014)}]{komabayashi_thermodynamics_2014}
Komabayashi, T. 2014, Journal of Geophysical Research: Solid Earth, 119, 4164, \dodoi{10.1002/2014JB010980}

\bibitem[{Komabayashi {et~al.}(2009)Komabayashi, Fei, Meng, \& Prakapenka}]{komabayashi_-situ_2009}
Komabayashi, T., Fei, Y., Meng, Y., \& Prakapenka, V. 2009, Earth and Planetary Science Letters, 282, 252, \dodoi{10.1016/j.epsl.2009.03.025}

\bibitem[{Kono {et~al.}(2015)Kono, Kenney-Benson, Shibazaki, Park, Shen, \& Wang}]{kono_high-pressure_2015}
Kono, Y., Kenney-Benson, C., Shibazaki, Y., {et~al.} 2015, Physics of the Earth and Planetary Interiors, 241, 57, \dodoi{10.1016/j.pepi.2015.02.006}

\bibitem[{Konôpková {et~al.}(2011)Konôpková, Lazor, Goncharov, \& Struzhkin}]{konopkova_thermal_2011}
Konôpková, Z., Lazor, P., Goncharov, A.~F., \& Struzhkin, V.~V. 2011, High Pressure Research, 31, 228, \dodoi{10.1080/08957959.2010.545059}

\bibitem[{Kraus {et~al.}(2022)Kraus, Hemley, Ali, Belof, Benedict, Bernier, Braun, Cohen, Collins, Coppari, Desjarlais, Fratanduono, Hamel, Krygier, Lazicki, Mcnaney, Millot, Myint, Newman, Rygg, Sterbentz, Stewart, Stixrude, Swift, Wehrenberg, \& Eggert}]{kraus_measuring_2022}
Kraus, R.~G., Hemley, R.~J., Ali, S.~J., {et~al.} 2022, Science, 375, 202, \dodoi{10.1126/science.abm1472}

\bibitem[{Kraut \& Kennedy(1966)}]{kraut_new_1966}
Kraut, E.~A., \& Kennedy, G.~C. 1966, Physical Review Letters, 16, 608, \dodoi{10.1103/PhysRevLett.16.608}

\bibitem[{Kulka {et~al.}(2020)Kulka, Dolinschi, Leinenweber, Prakapenka, \& Shim}]{kulka_bridgmaniteakimotoitemajorite_2020}
Kulka, B.~L., Dolinschi, J.~D., Leinenweber, K.~D., Prakapenka, V.~B., \& Shim, S.-H. 2020, Minerals, 10, 67, \dodoi{10.3390/min10010067}

\bibitem[{Kuroda {et~al.}(2000)Kuroda, Irifune, Inoue, Nishiyama, Miyashita, Funakoshi, \& Utsumi}]{kuroda_determination_2000}
Kuroda, K., Irifune, T., Inoue, T., {et~al.} 2000, Physics and Chemistry of Minerals, 27, 523, \dodoi{10.1007/s002690000096}

\bibitem[{Kuwayama {et~al.}(2005)Kuwayama, Hirose, Sata, \& Ohishi}]{kuwayama_pyrite-type_2005}
Kuwayama, Y., Hirose, K., Sata, N., \& Ohishi, Y. 2005, Science, 309, 923.
\newblock \url{http://www.jstor.org/stable/3842602}

\bibitem[{Kuwayama {et~al.}(2008)Kuwayama, Hirose, Sata, \& Ohishi}]{kuwayama_phase_2008}
---. 2008, Earth and Planetary Science Letters, 273, 379, \dodoi{10.1016/j.epsl.2008.07.001}

\bibitem[{Kuwayama {et~al.}(2020)Kuwayama, Morard, Nakajima, Hirose, Baron, Kawaguchi, Tsuchiya, Ishikawa, Hirao, \& Ohishi}]{kuwayama_equation_2020}
Kuwayama, Y., Morard, G., Nakajima, Y., {et~al.} 2020, Physical Review Letters, 124, 165701, \dodoi{10.1103/PhysRevLett.124.165701}

\bibitem[{Lainey {et~al.}(2017)Lainey, Jacobson, Tajeddine, Cooper, Murray, Robert, Tobie, Guillot, Mathis, Remus, Desmars, Arlot, De~Cuyper, Dehant, Pascu, Thuillot, Poncin-Lafitte, \& Zahn}]{lainey_new_2017}
Lainey, V., Jacobson, R.~A., Tajeddine, R., {et~al.} 2017, Icarus, 281, 286, \dodoi{10.1016/j.icarus.2016.07.014}

\bibitem[{Lewis(2004)}]{lewis_physics_2004}
Lewis, J.~S. 2004, Physics and chemistry of the solar system, 2nd edn., This is volume 87 in the {International} geophysics series (Amsterdam ; Boston: Elsevier Academic Press)

\bibitem[{{Lewis H. Cohen} \& {William Klement}(1967)}]{lewis_h_cohen_high-low_1967}
{Lewis H. Cohen}, \& {William Klement}. 1967, Journal of Geophysical Research, 72, 4245, \dodoi{https://doi.org/10.1029/JZ072i016p04245}

\bibitem[{Lherm {et~al.}(2024)Lherm, Nakajima, \& Blackman}]{lherm_thermal_2024}
Lherm, V., Nakajima, M., \& Blackman, E.~G. 2024, Physics of the Earth and Planetary Interiors, 356, 107267, \dodoi{10.1016/j.pepi.2024.107267}

\bibitem[{Li {et~al.}(2020)Li, Wu, Li, Xue, Tan, Zhou, Zhang, Xiong, Gao, \& Sekine}]{li_shock_2020}
Li, J., Wu, Q., Li, J., {et~al.} 2020, Geophysical Research Letters, 47, e2020GL087758, \dodoi{10.1029/2020GL087758}

\bibitem[{Li {et~al.}(2017)Li, Li, Lange, Liu, \& Militzer}]{li_determination_2017}
Li, Z., Li, J., Lange, R., Liu, J., \& Militzer, B. 2017, Earth and Planetary Science Letters, 457, 395, \dodoi{10.1016/j.epsl.2016.10.027}

\bibitem[{Liebske \& Frost(2012)}]{liebske_melting_2012}
Liebske, C., \& Frost, D.~J. 2012, Earth and Planetary Science Letters, 345-348, 159, \dodoi{10.1016/j.epsl.2012.06.038}

\bibitem[{Lindemann(1910)}]{lindemann_uber_1910}
Lindemann, F.~A. 1910, Zeitschrift für Physik, 11, 609

\bibitem[{Lindsley {et~al.}(1964)Lindsley, Davis, \& Macgregor}]{lindsley_ferrosilite_1964}
Lindsley, D.~H., Davis, B.~T., \& Macgregor, I.~D. 1964, Science, 144, 73, \dodoi{10.1126/science.144.3614.73}

\bibitem[{Liu \& Bassett(1975)}]{liu_melting_1975}
Liu, L.-G., \& Bassett, W.~A. 1975, Journal of Geophysical Research, 80, 3777, \dodoi{10.1029/JB080i026p03777}

\bibitem[{Liu \& Bassett(1986)}]{liu_elements_1986}
Liu, L.-g., \& Bassett, W.~A. 1986, Elements, {Oxides}, {Silicates}: {High} {Pressure} {Phases} {With} {Implications} for the {Earth}'s {Interior} (New York: Oxford University Press)

\bibitem[{Luo {et~al.}(2003)Luo, Ahrens, \& Asimow}]{luo_polymorphism_2003}
Luo, S., Ahrens, T.~J., \& Asimow, P.~D. 2003, Journal of Geophysical Research: Solid Earth, 108, 2002JB002317, \dodoi{10.1029/2002JB002317}

\bibitem[{Ma {et~al.}(2004)Ma, Somayazulu, Shen, Mao, Shu, \& Hemley}]{ma_situ_2004}
Ma, Y., Somayazulu, M., Shen, G., {et~al.} 2004, Physics of the Earth and Planetary Interiors, 143-144, 455, \dodoi{10.1016/j.pepi.2003.06.005}

\bibitem[{Mankovich \& Fuller(2021)}]{mankovich_diffuse_2021}
Mankovich, C.~R., \& Fuller, J. 2021, Nature Astronomy, 5, 1103, \dodoi{10.1038/s41550-021-01448-3}

\bibitem[{Mao {et~al.}(1967)Mao, Bassett, \& Takahashi}]{mao_effect_1967}
Mao, H.-K., Bassett, W.~A., \& Takahashi, T. 1967, Journal of Applied Physics, 38, 272, \dodoi{10.1063/1.1708965}

\bibitem[{Mao {et~al.}(1987)Mao, Bell, \& Hadidiacos}]{manghnani_experimental_1987}
Mao, H.~K., Bell, P.~M., \& Hadidiacos, C. 1987, in Geophysical {Monograph} {Series}, ed. M.~H. Manghnani \& Y.~Syono, Vol.~39 (Washington, D. C.: American Geophysical Union), 135--138, \dodoi{10.1029/GM039p0135}

\bibitem[{Mao {et~al.}(2018)Mao, Chen, Ding, Li, \& Wang}]{mao_solids_2018}
Mao, H.-K., Chen, X.-J., Ding, Y., Li, B., \& Wang, L. 2018, Reviews of Modern Physics, 90, 015007, \dodoi{10.1103/RevModPhys.90.015007}

\bibitem[{Mazevet {et~al.}(2019)Mazevet, Musella, \& Guyot}]{mazevet_fate_2019}
Mazevet, S., Musella, R., \& Guyot, F. 2019, Astronomy \& Astrophysics, 631, L4, \dodoi{10.1051/0004-6361/201936288}

\bibitem[{McNally {et~al.}(1961)McNally, Peters, \& Ribbe}]{mcnally_laboratory_1961}
McNally, R.~N., Peters, F.~I., \& Ribbe, P.~H. 1961, Journal of the American Ceramic Society, 44, 491, \dodoi{10.1111/j.1151-2916.1961.tb13711.x}

\bibitem[{McWilliams {et~al.}(2012)McWilliams, Spaulding, Eggert, Celliers, Hicks, Smith, Collins, \& Jeanloz}]{mcwilliams_phase_2012}
McWilliams, R.~S., Spaulding, D.~K., Eggert, J.~H., {et~al.} 2012, Science, 338, 1330, \dodoi{10.1126/science.1229450}

\bibitem[{Militzer \& Hubbard(2023)}]{militzer_relation_2023}
Militzer, B., \& Hubbard, W.~B. 2023, The Planetary Science Journal, 4, 95, \dodoi{10.3847/PSJ/acd2cd}

\bibitem[{Militzer \& Hubbard(2024)}]{militzer_study_2024}
---. 2024, Icarus, 411, 115955, \dodoi{10.1016/j.icarus.2024.115955}

\bibitem[{Millot {et~al.}(2019)Millot, Coppari, Rygg, Correa~Barrios, Hamel, Swift, \& Eggert}]{millot_nanosecond_2019}
Millot, M., Coppari, F., Rygg, J.~R., {et~al.} 2019, Nature, 569, 251, \dodoi{10.1038/s41586-019-1114-6}

\bibitem[{Mirwald \& Massonne(1980)}]{mirwald_lowhigh_1980}
Mirwald, P.~W., \& Massonne, H. 1980, Journal of Geophysical Research: Solid Earth, 85, 6983, \dodoi{10.1029/JB085iB12p06983}

\bibitem[{Miyagoshi {et~al.}(2015)Miyagoshi, Kameyama, \& Ogawa}]{miyagoshi_thermal_2015}
Miyagoshi, T., Kameyama, M., \& Ogawa, M. 2015, Journal of Geophysical Research: Planets, 120, 1267, \dodoi{10.1002/2015JE004793}

\bibitem[{Miyazaki \& Korenaga(2019)}]{miyazaki_timescale_2019}
Miyazaki, Y., \& Korenaga, J. 2019, Journal of Geophysical Research: Solid Earth, 124, 3382, \dodoi{10.1029/2018JB016932}

\bibitem[{Morard {et~al.}(2011)Morard, Bouchet, Valencia, Mazevet, \& Guyot}]{morard_melting_2011}
Morard, G., Bouchet, J., Valencia, D., Mazevet, S., \& Guyot, F. 2011, High Energy Density Physics, 7, 141, \dodoi{10.1016/j.hedp.2011.02.001}

\bibitem[{Morard {et~al.}(2018)Morard, Boccato, Rosa, Anzellini, Miozzi, Henry, Garbarino, Mezouar, Harmand, Guyot, Boulard, Kantor, Irifune, \& Torchio}]{morard_solving_2018}
Morard, G., Boccato, S., Rosa, A.~D., {et~al.} 2018, Geophysical Research Letters, 45, \dodoi{10.1029/2018GL079950}

\bibitem[{Murakami {et~al.}(2003)Murakami, Hirose, Ono, \& Ohishi}]{murakami_stability_2003}
Murakami, M., Hirose, K., Ono, S., \& Ohishi, Y. 2003, Geophysical Research Letters, 30, 2002GL016722, \dodoi{10.1029/2002GL016722}

\bibitem[{Musella {et~al.}(2019)Musella, Mazevet, \& Guyot}]{musella_physical_2019}
Musella, R., Mazevet, S., \& Guyot, F. 2019, Physical Review B, 99, 064110, \dodoi{10.1103/PhysRevB.99.064110}

\bibitem[{Müller {et~al.}(2020)Müller, Helled, \& Cumming}]{muller_challenge_2020}
Müller, S., Helled, R., \& Cumming, A. 2020, Astronomy \& Astrophysics, 638, A121, \dodoi{10.1051/0004-6361/201937376}

\bibitem[{Nakajima {et~al.}(2021)Nakajima, Golabek, Wünnemann, Rubie, Burger, Melosh, Jacobson, Manske, \& Hull}]{nakajima_scaling_2021}
Nakajima, M., Golabek, G.~J., Wünnemann, K., {et~al.} 2021, Earth and Planetary Science Letters, 568, 116983, \dodoi{10.1016/j.epsl.2021.116983}

\bibitem[{Nettelmann {et~al.}(2016)Nettelmann, Wang, Fortney, Hamel, Yellamilli, Bethkenhagen, \& Redmer}]{nettelmann_uranus_2016}
Nettelmann, N., Wang, K., Fortney, J., {et~al.} 2016, Icarus, 275, 107, \dodoi{10.1016/j.icarus.2016.04.008}

\bibitem[{Nishihara {et~al.}(2005)Nishihara, Nakayama, Takahashi, Iguchi, \& Funakoshi}]{nishihara_p-v-t_2005}
Nishihara, Y., Nakayama, K., Takahashi, E., Iguchi, T., \& Funakoshi, K.-i. 2005, Physics and Chemistry of Minerals, 31, 660, \dodoi{10.1007/s00269-004-0426-7}

\bibitem[{Niu {et~al.}(2015)Niu, Oganov, Chen, \& Li}]{niu_prediction_2015}
Niu, H., Oganov, A.~R., Chen, X.-Q., \& Li, D. 2015, Scientific Reports, 5, 18347, \dodoi{10.1038/srep18347}

\bibitem[{Nomura {et~al.}(2010)Nomura, Hirose, Sata, \& Ohishi}]{nomura_precise_2010}
Nomura, R., Hirose, K., Sata, N., \& Ohishi, Y. 2010, Physics of the Earth and Planetary Interiors, 183, 104, \dodoi{10.1016/j.pepi.2010.08.004}

\bibitem[{Oganov \& Ono(2004)}]{oganov_theoretical_2004}
Oganov, A.~R., \& Ono, S. 2004, Nature, 430, 445

\bibitem[{Ohnishi {et~al.}(2017)Ohnishi, Kuwayama, \& Inoue}]{ohnishi_melting_2017}
Ohnishi, S., Kuwayama, Y., \& Inoue, T. 2017, Physics and Chemistry of Minerals, 44, 445, \dodoi{10.1007/s00269-017-0871-8}

\bibitem[{Ohta {et~al.}(2016)Ohta, Kuwayama, Hirose, Shimizu, \& Ohishi}]{ohta_experimental_2016}
Ohta, K., Kuwayama, Y., Hirose, K., Shimizu, K., \& Ohishi, Y. 2016, Nature, 534, 95, \dodoi{10.1038/nature17957}

\bibitem[{Okuda {et~al.}(2024)Okuda, Hirose, Ohta, Kawaguchi-Imada, \& Oka}]{okuda_electrical_2024}
Okuda, Y., Hirose, K., Ohta, K., Kawaguchi-Imada, S., \& Oka, K. 2024, Geophysical Research Letters, 51, e2024GL109741, \dodoi{https://doi.org/10.1029/2024GL109741}

\bibitem[{Ono(2006)}]{ono_equation_2006}
Ono, S. 2006, American Mineralogist, 91, 475, \dodoi{10.2138/am.2006.2118}

\bibitem[{Ono {et~al.}(2002)Ono, Hirose, Murakami, \& Isshiki}]{ono_post-stishovite_2002}
Ono, S., Hirose, K., Murakami, M., \& Isshiki, M. 2002, Earth and Planetary Science Letters, 197, 187, \dodoi{10.1016/S0012-821X(02)00479-X}

\bibitem[{Ono {et~al.}(2017{\natexlab{a}})Ono, Kikegawa, \& Higo}]{ono_reaction_2017}
Ono, S., Kikegawa, T., \& Higo, Y. 2017{\natexlab{a}}, Physics and Chemistry of Minerals, 44, 425, \dodoi{10.1007/s00269-016-0869-7}

\bibitem[{Ono {et~al.}(2018)Ono, Kikegawa, \& Higo}]{ono_decomposition_2018}
---. 2018, American Mineralogist, 103, 1512, \dodoi{10.2138/am-2018-6313CCBY}

\bibitem[{Ono {et~al.}(2017{\natexlab{b}})Ono, Kikegawa, Higo, \& Tange}]{ono_precise_2017}
Ono, S., Kikegawa, T., Higo, Y., \& Tange, Y. 2017{\natexlab{b}}, Physics of the Earth and Planetary Interiors, 264, 1, \dodoi{10.1016/j.pepi.2017.01.003}

\bibitem[{Ono \& Oganov(2005)}]{ono_situ_2005}
Ono, S., \& Oganov, A. 2005, Earth and Planetary Science Letters, 236, 914, \dodoi{10.1016/j.epsl.2005.06.001}

\bibitem[{Ono {et~al.}(2001)Ono, Katsura, Ito, Kanzaki, Yoneda, Walter, Urakawa, Utsumi, \& Funakoshi}]{ono_situ_2001}
Ono, S., Katsura, T., Ito, E., {et~al.} 2001, Geophysical Research Letters, 28, 835, \dodoi{10.1029/1999GL008446}

\bibitem[{Ozawa {et~al.}(2018)Ozawa, Anzai, Hirose, Sinmyo, \& Tateno}]{ozawa_experimental_2018}
Ozawa, K., Anzai, M., Hirose, K., Sinmyo, R., \& Tateno, S. 2018, Geophysical Research Letters, 45, 9552, \dodoi{10.1029/2018GL079313}

\bibitem[{Pacalo \& Gasparik(1990)}]{pacalo_reversals_1990}
Pacalo, R. E.~G., \& Gasparik, T. 1990, Journal of Geophysical Research: Solid Earth, 95, 15853, \dodoi{10.1029/JB095iB10p15853}

\bibitem[{Pedregosa {et~al.}(2011)Pedregosa, Varoquaux, Gramfort, Michel, Thirion, Grisel, Blondel, Prettenhofer, Weiss, Dubourg, Vanderplas, Passos, Cournapeau, Brucher, Perrot, \& Duchesnay}]{pedregosa_scikit-learn_2011}
Pedregosa, F., Varoquaux, G., Gramfort, A., {et~al.} 2011, {J}ournal of {M}achine {L}earning {R}esearch, 12, 2825

\bibitem[{Perryman(2018)}]{perryman_exoplanet_2018}
Perryman, M. 2018, The {Exoplanet} {Handbook}, 2nd edn. (Cambridge University Press), \dodoi{10.1017/9781108304160}

\bibitem[{Petrenko \& Whitworth(2002)}]{petrenko_physics_2002}
Petrenko, V.~F., \& Whitworth, R.~W. 2002, Physics of {Ice} (Oxford University Press), \dodoi{10.1093/acprof:oso/9780198518945.001.0001}

\bibitem[{Pierru {et~al.}(2022)Pierru, Pison, Mathieu, Gardés, Garbarino, Mezouar, Hennet, \& Andrault}]{PIERRU2022117770}
Pierru, R., Pison, L., Mathieu, A., {et~al.} 2022, Earth and Planetary Science Letters, 595, 117770, \dodoi{https://doi.org/10.1016/j.epsl.2022.117770}

\bibitem[{Pigott {et~al.}(2015)Pigott, Ditmer, Fischer, Reaman, Hrubiak, Meng, Davis, \& Panero}]{pigott_highpressure_2015}
Pigott, J.~S., Ditmer, D.~A., Fischer, R.~A., {et~al.} 2015, Geophysical Research Letters, 42, \dodoi{10.1002/2015GL066577}

\bibitem[{Pommier(2018)}]{pommier_influence_2018}
Pommier, A. 2018, Earth and Planetary Science Letters, 496, 37, \dodoi{10.1016/j.epsl.2018.05.032}

\bibitem[{Prakapenka {et~al.}(2004)Prakapenka, Shen, Dubrovinsky, Rivers, \& Sutton}]{prakapenka_high_2004}
Prakapenka, V., Shen, G., Dubrovinsky, L., Rivers, M., \& Sutton, S. 2004, Journal of Physics and Chemistry of Solids, 65, 1537, \dodoi{10.1016/j.jpcs.2003.12.019}

\bibitem[{Prakapenka {et~al.}(2021)Prakapenka, Holtgrewe, Lobanov, \& Goncharov}]{prakapenka_structure_2021}
Prakapenka, V.~B., Holtgrewe, N., Lobanov, S.~S., \& Goncharov, A.~F. 2021, Nature Physics, 17, 1233, \dodoi{10.1038/s41567-021-01351-8}

\bibitem[{Presnall \& Gasparik(1990)}]{presnall_melting_1990}
Presnall, D.~C., \& Gasparik, T. 1990, Journal of Geophysical Research: Solid Earth, 95, 15771, \dodoi{10.1029/JB095iB10p15771}

\bibitem[{Riley(1966)}]{riley_determination_1966}
Riley, B. 1966, Revue internationale des hautes températures et des réfractaires, 3, 327

\bibitem[{Ronchi \& Sheindlin(2001)}]{ronchi_melting_2001}
Ronchi, C., \& Sheindlin, M. 2001, Journal of Applied Physics, 90, 3325, \dodoi{10.1063/1.1398069}

\bibitem[{Root {et~al.}(2015)Root, Shulenburger, Lemke, Dolan, Mattsson, \& Desjarlais}]{root_shock_2015}
Root, S., Shulenburger, L., Lemke, R.~W., {et~al.} 2015, Physical Review Letters, 115, 198501, \dodoi{10.1103/PhysRevLett.115.198501}

\bibitem[{Rutter {et~al.}(2002)Rutter, Secco, Liu, Uchida, Rivers, Sutton, \& Wang}]{rutter_viscosity_2002}
Rutter, M.~D., Secco, R.~A., Liu, H., {et~al.} 2002, Physical Review B, 66, 060102, \dodoi{10.1103/PhysRevB.66.060102}

\bibitem[{Rymer {et~al.}(2021)Rymer, Clyde, \& Runyon}]{rymer_neptune_2021}
Rymer, A., Clyde, B., \& Runyon, K. 2021, Neptune {Odyssey}: {Mission} to the {Neptune}-{Triron} {System}, Tech. rep., National Aeronautics and Space Administration (NASA).
\newblock \url{https://smd-cms.nasa.gov/wp-content/uploads/2023/05/NeptuneOdyssey.pdf}

\bibitem[{Sawamoto(1987)}]{manghnani_phase_1987}
Sawamoto, H. 1987, in Geophysical {Monograph} {Series}, ed. M.~H. Manghnani \& Y.~Syono, Vol.~39 (Washington, D. C.: American Geophysical Union), 209--219, \dodoi{10.1029/GM039p0209}

\bibitem[{Saxena {et~al.}(1994)Saxena, Shen, \& Lazor}]{saxena_temperatures_1994}
Saxena, S.~K., Shen, G., \& Lazor, P. 1994, Science, 264

\bibitem[{Schubert {et~al.}(2001)Schubert, Turcotte, \& Olson}]{schubert_mantle_2001}
Schubert, G., Turcotte, D.~L., \& Olson, P. 2001, Mantle {Convection} in the {Earth} and {Planets}, 1st edn. (Cambridge University Press), \dodoi{10.1017/CBO9780511612879}

\bibitem[{Seager \& Dotson(2010)}]{seager_exoplanets_2010}
Seager, S., \& Dotson, R. 2010, Exoplanets, The {University} of {Arizona} space science series (Tucson: University of Arizona press)

\bibitem[{Secco(2017)}]{secco_thermal_2017}
Secco, R.~A. 2017, Physics of the Earth and Planetary Interiors, 265, 23, \dodoi{10.1016/j.pepi.2017.01.005}

\bibitem[{Secco \& Schloessin(1989)}]{secco_electrical_1989}
Secco, R.~A., \& Schloessin, H.~H. 1989, Journal of Geophysical Research: Solid Earth, 94, 5887, \dodoi{10.1029/JB094iB05p05887}

\bibitem[{Serghiou {et~al.}(1995)Serghiou, Zerr, Chudinovskikh, \& Boehler}]{serghiou_coesitestishovite_1995}
Serghiou, G., Zerr, A., Chudinovskikh, L., \& Boehler, R. 1995, Geophysical Research Letters, 22, 441, \dodoi{10.1029/94GL02692}

\bibitem[{Shahnas \& Pysklywec(2021)}]{shahnas_focused_2021}
Shahnas, M.~H., \& Pysklywec, R.~N. 2021, Geochemistry, Geophysics, Geosystems, 22, e2021GC009910, \dodoi{10.1029/2021GC009910}

\bibitem[{Shahnas {et~al.}(2018)Shahnas, Pysklywec, \& Yuen}]{shahnas_penetrative_2018}
Shahnas, M.~H., Pysklywec, R.~N., \& Yuen, D.~A. 2018, Journal of Geophysical Research: Planets, 123, 2162, \dodoi{10.1029/2018JE005633}

\bibitem[{Shen \& Lazor(1995)}]{shen_measurement_1995}
Shen, G., \& Lazor, P. 1995, Journal of Geophysical Research: Solid Earth, 100, 17699, \dodoi{10.1029/95JB01864}

\bibitem[{Shen {et~al.}(1998)Shen, Mao, Hemley, Duffy, \& Rivers}]{shen_melting_1998}
Shen, G., Mao, H., Hemley, R.~J., Duffy, T.~S., \& Rivers, M.~L. 1998, Geophysical Research Letters, 25, 373, \dodoi{10.1029/97GL03776}

\bibitem[{Shen {et~al.}(2004)Shen, Prakapenka, Rivers, \& Sutton}]{shen_structure_2004}
Shen, G., Prakapenka, V.~B., Rivers, M.~L., \& Sutton, S.~R. 2004, Physical Review Letters, 92, 185701, \dodoi{10.1103/PhysRevLett.92.185701}

\bibitem[{Shieh {et~al.}(2005)Shieh, Duffy, \& Shen}]{shieh_x-ray_2005}
Shieh, S., Duffy, T., \& Shen, G. 2005, Earth and Planetary Science Letters, 235, 273, \dodoi{10.1016/j.epsl.2005.04.004}

\bibitem[{Shinmei {et~al.}(1999)Shinmei, Tomioka, Fujino, Kuroda, \& Irifune}]{shinmei_situ_1999}
Shinmei, T., Tomioka, N., Fujino, K., Kuroda, K., \& Irifune, T. 1999, American Mineralogist, 84, 1588, \dodoi{10.2138/am-1999-1012}

\bibitem[{Silber {et~al.}(2018)Silber, Secco, Yong, \& Littleton}]{silber_electrical_2018}
Silber, R.~E., Secco, R.~A., Yong, W., \& Littleton, J. A.~H. 2018, Scientific Reports, 8, 10758, \dodoi{10.1038/s41598-018-28921-w}

\bibitem[{Simon {et~al.}(2021)Simon, Nimmo, \& Anderson}]{simon_uranus_2021}
Simon, A., Nimmo, F., \& Anderson, R.~C. 2021, Uranus {Orbiter} and {Probe}: {Journey} to an {Ice} {Giant} {System}, Tech. rep., National Aeronautics and Space Administration (NASA).
\newblock \url{https://smd-cms.nasa.gov/wp-content/uploads/2023/10/uranus-orbiter-and-probe.pdf}

\bibitem[{Simon \& Glatzel(1929)}]{simon_bemerkungen_1929}
Simon, F., \& Glatzel, G. 1929, Zeitschrift für anorganische und allgemeine Chemie, 178, 309, \dodoi{10.1002/zaac.19291780123}

\bibitem[{Sinmyo {et~al.}(2019)Sinmyo, Hirose, \& Ohishi}]{sinmyo_melting_2019}
Sinmyo, R., Hirose, K., \& Ohishi, Y. 2019, Earth and Planetary Science Letters, 510, 45, \dodoi{10.1016/j.epsl.2019.01.006}

\bibitem[{Sola \& Alfè(2009)}]{sola_melting_2009}
Sola, E., \& Alfè, D. 2009, Physical Review Letters, 103, 078501, \dodoi{10.1103/PhysRevLett.103.078501}

\bibitem[{Solomatov(2007)}]{solomatov_magma_2007}
Solomatov, V. 2007, in Treatise on {Geophysics} (Elsevier), 91--119, \dodoi{10.1016/B978-044452748-6.00141-3}

\bibitem[{Soubiran \& Militzer(2020)}]{soubiran_anharmonicity_2020}
Soubiran, F., \& Militzer, B. 2020, Physical Review Letters, 125, 175701, \dodoi{10.1103/PhysRevLett.125.175701}

\bibitem[{Speziale {et~al.}(2001)Speziale, Zha, Duffy, Hemley, \& Mao}]{speziale_quasihydrostatic_2001}
Speziale, S., Zha, C., Duffy, T.~S., Hemley, R.~J., \& Mao, H. 2001, Journal of Geophysical Research: Solid Earth, 106, 515, \dodoi{10.1029/2000JB900318}

\bibitem[{Stevenson(1982)}]{stevenson_formation_1982}
Stevenson, D. 1982, Planetary and Space Science, 30, 755, \dodoi{10.1016/0032-0633(82)90108-8}

\bibitem[{Stevenson(1990)}]{newsom_fluid_1990}
Stevenson, D.~J. 1990, in Origin of the {Earth}, ed. H.~E. Newsom \& J.~H. Jones (Oxford University PressNew York, NY), 231--249, \dodoi{10.1093/oso/9780195066197.003.0014}

\bibitem[{Stixrude(2012)}]{stixrude_structure_2012}
Stixrude, L. 2012, Physical Review Letters, 108, 055505, \dodoi{10.1103/PhysRevLett.108.055505}

\bibitem[{Stixrude(2014)}]{stixrude_melting_2014}
---. 2014, Philosophical Transactions of the Royal Society A: Mathematical, Physical and Engineering Sciences, 372, 20130076, \dodoi{10.1098/rsta.2013.0076}

\bibitem[{Stixrude {et~al.}(2021)Stixrude, Baroni, \& Grasselli}]{stixrude_thermal_2021-1}
Stixrude, L., Baroni, S., \& Grasselli, F. 2021, The Planetary Science Journal, 2, 222, \dodoi{10.3847/PSJ/ac2a47}

\bibitem[{Stixrude \& Karki(2005)}]{stixrude_structure_2005}
Stixrude, L., \& Karki, B. 2005, Science, 310, 297, \dodoi{10.1126/science.1116952}

\bibitem[{Stixrude \& Lithgow-Bertelloni(2005)}]{stixrude_thermodynamics_2005}
Stixrude, L., \& Lithgow-Bertelloni, C. 2005, Geophysical Journal International, 162, 610, \dodoi{10.1111/j.1365-246X.2005.02642.x}

\bibitem[{Stixrude \& Lithgow-Bertelloni(2011)}]{stixrude_thermodynamics_2011}
---. 2011, Geophysical Journal International, 184, 1180, \dodoi{10.1111/j.1365-246X.2010.04890.x}

\bibitem[{Stixrude \& Lithgow-Bertelloni(2012)}]{stixrude_geophysics_2012}
---. 2012, Annual Review of Earth and Planetary Sciences, 40, 569, \dodoi{10.1146/annurev.earth.36.031207.124244}

\bibitem[{Stixrude \& Lithgow-Bertelloni(2021)}]{stixrude_thermal_2021}
---. 2021, Geophysical Journal International, 228, 1119, \dodoi{10.1093/gji/ggab394}

\bibitem[{Stixrude \& Lithgow-Bertelloni(2024)}]{stixrude_thermodynamics_2024}
---. 2024, Geophysical Journal International, 237, 1699, \dodoi{10.1093/gji/ggae126}

\bibitem[{Strong(1959)}]{strong_experimental_1959}
Strong, H.~M. 1959, Journal of Geophysical Research, 64, 653, \dodoi{10.1029/JZ064i006p00653}

\bibitem[{Strong {et~al.}(1973)Strong, Tuft, \& Hanneman}]{strong_iron_1973}
Strong, H.~M., Tuft, R.~E., \& Hanneman, R.~E. 1973, Metallurgical Transactions

\bibitem[{Suito(1977)}]{suito_phase_1977}
Suito, K. 1977, in High-{Pressure} {Research}: {Applications} in {Geophysics} (New York: Academic Press), 255--266

\bibitem[{Sun {et~al.}(2019)Sun, Shi, Mao, Zhou, \& Prakapenka}]{sun_high_2019}
Sun, N., Shi, W., Mao, Z., Zhou, C., \& Prakapenka, V.~B. 2019, Journal of Geophysical Research: Solid Earth, 124, 12620, \dodoi{10.1029/2019JB017853}

\bibitem[{Sun {et~al.}(2023)Sun, Mendelev, Zhang, Liu, Da, Wang, Wentzcovitch, \& Ho}]{sun_ab_2023}
Sun, Y., Mendelev, M.~I., Zhang, F., {et~al.} 2023, Geophysical Research Letters, 50, e2022GL102447, \dodoi{10.1029/2022GL102447}

\bibitem[{Swain {et~al.}(2024)Swain, Hasegawa, Thorngren, \& Roudier}]{swain_planet_2024}
Swain, M.~R., Hasegawa, Y., Thorngren, D.~P., \& Roudier, G.~M. 2024, Space Science Reviews, 220, 61, \dodoi{10.1007/s11214-024-01098-7}

\bibitem[{Tackley {et~al.}(2013)Tackley, Ammann, Brodholt, Dobson, \& Valencia}]{tackley_mantle_2013}
Tackley, P., Ammann, M., Brodholt, J., Dobson, D., \& Valencia, D. 2013, Icarus, 225, 50, \dodoi{10.1016/j.icarus.2013.03.013}

\bibitem[{Tange {et~al.}(2012)Tange, Kuwayama, Irifune, Funakoshi, \& Ohishi}]{tange_pvt_2012}
Tange, Y., Kuwayama, Y., Irifune, T., Funakoshi, K., \& Ohishi, Y. 2012, Journal of Geophysical Research: Solid Earth, 117, 2011JB008988, \dodoi{10.1029/2011JB008988}

\bibitem[{Taniuchi \& Tsuchiya(2018)}]{taniuchi_melting_2018}
Taniuchi, T., \& Tsuchiya, T. 2018, Journal of Physics: Condensed Matter, 30, 114003, \dodoi{10.1088/1361-648X/aaac96}

\bibitem[{Tateno {et~al.}(2014)Tateno, Hirose, \& Ohishi}]{tateno_melting_2014}
Tateno, S., Hirose, K., \& Ohishi, Y. 2014, Journal of Geophysical Research: Solid Earth, 119, 4684, \dodoi{10.1002/2013JB010616}

\bibitem[{Tateno {et~al.}(2010)Tateno, Hirose, Ohishi, \& Tatsumi}]{tateno_structure_2010}
Tateno, S., Hirose, K., Ohishi, Y., \& Tatsumi, Y. 2010, Science, 330, 359, \dodoi{10.1126/science.1194662}

\bibitem[{Tateno {et~al.}(2009)Tateno, Hirose, Sata, \& Ohishi}]{tateno_determination_2009}
Tateno, S., Hirose, K., Sata, N., \& Ohishi, Y. 2009, Earth and Planetary Science Letters, 277, 130, \dodoi{10.1016/j.epsl.2008.10.004}

\bibitem[{Terasaki {et~al.}(2002)Terasaki, Kato, Urakawa, Funakoshi, Sato, Suzuki, \& Okada}]{terasaki_viscosity_2002}
Terasaki, H., Kato, T., Urakawa, S., {et~al.} 2002, Geophysical Research Letters, 29, \dodoi{10.1029/2001GL014321}

\bibitem[{Thomson {et~al.}(2014)Thomson, Walter, Lord, \& Kohn}]{thomson_experimental_2014}
Thomson, A.~R., Walter, M.~J., Lord, O.~T., \& Kohn, S.~C. 2014, American Mineralogist, 99, 1544, \dodoi{10.2138/am.2014.4735}

\bibitem[{Tonks \& Melosh(1990)}]{newsom_physics_1990}
Tonks, W.~B., \& Melosh, H.~J. 1990, in Origin of the {Earth}, ed. H.~E. Newsom \& J.~H. Jones (Oxford University PressNew York, NY), 151--174, \dodoi{10.1093/oso/9780195066197.003.0010}

\bibitem[{Tonks \& Melosh(1993)}]{tonks_magma_1993}
---. 1993, Journal of Geophysical Research: Planets, 98, 5319, \dodoi{10.1029/92JE02726}

\bibitem[{Tsuchiya \& Tsuchiya(2011)}]{tsuchiya_prediction_2011}
Tsuchiya, T., \& Tsuchiya, J. 2011, Proceedings of the National Academy of Sciences, 108, 1252, \dodoi{10.1073/pnas.1013594108}

\bibitem[{Umemoto {et~al.}(2017)Umemoto, Wentzcovitch, Wu, Ji, Wang, \& Ho}]{umemoto_phase_2017}
Umemoto, K., Wentzcovitch, R.~M., Wu, S., {et~al.} 2017, Earth and Planetary Science Letters, 478, 40, \dodoi{10.1016/j.epsl.2017.08.032}

\bibitem[{Usui \& Tsuchiya(2010)}]{usui_ab_2010}
Usui, Y., \& Tsuchiya, T. 2010, Journal of Earth Science, 21, 801, \dodoi{10.1007/s12583-010-0126-9}

\bibitem[{Utsumi {et~al.}(1998)Utsumi, Weidner, \& Liebermann}]{manghnani_volume_1998}
Utsumi, W., Weidner, D.~J., \& Liebermann, R.~C. 1998, in Geophysical {Monograph} {Series}, ed. M.~H. Manghnani \& T.~Yagi, Vol. 101 (Washington, D. C.: American Geophysical Union), 327--333, \dodoi{10.1029/GM101p0327}

\bibitem[{Vazan {et~al.}(2022)Vazan, Sari, \& Kessel}]{vazan_new_2022}
Vazan, A., Sari, R., \& Kessel, R. 2022, The Astrophysical Journal, 926, 150, \dodoi{10.3847/1538-4357/ac458c}

\bibitem[{Wahl \& Militzer(2015)}]{wahl_high-temperature_2015}
Wahl, S.~M., \& Militzer, B. 2015, Earth and Planetary Science Letters, 410, 25, \dodoi{10.1016/j.epsl.2014.11.014}

\bibitem[{Wang {et~al.}(2012)Wang, Tange, Irifune, \& Funakoshi}]{wang_pvt_2012}
Wang, F., Tange, Y., Irifune, T., \& Funakoshi, K. 2012, Journal of Geophysical Research: Solid Earth, 117, 2011JB009100, \dodoi{10.1029/2011JB009100}

\bibitem[{Weck {et~al.}(2022)Weck, Queyroux, Ninet, Datchi, Mezouar, \& Loubeyre}]{weck_evidence_2022}
Weck, G., Queyroux, J.-A., Ninet, S., {et~al.} 2022, Physical Review Letters, 128, 165701, \dodoi{10.1103/PhysRevLett.128.165701}

\bibitem[{Wentzcovitch \& Stixrude(2010)}]{wentzcovitch_theoretical_2010}
Wentzcovitch, R.~M., \& Stixrude, L., eds. 2010, Theoretical and {Computational} {Methods} in {Mineral} {Physics}: {Geophysical} {Applications} (De Gruyter), \dodoi{10.1515/9781501508448}

\bibitem[{Wicks {et~al.}(2024)Wicks, Singh, Millot, Fratanduono, Coppari, Gorman, Ye, Rygg, Hari, Eggert, Duffy, \& Smith}]{wicks_b1-b2_2024}
Wicks, J.~K., Singh, S., Millot, M., {et~al.} 2024, Science Advances, 10, eadk0306, \dodoi{10.1126/sciadv.adk0306}

\bibitem[{Williams {et~al.}(1987)Williams, Jeanloz, Bass, Svendsen, \& Ahrens}]{williams_melting_1987}
Williams, Q., Jeanloz, R., Bass, J., Svendsen, B., \& Ahrens, T.~J. 1987, Science

\bibitem[{Wisesa {et~al.}(2023)Wisesa, Andolina, \& Saidi}]{wisesa_machine-learning_2023}
Wisesa, P., Andolina, C.~M., \& Saidi, W.~A. 2023, The Journal of Physical Chemistry Letters, 14, 8741, \dodoi{10.1021/acs.jpclett.3c02424}

\bibitem[{Yagi \& Akimoto(1976)}]{yagi_direct_1976}
Yagi, T., \& Akimoto, S.-I. 1976, Tectonophysics, 35, 259, \dodoi{10.1016/0040-1951(76)90042-1}

\bibitem[{Yamazaki {et~al.}(2014)Yamazaki, Ito, Yoshino, Tsujino, Yoneda, Guo, Xu, Higo, \& Funakoshi}]{yamazaki_over_2014}
Yamazaki, D., Ito, E., Yoshino, T., {et~al.} 2014, Physics of the Earth and Planetary Interiors, 228, 262, \dodoi{10.1016/j.pepi.2014.01.013}

\bibitem[{Yao {et~al.}(2021)Yao, Frost, \& Steinle‐Neumann}]{yao_lower_2021}
Yao, J., Frost, D.~J., \& Steinle‐Neumann, G. 2021, Journal of Geophysical Research: Solid Earth, 126, e2021JB022568, \dodoi{10.1029/2021JB022568}

\bibitem[{Ye {et~al.}(2017)Ye, Prakapenka, Meng, \& Shim}]{ye_intercomparison_2017}
Ye, Y., Prakapenka, V., Meng, Y., \& Shim, S. 2017, Journal of Geophysical Research: Solid Earth, 122, 3450, \dodoi{10.1002/2016JB013811}

\bibitem[{Yong {et~al.}(2019)Yong, Secco, Littleton, \& Silber}]{yong_iron_2019}
Yong, W., Secco, R.~A., Littleton, J. A.~H., \& Silber, R.~E. 2019, Geophysical Research Letters, 46, 11065, \dodoi{10.1029/2019GL084485}

\bibitem[{Yoo {et~al.}(1995)Yoo, Akella, Campbell, Mao, \& Hemley}]{yoo_phase_1995}
Yoo, C.~S., Akella, J., Campbell, A.~J., Mao, H.~K., \& Hemley, R.~J. 1995, Science, 270, 1473

\bibitem[{Yoshiasa {et~al.}(2013)Yoshiasa, Nakatsuka, Okube, \& Katsura}]{yoshiasa_single-crystal_2013}
Yoshiasa, A., Nakatsuka, A., Okube, M., \& Katsura, T. 2013, Acta Crystallographica Section B Structural Science, Crystal Engineering and Materials, 69, 541, \dodoi{10.1107/S2052519213028248}

\bibitem[{Young(1991)}]{young_phase_1991}
Young, D.~A. 1991, Phase diagrams of the elements (Berkeley, Calif.: Univ. of California Press)

\bibitem[{Zerr \& Boehler(1993)}]{zerr_melting_1993}
Zerr, A., \& Boehler, R. 1993, Science, 262, 553, \dodoi{10.1126/science.262.5133.553}

\bibitem[{Zerr \& Boehler(1994)}]{zerr_constraints_1994}
---. 1994, {N}ature, 371

\bibitem[{Zhang {et~al.}(2016)Zhang, Jackson, Zhao, Sturhahn, Alp, Hu, Toellner, Murphy, \& Prakapenka}]{zhang_temperature_2016}
Zhang, D., Jackson, J.~M., Zhao, J., {et~al.} 2016, Earth and Planetary Science Letters, 447, 72, \dodoi{10.1016/j.epsl.2016.04.026}

\bibitem[{Zhang {et~al.}(1996)Zhang, Li, Utsumi, \& Liebermann}]{zhang_situ_1996}
Zhang, J., Li, B., Utsumi, W., \& Liebermann, R. 1996, Physics and Chemistry of Minerals, 23, \dodoi{10.1007/BF00202987}

\bibitem[{Zhang {et~al.}(1993)Zhang, Liebermann, Gasparik, Herzberg, \& Fei}]{zhang_melting_1993}
Zhang, J., Liebermann, R.~C., Gasparik, T., Herzberg, C.~T., \& Fei, Y. 1993, Journal of Geophysical Research: Solid Earth, 98, 19785, \dodoi{10.1029/93JB02218}

\bibitem[{Zhang \& Fei(2008)}]{zhang_melting_2008}
Zhang, L., \& Fei, Y. 2008, Geophysical Research Letters, 35, 2008GL034585, \dodoi{10.1029/2008GL034585}

\bibitem[{Zhang {et~al.}(2020)Zhang, Hou, Liu, Zhang, Prakapenka, Greenberg, Fei, Cohen, \& Lin}]{zhang_reconciliation_2020}
Zhang, Y., Hou, M., Liu, G., {et~al.} 2020, Physical Review Letters, 125, 078501, \dodoi{10.1103/PhysRevLett.125.078501}

\bibitem[{Zhou {et~al.}(2021)Zhou, Gréaux, Liu, Higo, Arimoto, \& Irifune}]{zhou_sound_2021}
Zhou, C., Gréaux, S., Liu, Z., {et~al.} 2021, Geophysical Research Letters, 48, e2021GL093499, \dodoi{10.1029/2021GL093499}

\bibitem[{Zurkowski {et~al.}(2022)Zurkowski, Yang, Chariton, Prakapenka, \& Fei}]{zurkowski_synthesis_2022}
Zurkowski, C.~C., Yang, J., Chariton, S., Prakapenka, V.~B., \& Fei, Y. 2022, Journal of Geophysical Research: Planets, 127, e2022JE007344, \dodoi{10.1029/2022JE007344}

\end{thebibliography}
\bibliographystyle{aasjournal}


\end{CJK*}
\end{document}